\documentclass[twocolumn,trackchanges]{aastex63}

\received{2021 May 6}
\revised{2022 January 21}
\accepted{}
\submitjournal{ApJ}

\shorttitle{Radio constraints on Coma with reacceleration}
\shortauthors{Nishiwaki, Asano \& Murase}

\begin{document}

\title{Particle Reacceleration by Turbulence and Radio Constraints on Multi-Messenger High-Energy Emission from the Coma Cluster}

\author{Kosuke Nishiwaki}
\affiliation{Institute for Cosmic Ray Research, The University of Tokyo, 5-1-5 Kashiwanoha, Kashiwa, Chiba 277-8582, Japan}

\author{Katsuaki Asano}
\affiliation{Institute for Cosmic Ray Research, The University of Tokyo, 5-1-5 Kashiwanoha, Kashiwa, Chiba 277-8582, Japan}


\author{Kohta Murase}
\affiliation{Department of Physics, The Pennsylvania State University, University Park, Pennsylvania 16802, USA}
\affiliation{Department of Astronomy \& Astrophysics, The Pennsylvania State University, University Park, Pennsylvania 16802, USA}
\affiliation{Center for Multimessenger Astrophysics, Institute for Gravitation and the Cosmos, The Pennsylvania State University, University Park,Pennsylvania 16802, USA}
\affiliation{Yukawa Institute for Theoretical Physics, Kyoto University, Kyoto 606-8502, Japan}




\begin{abstract}
Galaxy clusters are considered to be gigantic reservoirs of cosmic rays (CRs).  
Some of the clusters are found with extended radio emission, which provides evidence for the existence of magnetic fields and CR electrons in the intra-cluster medium (ICM). 
The mechanism of radio halo (RH) emission is still under debate, and it has been believed that turbulent reacceleration plays an important role. 
In this paper, we study the reacceleration of CR protons and electrons in detail by numerically solving the Fokker-Planck equation, and show how radio and gamma-ray observations can be used to constrain CR distributions and resulting high-energy emission for the Coma cluster. We take into account the radial diffusion of CRs and follow the time evolution of their one-dimensional distribution, by which we investigate the radial profile of the CR injection that is consistent with the observed RH surface brightness. 
We find that the required injection profile is non-trivial, depending on whether CR electrons have the primary or secondary origin. 
Although the secondary CR electron scenario predicts larger gamma-ray and neutrino fluxes, it is in tension with the observed RH spectrum for hard injection indexes, $\alpha<2.45$. This tension is relaxed if the turbulent diffusion of CRs is much less efficient than the fiducial model, or the reacceleration is more efficient for lower energy CRs. In both secondary and primary scenario, we find that galaxy clusters can make a sizable contribution to the all-sky neutrino intensity if the CR energy spectrum is nearly flat. 
\end{abstract}

\keywords{Galaxy clusters (584): Coma cluster (270) - Particle astrophysics (96): Cosmic rays (329) - Neutrino astronomy (1100)}

\section{Introduction\label{sec:intro}} 
The detection of cosmic background radiation of the high-energy neutrinos by the IceCube neutrino observatory is an observational milestone of high-energy astrophysics \citep{2013PhRvL.111b1103A,2013Sci...342E...1I}. 
The observed intensities around $\sim100$ TeV to $\sim1$ PeV are consistent with the Waxman-Bahcall bound \citep{Waxman1999}, which may indicate that high-energy neutrinos and ultrahigh-energy cosmic rays (UHECRs) come from the same source class \citep{Yoshida:2020div}. The majority of IceCube neutrinos is still unknown, but such neutrinos should be produced by the hadronic interactions like $pp$ or $p\gamma$ collisions of relativistic protons. Many candidate sources have been proposed, including starburst galaxies  \citep[e.g.,][]{2006JCAP...05..003L,Murase2013,Tamborra2014,2015ApJ...806...24S} and galaxy clusters  \citep[e.g.,][]{Berezinsky_1997,Murase2008,2009ApJ...707..370K,Murase2013,Zandanel2015,Fang2016,Hussain:2021dqp}.

Galaxy clusters are the latest and largest cosmological structure in the universe. A fraction of gravitational energy dissipated during the structure formation can be expended on accelerating CRs via shocks and turbulence \citep[e.g., ][]{1998A&A...332..395E,2003ApJ...584..190F,Brunetti2007}. Galaxy clusters are regarded as ``cosmic-ray reservoirs" \citep[e.g.,][]{Murase2013,2019SSRv..215...14B} since they can confine CRs ions up to cosmological time with their large volumes and turbulent magnetic fields. 
Cosmic-ray protons (CRPs) accumulated in the intra-cluster medium (ICM) undergo inelastic $pp$ collisions between thermal protons, which produce charged and neutral pions. 
Secondary particles including gamma-ray photons, neutrinos, and cosmic-ray electrons/positrons (CREs) are produced as decay products of those pions. \par
Radio observations have detected diffuse synchrotron emission from many clusters. Some are in the form of giant radio haloes (RHs), roundish emission extended over the X-ray emitting regions, and some others are radio relics, elongated emission often found in peripheral regions \citep[see][for an observational review]{2019SSRv..215...16V}. The large extension of those radio structures is a major challenge for the theoretical modeling, because the cooling time of radio-emitting CREs is far shorter than the time required to diffuse across the emission region. That naturally requires {\it in situ} injection or acceleration of CREs at the emission region \citep[see][for a theoretical review]{Brunetti2014}.
There are two possibilities for the origin of CREs in the ICM. One is the secondary origin, where CREs are born as secondaries produced via inelastic $pp$ collisions \citep[e.g.,][]{1980ApJ...239L..93D,Blasi1999,KW09}. The other is the primary origin, i.e., CREs are injected from the same sources as CRPs. The former scenario naturally explains the extension of RHs, since parent CRPs can diffuse over the halo volume until they collide with thermal protons. \par

The physical origin of primary CRs is still an open question, but the fact that diffuse radio emission is usually found in merging systems suggests the possible connection between the structure formation and CR acceleration \citep[e.g.,][]{2001A&A...369..441G,2007A&A...463..937V,Cassano2010,2013A&A...557A..99K}. The shock waves formed through the merger of clusters and the mass accretion could accelerate CRs through the first-order Fermi acceleration process \citep[e.g.,][]{2012ApJ...756...97K,Ryu_2019}. Internal sources such as ordinary galaxies, galaxy mergers, and active galactic nuclei (AGNs) are also considered to be the sources of CRs \citep[e.g.,][]{1997ApJ...477..560E,Berezinsky_1997,Kashiyama:2014rza,Yuan:2017dle}. 
In the accretion/merger shock scenario, the contribution from massive clusters at low redshifts is expected to be dominant, while in the internal accelerator scenario the contribution from low-mass clusters including high-redshift ones is important \citep{Murase2016,2018NatPh..14..396F}. 

The most plausible origin of RHs is the reacceleration of seed CREs. In the so-called turbulent reacceleration scenario, stochastic interactions between CREs and turbulence caused by the merger of clusters accelerate seed CREs up to $\sim$ GeV energies. The interactions between particles and waves that transfer energies from the turbulence to particles in the ICM have been studied in detail \citep[e.g.,][]{PhysRevLett.89.281102,Brunetti2007}. Alfv\'{e}nic turbulence exhibits the anisotropic cascade that makes the interaction between particles inefficient at smaller scales \citep{Goldreich1995,PhysRevLett.89.281102}, so a resonant interaction called transit-time damping (TTD) with isotropic fast modes is often considered as the mechanism of the reacceleration \citep[e.g.,][]{Brunetti2011,2019ApJ...877...71T}. \par
This scenario can reproduce various observational features of RHs. For example, it predicts that the lifetime of RHs is about $\sim100$ Myr, which can explain the bi-modality in the radio--X-ray luminosity relation \citep{2005MNRAS.357.1313C,2015A&A...580A..97C}. This timescale may correspond to the turbulence surviving timescale after the cluster merger. 
That can also explain the apparent break feature appearing in the spectrum of the Coma RH \citep[e.g.,][]{Pizzo2010,Brunetti2013} by the balance between radiative cooling and the reacceleration of the CREs around $\sim$ GeV. \par
It is also notable that gamma-ray observations by the {\it Fermi} satellite with its Large Area Telescope (LAT) give stringent constraints on the density of CRPs in the Coma cluster \citep[e.g.,][]{Ackermann2016}. \citet{2018PhRvD..98f3006X} reported the first detection of an extended gamma-ray source in the direction of the Coma with an analysis of {\it Fermi} data. More recently, the existence of a gamma-ray source, 4FGL~J1256.9+2736, is indicated in the updated 4GFL catalogue \citep{2020ApJS..247...33A,2020arXiv200511208B}. \citet{2021A&A...648A..60A} also found  a significant signal and discussed the CRP content in the ICM and its possible connection to the radio emission. \par

A number of theoretical works have discussed the origin of CREs in the Coma cluster \citep[e.g.,][]{1987AA...182...21S,1993ApJ...406..399G,Blasi1999,2002ApJ...577..658O,KW09,2014MNRAS.438..124Z,Brunetti2017,Pinzke2017}. The ratio between primary and secondary CREs in the seed population for the reacceleration was discussed in \citep[e.g.,][]{Brunetti2017,Pinzke2017,2021A&A...648A..60A}, but it is largely uncertain because of the parameter degeneracy in the reacceleration process. 
The diffusion of parent CRPs from primary accelerators has been of interest \citep[e.g.,][]{Keshet:2010pf,Keshet:2010aq}, which has also been separately investigated in the calculation of high-energy emission or escaping CRs \citep[e.g.,][]{2009ApJ...707..370K,Fang2016,2018NatPh..14..396F,Hussain:2021dqp}.\par

In this paper, we evaluate multi-wavelength radiation from radio to gamma-ray and the neutrino emission from the Coma cluster. We follow the time evolution of the CR distribution in the Coma cluster from the radio-quiet state to the radio-loud state. Concerning primary CRs, we present two extreme cases. One is the ``secondary-dominant model", where all CREs are injected as secondary products of $pp$ collisions. The other is the ``primary-dominant model", where most of CREs in the ICM are injected from the same source as primary CRPs. We also test two types of turbulent reacceleration: ``hard-sphere" and ``Kolmogorov" type. 

This paper is organized as follows. In Section \ref{sec:Methods} we introduce the basic formalism for the CR acceleration and evolution, in Section \ref{sec:results} we explain the procedure to put constraints on model parameters from the observational properties of the Coma RH and summarize resulting fluxes including cosmic rays, gamma rays and neutrinos. In Section \ref{sec:discussion}, we evaluate the intensity of the background emission and compare our results with earlier studies. Our main results are summarized in Section \ref{sec:conclusion}.
Throughout this paper, we adopt the $\Lambda$CDM model with $H_0$ = 100 $h$ km/s/Mpc, $h=0.7$, $\Omega_m=0.28$, and $\Omega_\Lambda=0.72$.

\section{Cosmic-ray distribution and evolution in the ICM \label{sec:Methods}}

In our calculation, the Coma cluster is considered to be a spherical gas cloud containing CRs. We first show the basic equations that describe the time evolution of the CR distribution in Section \ref{subsec:basic}. The physical processes considered here are radiative and collisional cooling (Section \ref{subsec:basic}), hadronic interactions to generate pions and secondary CRE injection from their decay (Section \ref{subsec:secondary}), and the spatial diffusion and acceleration due to the interaction with turbulence (Section \ref{subsec:acc}). The injection spectrum of the primary CRs is assumed to be a single power-law spectrum with a cutoff (Section \ref{subsec:primary}). The procedure to obtain observable quantities such as flux and surface brightness from the CR distribution functions is explained in Section \ref{subsec:emiss}. The initial condition is explained in Section \ref{subsec:initial}. Finally, we summarize our model parameters in Section \ref{subsec:para}. 

\subsection{Basic equations \label{subsec:basic}}

We assume spherical symmetry and define the distribution function of CRs in radial position $r$, momentum $p$, and time $t$ as $N^\mathrm{tot}_s(r,p,t)$ (where the index $s$ denotes particle species), which is related to the total particle number through $N^\mathrm{tot}_s(t)=\int dr\int dp N_s(r,p,t)$. The number density of the particle, $n_s(r,t)$, is then $n_s(r,t)=\int dp N_s(r,p,t)/(4\pi r^2)$.

To follow the time evolution of $N_s(r,p,t)$, we solve the isotropic one-dimensional Fokker-Planck (FP) equation. For protons, it takes the form
\begin{eqnarray}
\frac{\partial N_\mathrm{p}}{\partial t}&=&\frac{\partial}{\partial p}\left[N_\mathrm{p}\left(b^{(p)}_\mathrm{C}-\frac{1}{p^2}\frac{\partial}{\partial p}\left(p^2D_{pp}\right)\right)\right] \nonumber\\
& &+\frac{\partial^2}{\partial p^2}[D_{pp} N_\mathrm{p}]+\frac{\partial}{\partial r}\left[D_{rr}\frac{\partial N_\mathrm{p}}{\partial r}-\frac{2}{r}N_\mathrm{p} D_{rr}\right]\nonumber\\
& &+Q_\mathrm{p}(r,p)-\frac{N_\mathrm{p}}{\tau_{pp}(r,p)},
\label{eq:Np}
\end{eqnarray}
where $b^{(p)}_\mathrm{C}$ represents the momentum loss rate ($b\equiv -dp/dt$) due to the Coulomb collisions (Eq. (\ref{eq:pCoulomb})), $D_{rr}$ and $D_{pp}$ are the spatial and momentum diffusion coefficients due to interactions with turbulence (Eqs. (\ref{eq:Drr}) and (\ref{eq:Dpp})), and $Q_\mathrm{p}(r,p)$ denotes the injection of primary CRPs (Eq. (\ref{eq:pri})). The number of primary CRPs injected per unit volume per unit time per momentum interval can be expressed as $q_\mathrm{p}(r,p)=Q_\mathrm{p}(r,p)/(4\pi r^2)$. The value $\tau_{pp}$ denotes the $pp$ collision timescale (Eq. (\ref{eq:taupp})). For simplicity, we ignore the effect of repeated collisions of a CRP, so we do not follow an energy loss per collision. The cooling due to the $pp$ collision is expressed like escape as $-N_\mathrm{p}/\tau_{pp}$ in the FP equation. This term is smaller than other terms in Eq. (\ref{eq:Np}) and makes only a negligible effect in the evolution of the CRP spectrum, so we do not include the inelasticity coefficient $\kappa_{pp}\approx0.5$ in Eq.~(\ref{eq:Np}). This means that we neglect multiple $pp$ collisions experienced by a single CRP.

The momentum loss of a CRP due to the combined effect of CRP-p, CRP-e Coulomb interactions can be expressed as \citep{Petrosian2015}
\begin{eqnarray}
b^{(p)}_\mathrm{C}(r,p)&=&\frac{3}{2}(\sigma_\mathrm{T}n_{\mathrm{th}}c)m_\mathrm{e}c\frac{\ln\Lambda}{\beta_\mathrm{p}^2}\left(1-\frac{k_\mathrm{B}T/2}{E_\mathrm{p}-m_\mathrm{p}c^2}\right)\nonumber \\
&&\times\sum_{s=e,p}\frac{m_\mathrm{e}}{m_s}\left[\mathrm{Erf}(\sqrt{x_s})-\frac{2}{\sqrt{\pi}}\sqrt{x_s}e^{-x_s}\right] \label{eq:pCoulomb}, \nonumber\\
\simeq&3.5& \times 10^{-29} \mathrm{erg}\; \mathrm{cm}^{-1} \frac{1}{\beta_\mathrm{p}^2}\left(\frac{n_\mathrm{th}}{1\; \mathrm{cm}^{-3}}\right) \left(\frac{\ln\Lambda}{38}\right)\nonumber \\
&&\times \left[1-\frac{k_\mathrm{B}T/(2m_\mathrm{p}c^2)}{\sqrt{1+(p/(m_\mathrm{p}c))^2}-1}\right]\nonumber\\
&\times&\sum_{s=e,p}\frac{m_\mathrm{e}}{m_s}\left(\int_0^{\sqrt{x_s}}dye^{-y^2} - \sqrt{x_s}e^{-x_s}\right), 
\label{eq:pCoulomb2}
\end{eqnarray}
where $\sigma_\mathrm{T}$ is the Thomson cross section, $\Lambda$ is the Coulomb logarithm, $x_s \equiv\frac{ (E_\mathrm{p}-m_\mathrm{p}c^2)}{k_BT}\frac{m_s}{m_\mathrm{p}}$, where the index $s=$e, p stands for the species of target particles, $E_\mathrm{p}=\sqrt{m_\mathrm{p}^2c^4+p^2c^2}$, and $\beta_\mathrm{p}$ is the particle velocity in unit of $c$. The function $\mathrm{Erf}(x)$ in Eq. (\ref{eq:pCoulomb}) stands for the error function, and $n_{\mathrm{th}}$ and $T$ are the density and temperature of the thermal gas in ICM, respectively.

In this paper, we adopt the beta-model profile for the thermal electron density derived from the X-ray observation \citep{1992A&A...259L..31B};
\begin{eqnarray}
n_{\mathrm{th}}(r)=n_{\mathrm{th}}(0)\left\{1+\left(\frac{r}{r_\mathrm{c}}\right)^2\right\}^{-\frac{3}{2}\beta},
\label{eq:beta-model}
\end{eqnarray}
where, $n_{\mathrm{th}}(0)=3.42\times10^{-3}\:\mathrm{cm}^{-3}$, $\beta=0.75$, and the core radius of the Coma cluster is given by $r_c = 290$ kpc.\par
We also use the temperature profile following \citep{Bonamente_2009,Pinzke2017};
\begin{eqnarray}
k_\mathrm{B}T(r) = 8.25\; \mathrm{keV} \left[1+\left(\frac{2r}{r_{200}}\right)^2\right]^{-0.32},
\label{eq:temp}
\end{eqnarray}
where the virial radius of the Coma cluster is $r_{200} = 2.3$ Mpc \citep{Reiprich_2002}. Assuming that the turbulence responsible for the reacceleration is driven by a cluster merger, the terms proportional to $D_{pp}$ have finite values only after the merger (Sect. \ref{subsec:halo}, Sect. \ref{subsec:initial}). \par

For electrons and positrons, the FP equation becomes
\begin{eqnarray}
\frac{\partial N_\mathrm{e}}{\partial t}&=&\frac{\partial}{\partial p}\left[N_\mathrm{e}\left(b_{\mathrm{rad}}+b^{(e)}_\mathrm{C}-\frac{1}{p^2}\frac{\partial}{\partial p}\left(p^2D_{pp}\right)\right)\right] \nonumber\\
& &+\frac{\partial^2}{\partial p^2}[D_{pp} N_\mathrm{e}]+\frac{\partial}{\partial r}\left[D_{rr}\frac{\partial N_\mathrm{e}}{\partial r}-\frac{2}{r}N_\mathrm{e} D_{rr}\right] \nonumber \\
& &+Q_\mathrm{e}(r,p;N_\mathrm{p}).
\label{eq:Ne}
\end{eqnarray}
The energy loss rate of a CRE due to CRE-e collisions \footnote{The loss due to CRE-p collision is negligible, since the lightest particle contributes most to the stopping power of the plasma \citep[e.g.,][]{2009herb.book.....D}} in ICM is
\begin{eqnarray}
b^{(e)}_\mathrm{C}(r,p)&=& \frac{3}{2}(\sigma_\mathrm{T} c n_{\mathrm{th}})m_\mathrm{e}c\frac{1}{\beta_\mathrm{e}^2}B_\mathrm{rel},  \\
\simeq& 3.05& \times 10^{-29} \mathrm{erg}\; \mathrm{cm}^{-1} \frac{1}{\beta_\mathrm{e}^2}\left(\frac{n_\mathrm{th}}{1\;\mathrm{cm}^{-3}}\right)\nonumber \\ 
&&\times\left[1+\frac{1}{74.8}\ln\left(\frac{p/m_\mathrm{e}c}{\left(n_\mathrm{th}/1\;\mathrm{cm}^{-3}\right)}\right)\right], \label{eq:bC}
\end{eqnarray}
where $\beta_\mathrm{e}$ is the velocity of the CRE and $B_\mathrm{rel}$ is the dimensionless stopping number \citep[see][Eq. (5.5)]{GOULD1972145}. \par
The radiative momentum loss term, $b_\mathrm{rad}$, includes both synchrotron radiation and inverse-Compton scattering (ICS); $b_\mathrm{rad}=b_\mathrm{syn}+b_\mathrm{IC}$.The bremsstrahlung loss is negligible compared to $b^{(e)}_\mathrm{C}+b_\mathrm{rad}$ \citep{Sarazin1999}. Radio-emitting CREs in ICM also emit $\sim10$ keV photons due to the ICS with CMB photons. We use the formulae given in \citet{Rybicki:847173} and \citet{1996ApJ...463..555I} for these processes (see Eq. (\ref{eq:bIC}) for the ICS radiation). \par
The injection of CREs can be divided into primary and secondary injection; $Q_\mathrm{e}(r,p;N_\mathrm{p})=Q^{\mathrm{sec}}_\mathrm{e}(r,p,;N_{\rm p})+Q_\mathrm{e}^{\mathrm{pri}}(r,p)$ (see Eqs. (\ref{eq:Qe}) and (\ref{eq:Qepri})).

\subsection{Production of secondary electrons \label{subsec:secondary}}
Inelastic collisions between CRPs and thermal protons in the ICM lead to mesons that are primarily pions ($p+p\rightarrow \pi^{0,\pm} + X$), whose decay channels are
\begin{eqnarray}
\left\{
\begin{array}{l}
\pi^0 \rightarrow 2\gamma,\;\;\pi^\pm \rightarrow \mu^\pm +\nu_\mu (\bar{\nu}_\mu),   \\
\mu^\pm \rightarrow e^\pm+\bar{\nu}_\mu(\nu_\mu) +\nu_\mathrm{e} (\bar{\nu}_\mathrm{e}).
\end{array}
\right.
\label{eq:pi}
\end{eqnarray}
The collision timescale is written as
\begin{eqnarray}\label{eq:taupp}
\tau_{pp}(r,p)=\frac{1}{cn_{\mathrm{th}}(r)\sigma_{\rm inel}(p)}, 
\end{eqnarray}
where $\sigma_{\rm inel}(p)\approx 34$ mb is the total inelastic cross section, which is given in e.g., \citet{Kamae2006}. Here, we assume that the ICM is pure hydrogen plasma
and use Eq.~(\ref{eq:beta-model}) for the density of thermal protons, although the existence of helium nuclei can affect the $pp$ production rate in the ICM\footnote{As discussed in \citet{2020A&A...644A..70A}, the production rate of secondary-particles would be increased by a factor of $\sim$ 1.5, considering the helium mass fraction of 0.27.}.
Using the inclusive cross section for charged and neutral pion production $\sigma^{0,\pm}$, it can be written as $\sigma_{\rm inel}(p)=\sigma^{0}(p)+\sigma^{\pm}(p)$. \par

The injection rate of generated pions $Q_\pi(E_\pi)$ (which has the same dimension as $Q_\mathrm{p}$ and $Q_\mathrm{e}$ in Eqs. (\ref{eq:Np}) and (\ref{eq:Ne})) can be calculated from
\begin{eqnarray}
Q_\pi^{0,\pm}(E_\pi;N_\mathrm{p})&=&n_{\mathrm{th}}(r)c\int_{E_\mathrm{th}}^\infty dp\beta_\mathrm{p} N_{\rm p}(r,p,t) \nonumber \\
&& \times \sigma^{0,\pm}(p)\frac{F_\pi(E_\pi,E_{\rm p})}{E_{\rm p}},
\label{eq:Qpi}
\end{eqnarray}
where $E_{\mathrm{th}}\approx1.22$~GeV is the threshold energy for the pion production, and 
$F_\pi(E_\pi,E_p)$ is the spectrum of pions produced in a single $pp$ collision by a CRP of energy $E_\mathrm{p}$. 
The injection rate of pions per unit volume, $q_\pi^{0,\pm}$, is expressed as $q_\pi^{0,\pm}=Q_\pi^{0,\pm}/(4\pi r^2)$. We adopt the approximate expression of $F_\pi(E_\pi,E_{\rm p})$ given in \citet{Kelner2006} (see Eq.~(\ref{eq:QGSJET}) in the Appendix) for both neutral and charged pions. To distinguish the cross sections for neutral and charged pion productions, we adopt the inclusive cross section, $\sigma^{0,\pm}(p)$, given in \citet{Kamae2006,Kamae2007}. There is a slight ($\lesssim50\%$) difference in pion production rate around 100 MeV between our method and, for example, \citet{Brunetti2017}, where the isobaric model by \citet{1970Ap&SS...6..377S} and high-energy model by \citet{Kelner2006} 
are adopted at lower and higher energies, respectively. The uncertainty in secondary production rate, including that arising from the helium abundance noted above, is much less significant than the uncertainty in the CR injection rate or the turbulent reacceleration in our modeling (Sect.~\ref{sec:results}).

Using the injection rate of pions $Q_\pi$, the injection rate of secondary electron/positron can be written as \citep[e.g.,][]{Brunetti2017}
\begin{eqnarray}
Q_\mathrm{e}^{\mathrm{sec}}(r,p;N_\mathrm{p})&=&\int dE_\pi \int dE_\mu Q_\pi(E_\pi;N_\mathrm{p}) \nonumber \\
&\times& F_\mathrm{e}(E_\mathrm{e},E_\mu,E_{\pi^\pm})F_\mu(E_\mu,E_{\pi^\pm}),
\label{eq:Qe}
\end{eqnarray}
where $F_\mu(E_\mu,E_{\pi^\pm})$ is the spectrum of muons from decay of $\pi^\pm$ with energy $E_\pi$, which can be obtained with simple kinematics. Hereafter in this section, we omit the $\pm$ symbol on $\pi^\pm$. 
In the rest frame of a pion, the energy of secondary muon and muonic neutrino are $E_\mu'=\frac{m^2_{\pi}c^4+m_\mu^2 c^4}{2m_{\pi}c^2}\approx110\;\mathrm{GeV}, \;\; E'_{\nu_\mu}=\frac{m^2_{\pi}c^4-m^2_\mu c^4}{2m_{\pi}c^2}\approx29.9\;\mathrm{GeV},$
where we neglect the mass of neutrinos. The Lorentz factor and velocity of the muon are then $\gamma_\mu'=(m_\pi^2+m_\mu^2)/(2m_\pi m_\mu)\simeq1.039$ and $\beta_\mu'\simeq 0.2714$  \citep[e.g.,][]{Moskalenko1998}. Since a pion decays isotropically in its rest frame, the spectrum in the source frame is 
\begin{eqnarray} 
F_\mu(E_\mu,E_\pi)=\frac{m_\pi^2}{m_\pi^2-m_\mu^2}\frac{1}{\sqrt{E_\pi^2-m_\pi^2c^4}},
\end{eqnarray}
within the energy range $\gamma_\mu^-m_\mu c^2 \le E_\mu\le \gamma_\mu^+m_\mu c^2$, where $\gamma_\mu^\pm=\gamma_\pi\gamma_\mu'(1\pm\beta_\pi\beta_\mu')$ are the minimum and maximum Lorentz factor of the muon in the laboratory system, respectively. The function $F_{\mathrm{e}}(E_\mathrm{e},E_\mu,E_\pi)$ stands for the spectrum of secondary electrons and positrons from decay of the muon of energy $E_\mu$, which also depends on the energy of the parent pion of $E_\pi$ since the muons from $\pi^\pm \rightarrow \mu^\pm +\nu_\mu (\bar{\nu}_\mu)$ are fully polarized. 
The expression is given by \citep[e.g.,][]{1991crpp.book.....G,Blasi1999};
\begin{eqnarray}
&F_\mathrm{e}&(E_\mathrm{e},E_\mu,E_\pi)= \frac{4}{\beta_\mu E_\mu}\left[\frac{5}{12}-\frac{3}{4}\lambda^2+\frac{1}{3}\lambda^3 \right. \nonumber \\
&&\left. -\frac{P_\mu(E_\pi,E_\mu)}{2\beta_\mu}\left\{\frac{1}{6}-\left(\beta_\mu+\frac{1}{2}\right)\lambda^2+\left(\beta_\mu+\frac{1}{3}\right)\lambda^3\right\} \right], \nonumber \\
& \mathrm{for} & \left(\frac{1-\beta_\mu}{1+\beta_\mu}\le\lambda\le1\right),
\end{eqnarray}
and
\begin{eqnarray}
&F_\mathrm{e}&(E_\mathrm{e},E_\mu,E_\pi)=\frac{4}{\beta_\mu E_\mu} \nonumber\\
&\times&\left[\frac{\lambda^2\beta_\mu}{(1-\beta_\mu)^2}\left\{3-\frac{2}{3}\lambda\left(\frac{3+\beta_\mu^2}{1-\beta_\mu}\right)\right\} \right. \nonumber \\
&&\left. -\frac{P_\mu(E_\pi,E_\mu)}{1-\beta_\mu}\left\{ \lambda^2(1+\beta_\mu)-\frac{2\lambda^2}{1-\beta_\mu}\left[\frac{1}{2}+\lambda(1+\beta_\mu)\right]\right.\right. \nonumber \\
&&\;\;\;\left. \left.+\frac{2\lambda^3(3+\beta_\mu^2)}{3(1-\beta_\mu)^2}\right\} \right],  \nonumber \\
& \mathrm{for}&\left(0\le\lambda\le\frac{1-\beta_\mu}{1+\beta_\mu}\right),
\end{eqnarray}
where
\begin{eqnarray}
\lambda=\frac{E_\mathrm{e}}{E_\mu}, \;\; \beta_\mu=\sqrt{1-\frac{m_\mu^2c^4}{E_\mu^2}},
\end{eqnarray}
and 
\begin{eqnarray}
P_\mu(E_\pi,E_\mu)&=&\frac{1}{\beta_\mu}\left[\frac{2E_\pi\xi}{E_\mu(1-\xi)}-\frac{1+\xi}{1-\xi}\right], \\
\xi&=&\frac{m_\mu^2}{m_\pi^2}.
\end{eqnarray}
Following \citet{Brunetti2005}, we assume that the muon spectrum is approximated by the delta function at the energy $E_\mu=\frac{1}{2}(E_\mu^\mathrm{min}+E_\mu^\mathrm{max})$ in the laboratory system, that is,
\begin{eqnarray}
F_\mu(E_\mu,E_\pi)=\delta\left(E_\mu-\frac{m_\pi^2-m_\mu^2}{m_\pi^2}\frac{E_\pi}{2\beta_\mu'}\right).
\end{eqnarray}
In this case, the integral with respect to $E_\mu$ in Eq. (\ref{eq:Qe}) can be performed self-evidently. The resulting injection spectrum of secondary CREs has a power-law index of $\alpha_\mathrm{e}=\alpha+\Delta$ with $\Delta\sim0.05-0.1$ for $N_\mathrm{p}\propto p^{-\alpha}$ for parent CRPs \citep[e.g.,][]{Kamae2006,Kelner2006}.

\subsection{Injection of primary cosmic-rays \label{subsec:primary}}

Galaxy clusters can be regarded as reservoirs of CRs because they can confine accelerated particles for a cosmological timescale~\citep{Murase:2012rd}. The candidate CR accelerators include structure formation shocks, cluster mergers, AGN, ordinary galaxies, and galaxy mergers. 
\par

Among these, structure formation shocks should have a connection with the occurrence of RHs because they are usually found in merging systems \citep[e.g.,][]{Cassano2010}. During a merger between two clusters with the same mass and radius of $M\sim 10^{15}M_\sun$ and $R\sim2$ Mpc, respectively, the gravitational energy of $E\sim GM^2/R\sim 10^{64}$ erg dissipates within the dynamical timescale of $t_\mathrm{dyn}\sim 10^{9}$ yr. Assuming that $\sim1$ \% of the energy is used to accelerate CRs, the injection power of primary CRs is evaluated as $L_\mathrm{CR}\sim10^{45}$ erg/s. \par

In a simple test particle regime, the diffusive shock acceleration (DSA) theory predicts the CR injection with a single power-law distribution in momentum, whose slope depends only on the shock Mach number. We assume that primary CRPs are injected with a single power-law spectrum with an exponential cutoff; 
\begin{eqnarray} 
Q_\mathrm{p} (r,p)=C^\mathrm{inj}_\mathrm{p}p^{-\alpha}\exp\left[-\frac{E_\mathrm{p}}{E_\mathrm{p}^{\mathrm{max}}}\right]K(r),
\label{eq:pri}
\end{eqnarray}
where $E_\mathrm{p}^\mathrm{max}$ stands for the maximum energy of primary CRPs. We adopt $E^\mathrm{max}_\mathrm{p}=100$ PeV as a reference value, though the energy up to the ankle ($\sim10^{18.5}$ eV) could be achieved by primary sources, such as AGNs~\citep[e.g.,][]{2009ApJ...707..370K,2018NatPh..14..396F} or strong shocks~\citep{Kang:1996rp,Inoue:2005vz,Inoue:2007kn}. The Larmor radius of CRPs with $E_\mathrm{p}^\mathrm{max}=100$ PeV in a $\mu$G magnetic field is $r_\mathrm{L}\sim100$ pc, which is much smaller than the typical scale of the accretion shocks, $\sim$Mpc. The minimum momentum of CRPs is taken to be $p=30$ MeV$/c$, which is about ten times larger than the momentum of the thermal protons.

The function $K(r)$ represents the radial dependence of the injection, which is to be determined to reproduce the observed surface brightness profile of the RH (Section \ref{subsec:halo}). Note that the luminosity of the injection, i.e., the normalization $C^\mathrm{inj}_\mathrm{p}$, is tuned to match the observed radio synchrotron flux at 350 MHz (Section \ref{subsec:halo}). \par
A certain amount of electrons should also be injected as primary CRs. The presence of primary CREs affects the relative strength of hadronic emission to leptonic ones. However, the ratio of primary to secondary CREs is usually uncertain in observations. We treat the primary CRE injection by introducing a parameter $f_\mathrm{ep}$;
\begin{eqnarray}
Q_\mathrm{e}^{\mathrm{pri}} (r,p) = f_{\mathrm{ep}} Q_\mathrm{p}(r,p).
\label{eq:Qepri}
\end{eqnarray}
Using this relation, we extrapolate Eq.~(\ref{eq:pri}), which is valid only above the minimum momentum of the CRP, $p=30$ MeV$/c$, to the minimum momentum of CREs, $p_\mathrm{e}=0.3m_\mathrm{e}c=150$ keV$/c$. The contribution of CREs below this energy is negligible due to the strong Coulomb cooling compared to the acceleration (Figure~\ref{fig:time} left). Concerning $f_\mathrm{ep}$, we consider two example cases: the secondary-dominant model ($f_\mathrm{ep}=0$) and the primary-dominant model ($f_\mathrm{ep}=0.01$). The possible radial dependence of $f_\mathrm{ep}$ \citep[e.g.,][]{2008MNRAS.385.1211P} is not considered in our calculation. \par
The former case, $f_\mathrm{ep}=0$, is motivated by the injection from AGNs~\citep[e.g.,][]{2018NatPh..14..396F}, where only high-energy ions can diffuse out from their radio lobes, while CREs lose their energies inside the lobes due energy losses during the expansion. Another possibility for that case is that CRs are accelerated at shock waves in the ICM with low Mach numbers~\citep[e.g.,][]{2020ApJ...892...86H}, where particles with smaller rigidities ($= pc/(Ze)$) are less likely to recross the shock front, and therefore the acceleration of electrons from thermal energies can be much inefficient compared to protons \citep{Brunetti2014}. \par
On the other hand, $f_\mathrm{ep}=0.01$ corresponds to the observed CRE to CRP ratio in our galaxy \citep[e.g.,][]{SchlickiBuch}. If CRs in the ICM are provided by the internal sources, that value may be the upper limit for $f_\mathrm{ep}$. 
Some numerical studies suggest that the fluctuations at shock vicinity, such as electrostatic or whistler waves, support the injection of CREs into the Fermi acceleration process \citep[e.g.,][]{Amano_2008,Riquelme_2011}, which could potentially increase the CRE to CRP ratio. However, the injection processes of CREs at weak shocks in high-beta plasma are still under debate \citep[e.g.,][]{Kang_2019}.

\subsection{Particle acceleration and diffusion in the ICM \label{subsec:acc}}

The magnetic field in the Coma cluster is well studied with the rotation measure (RM) measurements. Here we use the following scaling of the magnetic field strength with cluster thermal density: 
\begin{eqnarray}
B(r)=B_0\left(\frac{n_{\mathrm{th}}(r)}{n_{\mathrm{th}}(0)}\right)^{\eta_B},
\label{eq:B}
\end{eqnarray}
where $B_0=4.7\:\mu \mathrm{G}$ and $\eta_B=0.5$ are the best fit values for the RM data \citep{Bonafede2010}. The uncertainty in the magnetic field estimate and its impact on our results are discussed in Sect.~\ref{subsec:caveats}. \par

CRs in the ICM undergo acceleration and diffusion due to the interaction with MHD turbulences. We assume that the spatial diffusion is caused by the isotropic pitch angle scattering with Alfv\'{e}n waves. When the Larmor radius of a particle $r_\mathrm{L}$ is smaller than the maximum size of the turbulent eddy, $l^\mathrm{A}_\mathrm{c}\sim0.1$ Mpc, the propagation of the particle is in the diffusive regime. The diffusion coefficient in that regime can be written as \citep[e.g.,][]{Murase2013,Fang2016}
\begin{eqnarray}
D_{rr}(r,p)&=&\frac{1}{3}\left(\frac{B(r)}{\delta B}\right)^2\beta_\mathrm{p}cr_\mathrm{L}^{1/3}(l^\mathrm{A}_{\rm c})^{2/3}.
 \label{eq:Drr}
\end{eqnarray}
For reference, the typical value of $D_{rr}$ in our Galaxy is $D_{rr}\sim {\rm a~few}\times 10^{28}$ cm$^2$ sec$^{-1}$ for protons with 1 GeV \citep[e.g.,][]{2007ARNPS..57..285S}, which should be smaller than Eq. (\ref{eq:Drr}) due to the shorter coherent length in our Galaxy. 
We take $\delta B\sim B$ at $l^\mathrm{A}_\mathrm{c}=0.1$ Mpc and the Kolmogorov scaling for Alfv\'{e}nic turbulence, neglecting the possible $r$ dependence of these quantities for simplicity. Then, the spatial diffusion coefficient is written as
\begin{eqnarray}
D_{rr}(r,p)&\simeq &6.8\times10^{29}~\mathrm{cm}^2 \mathrm{s}^{-1} \beta_\mathrm{p}\left(\frac{l^\mathrm{A}_\mathrm{c}}{0.1\:\mathrm{Mpc}}\right)^{2/3} \nonumber\\
& &\times\left(\frac{p}{1\:\mathrm{GeV/c}}\right)^{1/3}\left(\frac{B(r)}{1~\mu G}\right)^{-1/3}. \label{eq:Drr2}
\end{eqnarray}
The similar value has often been used~\citep[e.g.,][]{Murase2008}, although larger values may also be possible~\citep{Keshet:2010aq}. 

The time required to diffuse a distance comparable to the value of the radial position of the particle $r$ can be estimated as 
\begin{eqnarray}
t_{\mathrm{diff}}(r,p) = \frac{r^2}{4D_{rr}(r,p)}\propto p^{-1/3}.  
\end{eqnarray}
The diffusion timescale for GeV electrons over the scale of RHs ($r\sim1$ Mpc) is much longer than the Hubble time. This requires that CREs are injected {\it in situ} in the emitting region. Moreover, $t_{\rm diff}$ for parent CRPs can also be $\sim$ Gyr, so the resulting spatial distribution of GeV electrons depends on the injection profile. 
Note that CRPs with $r_\mathrm{L}$ larger than $l^\mathrm{A}_{\rm c}$ are in the semi-diffusive regime, where $t_{\rm diff}\propto p^{-2}$. 
\par

We simply write the momentum diffusion coefficient as a power-law function of particle momentum with an exponential cutoff;
\begin{eqnarray}
D_{pp} (r,p) =\frac{p^2}{(q+2)\tau_\mathrm{acc}}\left(\frac{p}{1\; \mathrm{GeV/c}}\right)^{q-2}\nonumber \\
\times\exp\left[-\frac{E_\mathrm{p}}{E^{\mathrm{max}}_c(r)}\right]\exp\left(-\frac{m_sc}{p}\right),
\label{eq:Dpp}
\end{eqnarray}
where the exponential cutoffs at both the maximum energy $E^{\mathrm{max}}_c$ and the minimum momentum $m_sc$ are introduced, where the index $s$ denotes the particle species. For simplicity, we adopt $E^{\mathrm{max}}_\mathrm{c} (r) = qB(r) l^\mathrm{F}_\mathrm{c} \sim 9\times 10^{19}(B(r)/1~\mu G) (l^\mathrm{F}_\mathrm{c}/0.1 \mathrm{Mpc})$~eV for the maximum energy of CRs achieved by that acceleration; that is, the particles whose Larmor radius larger than $l^\mathrm{F}_\mathrm{c}$ cannot be accelerated efficiently, where $l^\mathrm{F}_\mathrm{c}$ is the maximum size of the turbulent eddy of compressible turbulence. We assume $l^\mathrm{F}_\mathrm{c} = 0.1$ Mpc as a reference, and this is somewhat lower than the Hillas limit with the system size ($\sim{\rm Mpc}$). 
\citet{2016MNRAS.463..655K} measured the power spectrum of the pressure fluctuation in Coma using the observation of the thermal Sunyaev-Zel'dovich effect (SZ), and found an injection scale of $\approx0.5$ Mpc. Considering the uncertainty in the measured scale, which is basically an interpolation between the SZ analysis and X-ray analysis by \citet{2012MNRAS.421.1123C}, our assumption of $l^\mathrm{F}_\mathrm{c} = 0.1$ Mpc is compatible with those observations. Our calculation is not sensitive to $E^{\rm max}_{\rm c}$, since the maximum energy of CRPs hardly reaches such high energies starting from $E_p^{\rm max}=100~{\rm PeV}$ of Eq.~(\ref{eq:pri}) (see also Sect.~\ref{subsec:esc}).
\par

The index $q$ is treated as a model parameter (Sect. \ref{subsec:para}). We examine two cases for $q$: $q = 2,\; 5/3$.  We call $q=2$ ``hard-sphere type" acceleration and call $q=5/3$ ``Kolmogorov type" acceleration. The parameter $\tau_\mathrm{acc}$ denotes the acceleration timescale of particles with momentum $p= 1$ GeV$/c$, which is constrained from the spectral shape of the RH (Section \ref{subsec:halo}). Note that we assume $\tau_\mathrm{acc}$ is constant with the radius for simplicity. \par

The acceleration time of CRs is estimated as
\begin{eqnarray}
t_{\mathrm{acc}}(r,p) = \frac{p^2}{(2+q)D_{pp}(r,p)} \propto p^{2-q}.
\label{eq:tacc}
\end{eqnarray}
This timescale is independent of $p$ in the hard-sphere case ($q=2$), while it is shorter for the smaller momentum in the Kolmogorov case ($q=5/3$). 
The momentum diffusion coefficient for the stochastic acceleration by the pitch angle scattering with Alfv\'{e}nic waves takes the form $D_{pp}\propto p^w$ (i.e., $q=w$), where the index $w$ has the same value as the slope of the turbulent spectrum: $w=\frac{5}{3}$ for the Kolmogorov scaling and $w=\frac{3}{2}$ for the Iroshnikov-Kraichnan (IK) scaling \citep[e.g.,][]{Becker2006}. \citet{Brunetti2007} self-consistently calculated $D_{pp}$ for TTD with compressive MHD mode via quasi-linear theory (QLT). In this case, the acceleration timescale (Eq. (\ref{eq:tacc})) for CR particles does not depend on particle momentum, because all CRs are assumed to interact with the turbulence at the cutoff scale. Thus, this mechanism has the same index $q$ as our hard-sphere model; $q=2$.\par

Assuming the IK scaling for the compressible turbulence, $\tau_\mathrm{acc}$ for the TTD acceleration can be estimated as \citep[e.g.,][]{Brunetti2015}
\begin{eqnarray}
\tau_\mathrm{acc}&\approx& \frac{4c\rho}{\pi}I^{-1}_\theta(x)\left[\int_{k_L}^{k_\mathrm{cut}}dkkW(k)\right]^{-1} \nonumber\\
&\simeq&300 \;\mathrm{Myr}\left(\frac{L}{300\;\mathrm{kpc}}\right)\left(\frac{\mathcal{M}_\mathrm{s}}{0.5}\right)^{-4}\left(\frac{c_\mathrm{s}}{10^8\;\mathrm{cm/s}}\right)^{-1},\nonumber\\
\label{eq:TTD}
\end{eqnarray}
where $\rho$ is the mass density of the ICM, $W(k)$ is the total energy spectrum of the compressible turbulence, $I_\theta(x)\equiv\int^{\arccos(x)}_0d\theta \frac{\sin^3\theta}{|\cos\theta|}\left[1-\left(\frac{x}{\cos\theta}\right)^2\right]$ with $x=c_\mathrm{s}/c$, $c_\mathrm{s}$ is the sound speed of the ICM, $k_L\equiv2\pi/L$ and $k_\mathrm{cut}$ are wave numbers corresponding to the injection scale and the cutoff scale, respectively, and $\mathcal{M}_\mathrm{s}$ is the Mach number of the turbulent velocity at the injection scale. Here the cutoff scale of the turbulence is determined by the dissipation due to the TTD interaction with thermal electrons \citep[see][]{Brunetti2007}.\par

\begin{figure*}[tb]	
		\centering
    	\plottwo{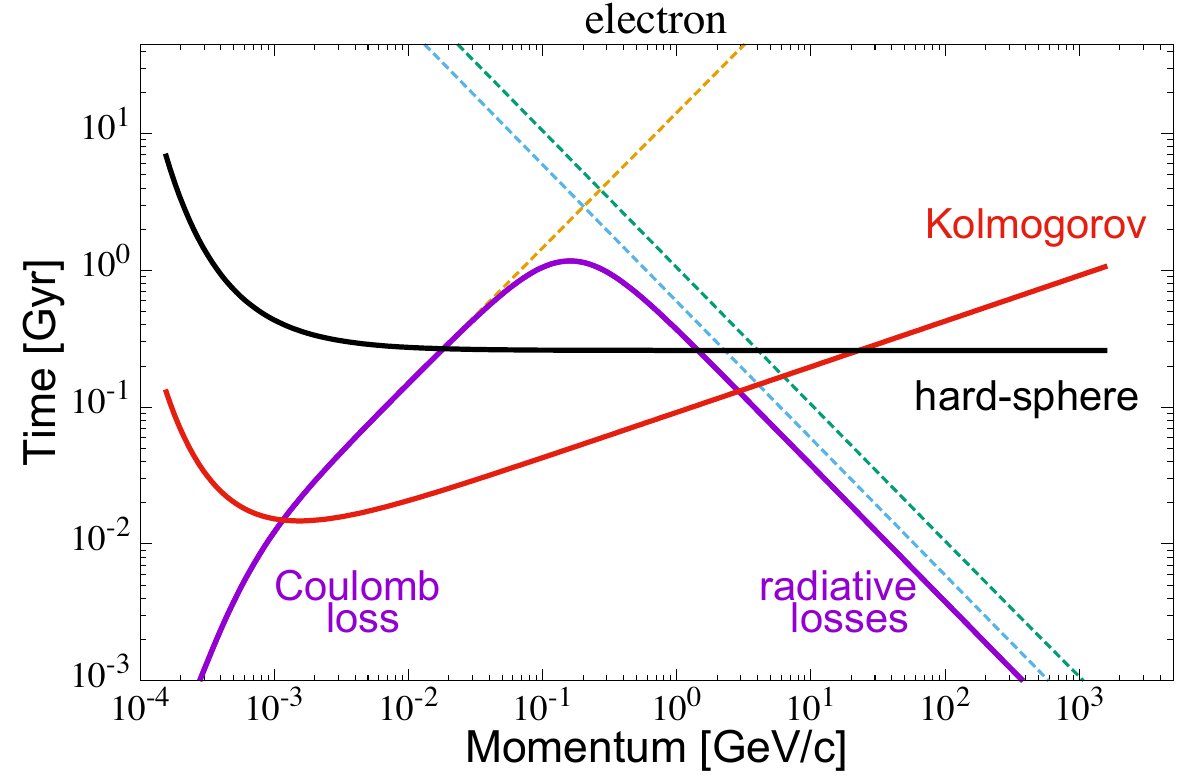}{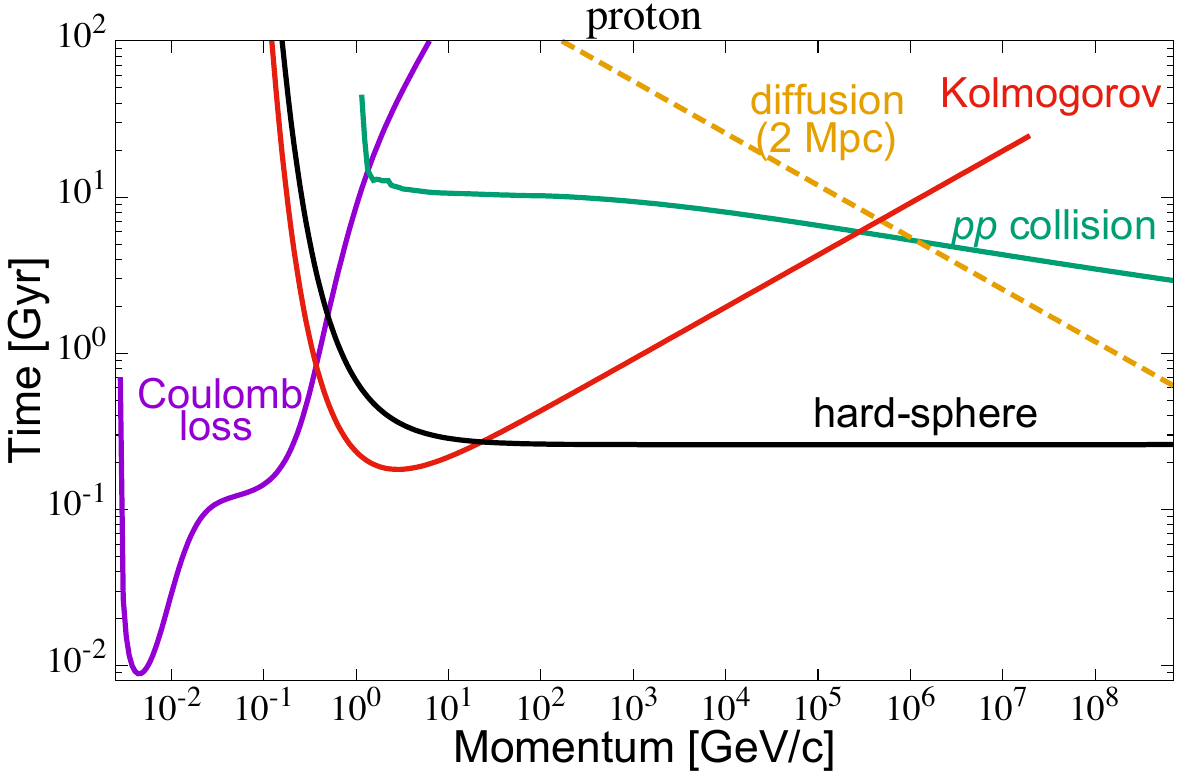}
	    \caption{{\it Left}: Physical timescales of CREs at $r=50$ kpc from the center of the cluster as functions of the momentum. The black line corresponds to hard-sphere type acceleration with $\tau_\mathrm{acc} = 260$ Myr, while the red line to Kolmogorov type with $\tau_\mathrm{acc} = 100$ Myr. The acceleration timescale becomes comparable to the cooling timescale of CREs around $E_\mathrm{e} \sim 1$ GeV (see the main text for detail). Dashed lines show the cooling timescale of the Coulomb collision (orange), synchrotron radiation (cyan), and ICS (green). The total cooling timescale is shown with solid magenta line. {\it Right}: Physical timescales of CRPs. The dashed orange line and the green line show the diffusion timescale on 2 Mpc scale and the $pp$ collision timescale with $n_\mathrm{th}=n_{0}$, respectively. \label{fig:time}	}
\end{figure*}

Figure \ref{fig:time} shows physical timescales of various terms in Eqs. (\ref{eq:Np}) and (\ref{eq:Ne}). Note again that we assume $\tau_\mathrm{acc}$ to be constant with radius. Following \citet{Brunetti2017}, we choose $\tau_{\mathrm{acc}}\sim300$ Myr, which is a typical value to explain the break in the spectrum around 1.4 GHz (see Section {\ref{subsec:halo}}). We do not solve the decay of turbulence, so $\tau_\mathrm{acc}$ remains a constant for several 100 Myr.   \par

Radio surveys have revealed that clusters with similar X-ray luminosity can be divided into two populations: ``radio-loud" clusters hosting RHs and ``radio-quiet'' clusters do not show any sign of cluster-scale radio emission. According to the Extended Giant Metrewave Radio Telescope (GMRT) Radio Halo Survey (EGRHS), the fraction of radio-loud clusters is about 30\% \citep{2013A&A...557A..99K}. 
Moreover, intermediate clusters between these two states are hardly detected. This clear bi-modality implies the existence of a mechanism that quickly turns on and off the RH. 
The timescale for clusters staying in the intermediate state, $\tau_\mathrm{in}$, can be estimated as the time between formation and observation of clusters times the fraction of clusters in that region; $\tau_\mathrm{in}\sim 100$ Myr \citep{2009A&A...507..661B}. Considering that RHs tend to be found in merging systems, the turbulence generated during cluster mergers could rapidly accelerate CRs within a few~$\times 100$~Myr \citep{Brunetti2014}. The damping of MHD waves might play a role in turning off RHs, as it enables super-Alfv\'{e}nic streaming of CRPs \citep{2013MNRAS.434.2209W}. In our calculation, we first prepare an initial distribution of CRs that corresponds to the radio-quiet state as explained in Sect. \ref{subsec:initial}, and then turn on the reacceleration and follow the evolution of the CR spectra.

\subsection{Emissivities and radiative transfer \label{subsec:emiss}}

In this section, we describe how to calculate the fluxes of electromagnetic waves and neutrinos from the given $N_\mathrm{p}$ and $N_\mathrm{e}$. We here identify two types of radiation: {\it leptonic} radiation is synchrotron radiation and the one from ICS with CMB photons, while {\it hadronic} ones are associate with the decay of pions produced by the $pp$ collision. The emissivity is defined as the energy emitted per frequency interval per unit volume per unit time per unit solid angle. The emissivity of the ICS radiation is shown in the Appendix (see Eq. (\ref{eq:eIC})). The emissivity of the hadronic gamma rays is calculated with the $\pi^0$ injection spectrum $q_\pi^0$ (the number of pions injected per logarithmic energy per unit volume per unit time);
\begin{eqnarray}
\varepsilon_{\gamma}(E_\gamma)&=&2E_\gamma\int_{E_{\mathrm{min}}(E_\gamma)}\frac{q_{\pi}^0(E_\pi)}{\sqrt{E_{\pi}^2-m_{\pi}^2c^4}}dE_\pi, 
\label{eq:egamma}
\end{eqnarray}
where $E_{\mathrm{min}}(E_\gamma)=E_\gamma+(m_\pi^2c^4)/(4E_\gamma)$ and $q_\pi^{0,\pm}=Q_\pi^{0,\pm}/(4\pi r^2)$. The gamma-ray photons produced by the  decay of $\pi^0$ of energy $E_\pi$ are distributed in the energy range of $\frac{1}{2}E_\pi(1-\beta_\pi)\le E_\gamma\le\frac{1}{2}E_\pi(1+\beta_\pi)$ with an equal probability. 

Muonic neutrinos are produced from the decay of both charged pions and secondary muons. Using the injection spectrum of charged pions, $q_\pi^\pm$, the emissivity is written as
\begin{eqnarray}
&\varepsilon_{\nu_\mu}&(E_{\nu_\mu})\nonumber \\
&&=2E_{\nu_\mu}\int_0^1(f_{\nu^{(1)}_\mu}(x)+f_{\nu^{(2)}_\mu}(x))q_\pi^\pm(E_\pi)\frac{dx}{x},\\
&&= 2E_{\nu_\mu}\left[\frac{1}{\zeta}\int_0^\zeta q_\pi^\pm\left(\frac{E_{\nu_\mu}}{x}\right)\frac{dx}{x}\right. \nonumber\\
&&\left. \;\;\;\;\;\; +\int_0^1f_{\nu^{(2)}_\mu}(x)q_\pi^\pm\left(\frac{E_{\nu_\mu}}{x}\right)\frac{dx}{x}\right],
\label{eq:enumu}
\end{eqnarray}
where $x=E_{\nu_\mu}/E_\pi$, $\zeta=1-m_\mu^2/m_\pi^2 $, and $f_{\nu^{(1)}_\mu}(x)$ and $f_{\nu^{(2)}_\mu}(x)$ are spectra of muonic neutrinos produced by the decay of pions and secondary muons, respectively. The function $f_{\nu^{(2)}_\mu}(x)$ is normalized as $\int_0^1f_{\nu^{(2)}_\mu}(x)dx=1$. The muonic neutrinos from the decay of an ultra-relativistic pion are evenly distributed within $0< E_{\nu_\mu}\le\zeta E_\pi$.
                                                            
Similarly, for electronic neutrinos, we have
\begin{eqnarray}
\varepsilon_{\nu_\mathrm{e}}(E_{\nu_\mathrm{e}})=2E_{\nu_\mathrm{e}}\int_0^1f_{\nu_\mathrm{e}}(x)q_\pi^\pm\left(\frac{E_{\nu_\mathrm{e}}}{x}\right)\frac{dx}{x},
\label{eq:enue}
\end{eqnarray}
where $x=E_{\nu_\mathrm{e}}/E_\pi$. The approximate expressions for $f_{\nu^{(1)}_\mu}(x)$, $f_{\nu^{(2)}_\mu}(x)$ and $f_{\nu_\mathrm{e}}(x)$ are given in \citet{Kelner2006} (see Appendix A). Note that those expressions only valid for the decay of relativistic pions ($E_\pi\gg m_\pi c^2$). Since we are interested in the reacceleration of $\sim$100 MeV CREs, we do not use them for the injection of secondary electrons/positions (Eq.~(\ref{eq:Qe})) but for the emission of high-energy neutrinos.

Surface brightness or intensity for an optically thin source is obtained by integrating the emissivity along the line of sight. Now we assume that the Coma cluster has spherical symmetry, the observed surface brightness at projected radius $r$ is written as~\citep[see, e.g., Eq.~2.12 of][]{Murase:2012rd}
\begin{eqnarray}
B_\nu(\nu,r) =\frac{2}{(1+z)^3}\int_{r}^{r_{200}}\frac{dr'}{\sqrt{1-r^2/r'^2}}\varepsilon(\nu',r'),
\label{eq:intensity}
\end{eqnarray}
where $\nu'=(1+z)\nu$, and $z$ is the redshift of the source.
The observed flux is obtained by integrating the intensity with respect to the solid angle, 
\begin{eqnarray}
F_\nu(\nu)=2\pi\int^{\theta_0}_0 d\theta \sin\theta\cos\theta B_\nu(\nu,r(\theta)),
\end{eqnarray} 
where $\theta=r/D_\mathrm{A}$ is the angular size corresponding to radius $r$, $\theta_0=r_\mathrm{ap}/D_\mathrm{A}$, $D_\mathrm{A}$ is the angular diameter distance of the source, and $r_\mathrm{ap}$ is the assumed aperture radius at each wavelength. We take $r_\mathrm{ap}=0.525, 1.2, 2.0$ Mpc for radio \citep[e.g.,][]{Pizzo2010,Brunetti2017}, non-thermal X-ray \citep{Wik_2011}, and gamma-ray \citep{Ackermann2016}, respectively. We take $r_\mathrm{ap}=r_{200}\approx2.3$ ${\rm Mpc}$ \citep{Reiprich_2002} for neutrinos. We always use $r_{200}$ as the maximum value for the integral region of Eq. (\ref{eq:intensity}), regardless of $r_{\rm ap}$. Total radiated luminosity becomes
\begin{eqnarray}
L= 4\pi D_\mathrm{L}^2 \int d\nu F_\nu,
\end{eqnarray}
where $D_\mathrm{L}=(1+z)^2D_\mathrm{A}$ is the luminosity distance, and $z=0.0232$ and $D_\mathrm{L}\approx103$ Mpc for the Coma cluster \citep[e.g.,][]{1989ApJS...70....1A}.
Gamma-ray photons that travel across the cosmological distance interact with the extragalactic background light (EBL) photons at IR and optical wavelengths, and they are attenuated by the $\gamma+\gamma \rightarrow e^++e^-$ process. The optical depth of $\gamma\gamma$ interaction, $\tau_{\gamma\gamma}(E_\gamma,z)$ depends on the energy of gamma-ray photon and redshift of the source. The observed gamma-ray flux becomes,
\begin{eqnarray}
F_\gamma^{\mathrm{obs}}=F_\gamma^\mathrm{int}\exp(-\tau_{\gamma\gamma}(E_\gamma,z)),
\end{eqnarray}
where $F_\gamma^\mathrm{int}$ is the intrinsic gamma-ray flux. We adopt the table for $\tau_{\gamma\gamma}(E_\gamma,z)$ provided in \citet{2011MNRAS.410.2556D}.

\subsection{Initial condition\label{subsec:initial}}
In our calculation, we first prepare an initial CR distribution, which corresponds to the ``radio-quiet'' state, where the radio flux is too faint to be observed. 
To prepare an initial quiet state for each model, we integrate the FP equations (Eqs. (\ref{eq:Np}) and (\ref{eq:Ne})) without the reacceleration (i.e. $D_{pp}=0$) for a duration of $t_0$. In this paper, we take $t_0=4$ Gyr, regardless of the model parameters. This corresponds to the assumption that the injection starts from $z\sim0.45$ and the amount of CRs before that epoch is negligible. The value of $t_0$ affects the injection rate of primary CRs. The cooling timescale of CREs takes maximum value of $\approx1$ Gyr at $E\approx100$ MeV, and $t_0$ should be longer than that timescale to obtain a relaxed spectrum of seed CREs in the ``radio-quiet" state. Besides, $t_0$ should be smaller than the age of a cluster; $t_0\lesssim10$ Gyr.  \par

After this injection phase, the reacceleration in the ICM is switched on and lasts until the current ``radio-loud" state is achieved. We use $t_\mathrm{R}$ as the elapsed time after the reacceleration is switched on. We assume that the present luminosity of the RH is still increasing, so $t=t_0+t_\mathrm{R}$ corresponds to the current state. Considering the bi-modality of the cluster population (Section \ref{sec:intro}), we refer to the state at $t\le t_0$ as the ``quiet'' state and the state at $t_0<t< t_0 +t_\mathrm{R}$ as the ``loud'' state. The primary injections, $Q_\mathrm{p}$ and $Q_\mathrm{e}^\mathrm{pri}$, are assumed to be constant throughout the calculation. In this work, we focus on the evolution of the CR distribution through the diffusion and resulting emission at the current state of the cluster, so we fix the properties of the cluster, e.g., the magnetic field (Eq.~(\ref{eq:B})) and thermal gas density (Eq.~(\ref{eq:beta-model})), although they would be considerably disturbed by the merger activity.

\subsection{Model parameters\label{subsec:para}}
\begin{table}[tbh]
\centering
\caption{Model parameters\label{tab:para}}
\begin{tabular}{cccc}
\hline\hline
Parameter& Symbol&Definition\\ 
\hline
reacceleration index& $q$& Eq. (\ref{eq:Dpp}) \\
Duration of the reacceleration&$t_\mathrm{R}$&Section \ref{subsec:initial}\\ 
Primary electron ratio&$f_\mathrm{ep}$&Eq. (\ref{eq:Qepri})\\
Normalization of the injection &$C_\mathrm{p}^\mathrm{inj}$& Eq. (\ref{eq:pri})\\
Injection spectral index&$\alpha$&Eq. (\ref{eq:pri})\\
Radial dependence of the injection& $K(r)$&Eq. (\ref{eq:pri})\\
\hline
\end{tabular}
\end{table}
In Table \ref{tab:para}, we summarize our model parameters. We test two types of the reacceleration with different $q$ as explained in the previous section. The duration of reacceleration, $t_\mathrm{R}$ or $t_\mathrm{R}/\tau_\mathrm{acc}$, affects the spectrum of both synchrotron and gamma-ray radiation (Section \ref{subsec:halo}). The parameter $t_\mathrm{R}$ should be chosen to explain the break appearing in the radio spectrum. 

In this paper, we fix $\tau_\mathrm{acc}=260$ Myr for the hard-sphere type reacceleration ($q=2$) and $\tau_\mathrm{acc}=100$ Myr for the Kolmogorov type reacceleration ($q=5/3$). As long as the acceleration timescale is in the range of $150\lesssim\tau_\mathrm{acc}\lesssim500$ Myr, the hard-sphere model can reproduce the  radio spectrum by tuning other parameters like $\alpha$, $C_\mathrm{p}^\mathrm{inj}$, and $t_\mathrm{R}$. \par

Concerning the injection of primary CRs, we have three parameters and one unknown function. We test two extreme cases for primary CREs: the primary-dominant case ($f_\mathrm{ep}=0.01$) and the secondary-dominant case ($f_\mathrm{ep}=0$). The energy spectrum of the primary CRs is modeled with Eq. (\ref{eq:pri}). We test four cases for the injection spectral index: $\alpha=2.0, 2.1, 2.2$, and $2.45$. The normalization $C_\mathrm{p}^\mathrm{inj}$ is determined from the absolute value of the observed flux at 350 MHz. The radial dependence of the injection $K(r)$ is constrained from the surface brightness profile of the RH. 
\par
We fix the strength and radial profile of the magnetic field, adopting the best fit value from \citet{Bonafede2010}; $(B_0,\eta_B)=(4.7\; \mu \mathrm{G},0.5)$.



\section{Results \label{sec:results}} 
In this section, we show the evolution of the CR distribution and non-thermal radiation from the Coma RH by integrating the FP equations. Example results for the time evolution of the spectra are shown in Section \ref{subsec:overview}. The constraints on the model parameters from radio and gamma-ray observations are discussed in Section \ref{subsec:halo}. The fluxes of high-energy radiations, including hard X-ray, gamma-ray, and neutrinos are shown in Section \ref{subsec:hard} for the hard-sphere type acceleration and in Section \ref{subsec:Kol} for the Kolmogorov type. We discuss the diffusive escape of high-energy CRPs in Section \ref{subsec:esc}.

\subsection{Overview of the time evolution of cosmic-ray spectra\label{subsec:overview}}

\begin{figure*}[tbh]
	 \plottwo{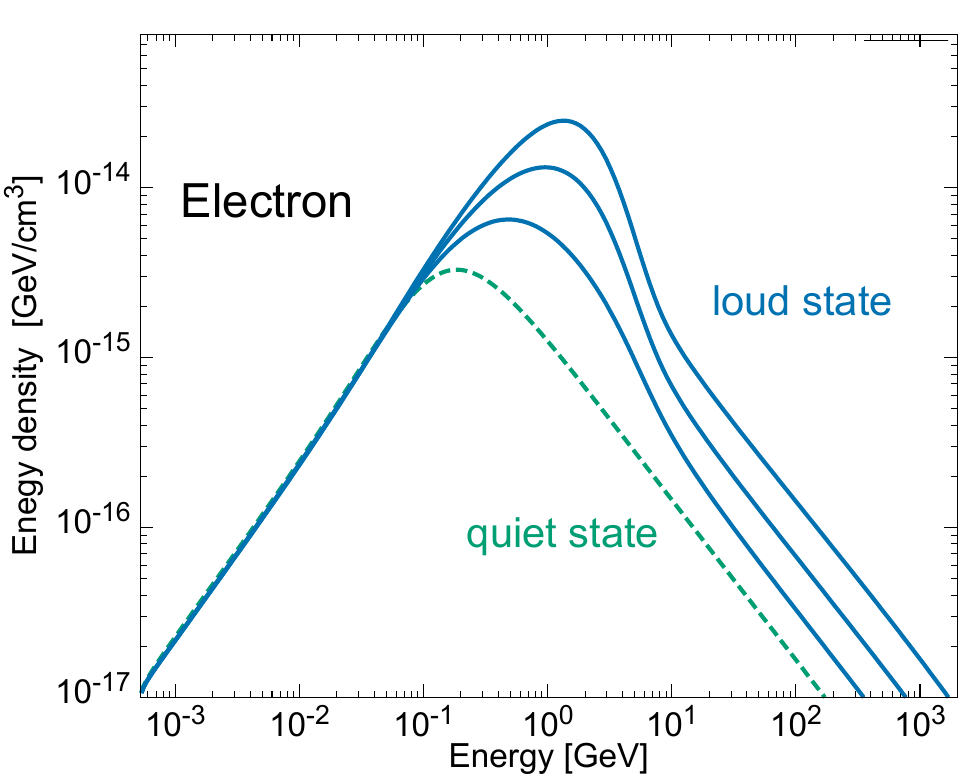}{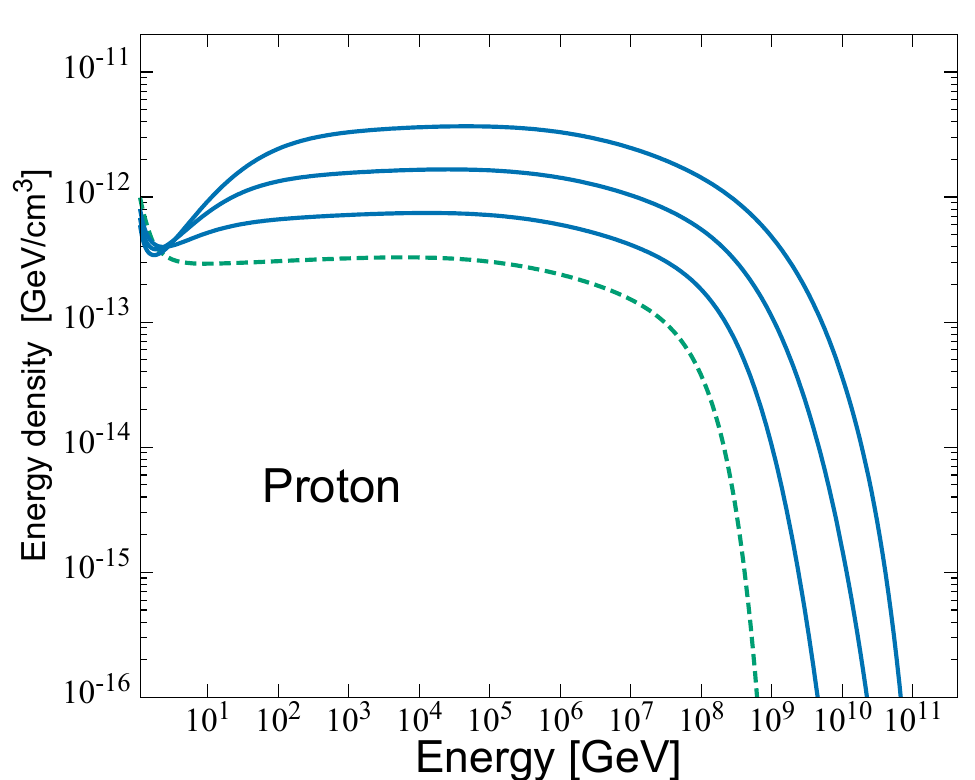}
	 \plottwo{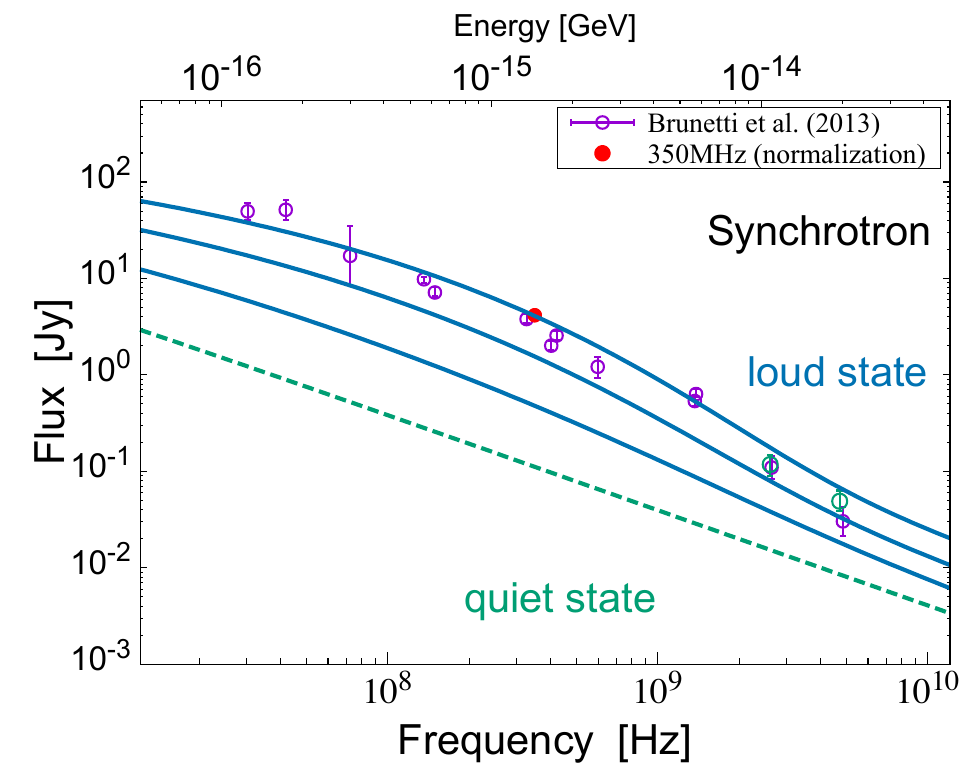}{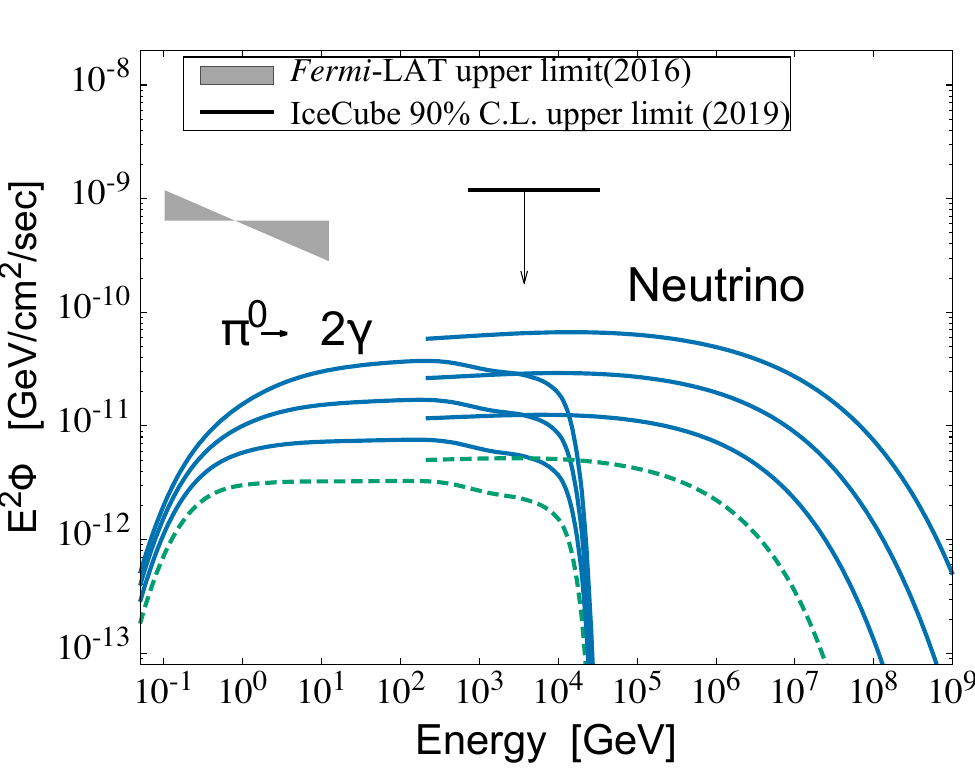}
     \caption{Time evolution of the CR energy distributions ({\it top}) and corresponding radiations ({\it bottom}) for an example case with $B_0=4.7$ $\mu$G, $\eta_B=0.5$, the hard-sphere type acceleration ($q=2$) and the injection spectral index $\alpha=2.0$ in primary-dominant scenario ($f_\mathrm{ep}=0.01$). The CR spectra are averaged within the core radius $r\le r_\mathrm{c} = 290$ kpc. The aperture radii $r_\mathrm{ap}$ used to calculate the emission fluxes are different for each radiation; $r_\mathrm{ap}=525$ kpc for synchrotron, $2.0$ Mpc for $\pi^0$ gamma-ray and $r_{200}$ for neutrinos. In each figure, from bottom to top, the spectra at $t_\mathrm{R} = 200$, 400 and 500 Myr are shown with solid curves. The spectra before the reacceleration ($t_\mathrm{R}=0$ Myr) are shown with dashed curves.  {\it Left} panels: The CRE and synchrotron spectra. 
{\it Right} panels: The spectra of CRPs and hadronic radiations. The neutrino fluxes are the sum of the ones of all flavors. In the right bottom panel, the results are compared with upper limits given by {\it Fermi}-LAT and IceCube (see the main text for detail). The neutrino spectra below 300 GeV are not shown to improve the visibility. \label{fig:main}}
\end{figure*}

Figure \ref{fig:main} shows example results for a model with a given injection profile $K(r)$ (we will explain this function is determined later on), the primary-dominant injection ($f_\mathrm{ep}=0.01$), the hard-sphere type reacceleration ($q=2$), and the injection spectral index of $\alpha=2.0$. Before the reacceleration starts, the CRE spectrum (dashed line in the left top panel) has a shape characterized by a single bump, reflecting the energy dependence of the cooling time shown in Figure \ref{fig:time}. The cooling of low-energy CREs is dominated by the Coulomb collisions with thermal particles (Eq. (\ref{eq:bC})), while radiative cooling dominates at higher energies. The maximum cooling time of CREs appears at energies $\sim$100 MeV. The radio data are taken from \citet{Brunetti2013}. The green empty points at 2.7 and 4.8 GHz show the flux corrected for the decrement due to thermal SZ effect. \par

The reacceleration lifts the bump up to the energy at which the acceleration balances the cooling. This energy can be found in Figure \ref{fig:time} as a cross-point of the timescales of those processes, and it is $\approx2$ GeV for the hard-sphere model with $\tau_\mathrm{acc} =260$ Myr. This shift of the bump shape affects the resulting radio spectrum (left bottom). The adequate choice of $\tau_\mathrm{acc}$ makes a break of the spectrum around a few GHz. In other words, $\tau_\mathrm{acc}\approx300$ Myr is required to fit the observed spectra when CRs are injected with a single power-law spectrum. \par

On the other hand, the spectral shape of CRPs remains to be the single power-law with an exponential cutoff, since CRPs do not suffer significant cooling and the acceleration timescale of the hard-sphere reacceleration does not depend on the momentum of particles (Section \ref{subsec:acc}). The maximum energy reaches $10^{19}$~eV within $t_\mathrm{R}\lesssim500$ Myr. The spectra of gamma-rays and neutrinos follow the evolution of the CRP spectrum and resulting fluxes are about one order of magnitude larger than those in the quiet state. \par

In the Kolmogorov model ($q=5/3$), the acceleration timescale is longer for higher energy particles (Eq. (\ref{eq:tacc})), so CRPs above $\sim1$ TeV are not efficiently reaccelerated within the reacceleration phase of $t_\mathrm{R}\lesssim1$~Gyr.
Thus, the predicted fluxes of hadronic emission and escaping high-energy CRPs become much lower than the hard-sphere model (see Section \ref{subsec:Kol}). \par
  
\subsection{Constraints on model parameters \label{subsec:halo}}
\begin{figure*}[tbh]
	\centering
		\plottwo{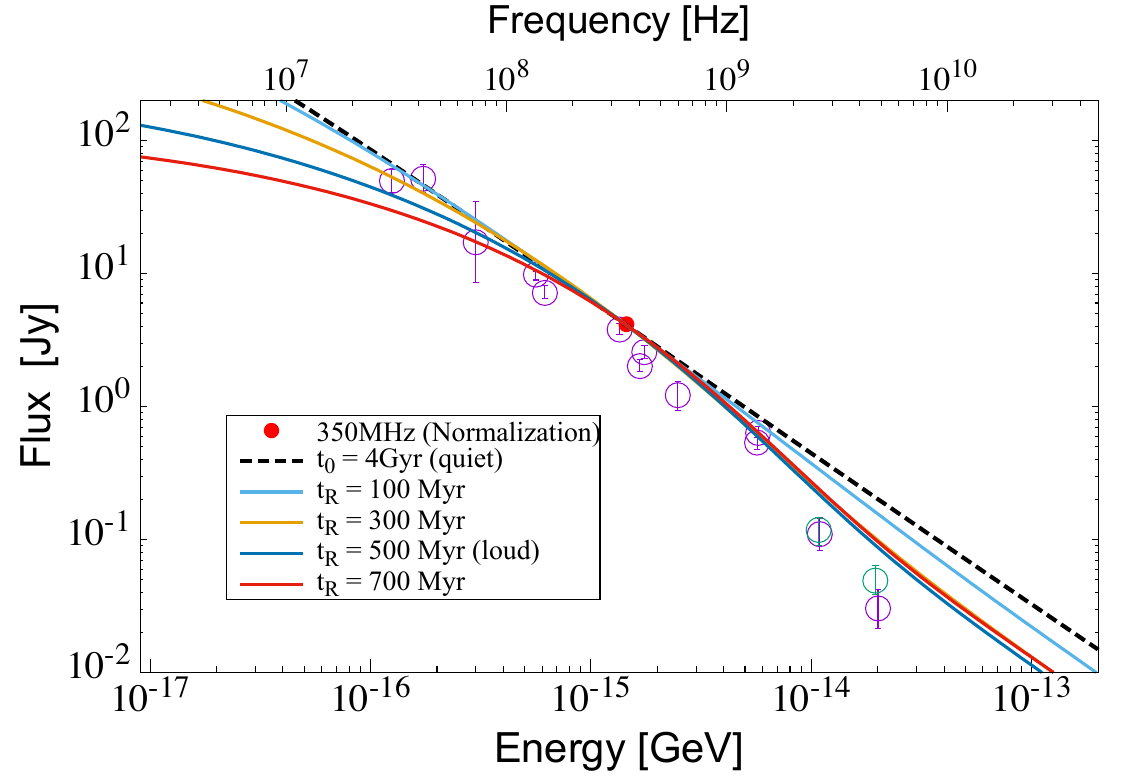}{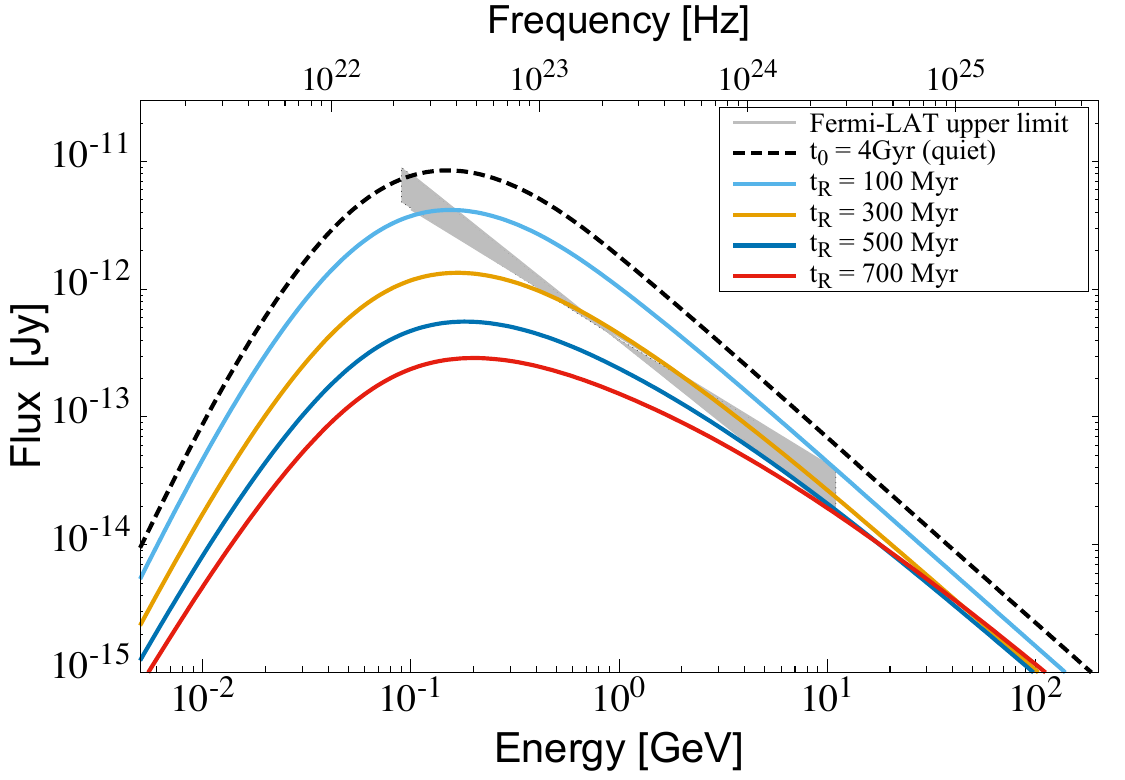}
	    \caption{Normalized spectra of the RH ({\it left}) and gamma rays from $\pi^0$ decay ({\it right}) for various $t_\mathrm{R}$ (the elapsed time of the reacceleration phase) for no primary case ($f_\mathrm{ep}=0$) with hard-sphere type reacceleration ($q=2$). The model parameters used here are same as Figure \ref{fig:main}. At each $t_\mathrm{R}$, the radio flux at 350 MHz is used to normalize the gamma-ray flux. The black dashed line shows the result of the pure-secondary model, where $t_\mathrm{R}=0$. The data points on the left panel are taken from \citet{Brunetti2017}. The gamma-ray fluxes on the right panel are compared with the {\it Fermi}-LAT limit from \citet{Ackermann2016}.\label{fig:halo}}
\end{figure*}

In the following, we show constraints on the duration of the reacceleration phase $t_\mathrm{R}$ and the radial profile of the primary CR injection, $K(r)$. 
In our calculation, $t_\mathrm{R}$ is mainly constrained from the RH spectrum. Figure \ref{fig:halo} (left) shows the $t_\mathrm{R}$ dependence of the shape of the synchrotron spectrum. In this figure, the primary CR injection rate is adjusted to realize the observed flux at 350 MHz for each $t_\mathrm{R}$ so that we can compare the shape of the spectrum for different $t_\mathrm{R}$. The flux, in practice, increases with time as shown in Figure \ref{fig:main} (bottom left). The dashed line in Figure \ref{fig:halo} represents the case without reacceleration (pure-secondary model), which is in tension with the observed break at 1 GHz. Note that the aperture radius assumed in radio data is not strictly constant with frequency. The tension in the single power-law model would be relaxed, when a much softer injection index ($\alpha\approx2.8$) is adopted and the normalization is not anchored to the flux at 350 MHz. \par
The parameters that affect the spectral shape are the injection index $\alpha$, parameters for the reacceleration, $\tau_\mathrm{acc}$ and $ t_\mathrm{R}$, and the amount of primary CREs $f_{\mathrm{ep}}$. The radio flux is mostly contributed by CREs in the core region, $r\le r_\mathrm{c}$, where the magnetic field is relatively strong. Hence the radial dependence of the injection, $K(r)$, dose not greatly affect the radio spectrum.

\begin{figure*}[tbh]
	\centering
	\plottwo{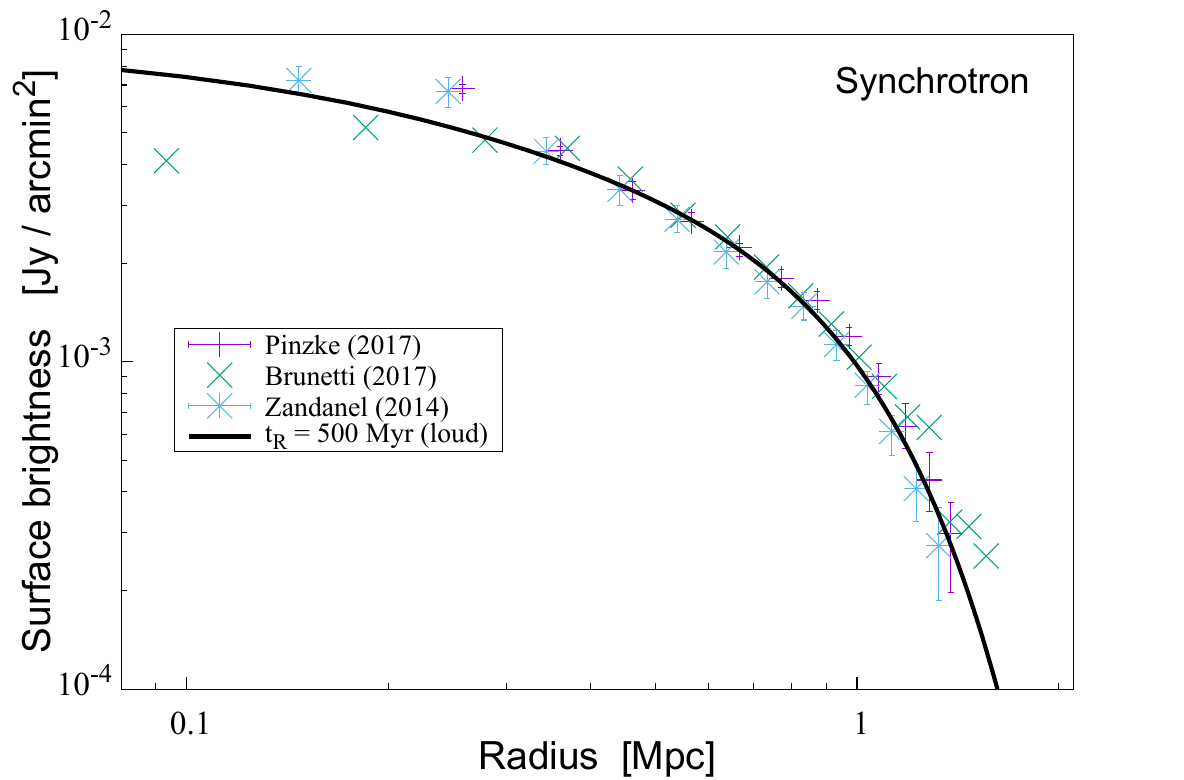}{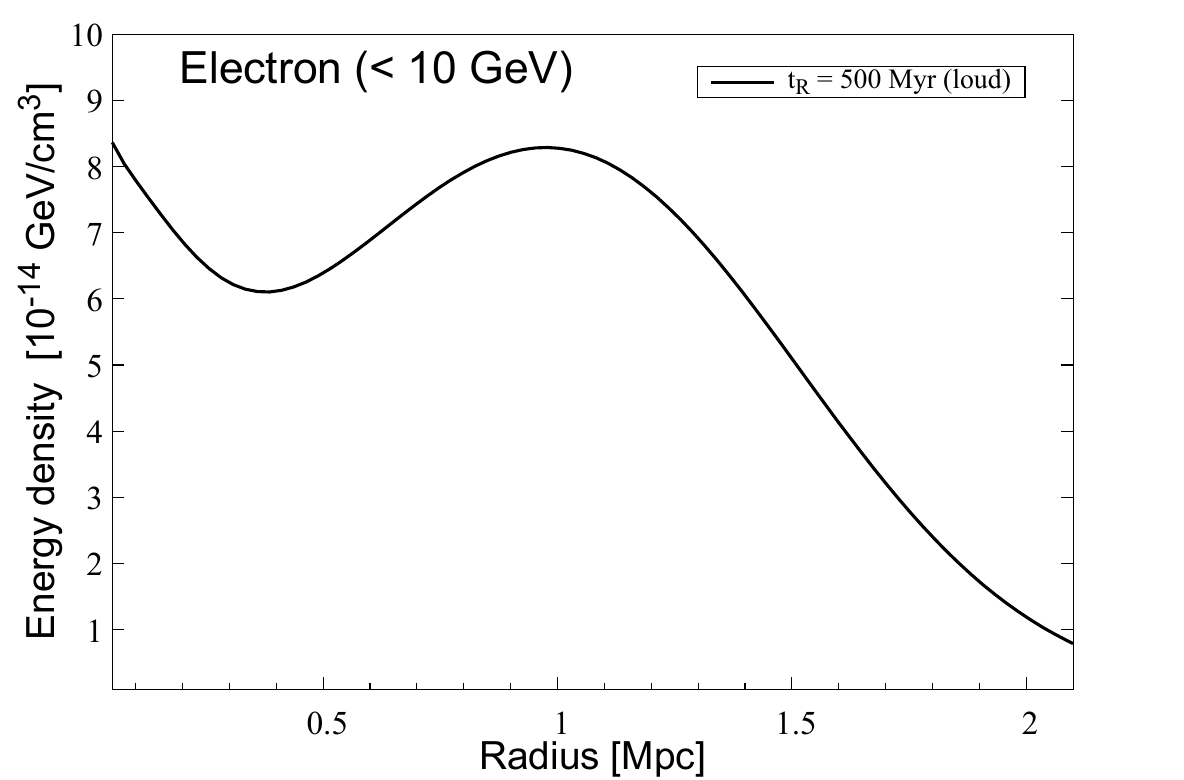}
	\caption{ {\it Left}: Radio surface brightness profile at 350MHz for the same model as Figure \ref{fig:main} and Figure \ref{fig:halo}, i.e., the secondary-dominant model with the hard-sphere reacceleration, $B_0=4.7$ $\mu$G and soft injection $\alpha=2.45$. The data points on the left panel are taken from various papers \citep{Brunetti2017,Pinzke2017,Zandanel2014b}, which are originally based on the same Westerbork Synthesis Radio Telescope (WSRT) observation \citep{Brown2011}. {\it Right}: Radial distribution of the CRE energy density in the same calculation as the left panel. To clarify the connection with radio emission, only CREs of energy less than 10 GeV are considered here.\label{fig:surface}}
\end{figure*}

The upper limit on gamma-ray flux gives another constraint on the $t_\mathrm{R}$. In Figure \ref{fig:halo} (right), we show the gamma-ray fluxes normalized by the radio flux of 350 MHz at each $t_\mathrm{R}$. As discussed in \citet{Brunetti2017}, the ``pure-secondary model" ($t_\mathrm{R}=0$ and $f_\mathrm{ep}=0$) is in tension with the limit from {\it Fermi}-LAT data, when $(B_0, \eta_B)$ = $(4.7\;\mu\mathrm{G}, 0.5)$. In order to relax the tension, the magnetic field is required to be $\sim$10 times larger than the one we adopted. In general, the spectrum of CREs above $\sim100$ MeV is softer than that of CRPs because high energy electrons quickly lose their energies through radiation, so the relative increase of the energy density of radio-emitting CREs due to reacceleration is lager than that for gamma-ray--emitting CRPs.
Because of this, the ratio between the fluxes of radio and gamma-ray, $F_{\mathrm{radio}}/F_\gamma$, increases with $t_\mathrm{R}/\tau_\mathrm{acc}$, so the upper limit on $F_\gamma$ gives the lower bound for $t_\mathrm{R}$. From this figure, for example, $t_\mathrm{R}\gtrsim400$ Myr is required from the gamma-ray upper limit. \par

Our model should also reproduce the observed surface brightness profile of the RH \citep[e.g.,][]{Brown2011}.
The current spatial distribution of CREs is the consequence of the combined effects of various processes: primary injection, spatial diffusion, and secondary injection from inelastic $pp$ collisions, so we need to follow the diffusive evolution of CR distributions with a modeled injection that persists over several Gyrs.\par
In the secondary-dominant model, the injection rate of CREs is proportional to the product of the densities of parental CRPs and thermal protons. \citet{Brunetti2017} pointed out that the ratio of the CRP energy density to the thermal energy density needs to increase with radius to reproduce the broad profile of the radio surface brightness of the Coma RH, considering the gamma-ray upper limit given by {\it Fermi}-LAT. This may suggest that the injection of CRs occurs at peripheral regions rather than the central region, where the thermal gas density is relatively large. \par

Since both Coulomb and synchrotron coolings are weaker at larger $r$, the relative increase of the synchrotron emissivity due to the reacceleration is more prominent at larger $r$. Thus, the surface brightness profile becomes broad with $t_\mathrm{R}$.
\par

\begin{figure*}[tbh]
    \centering
    \plottwo{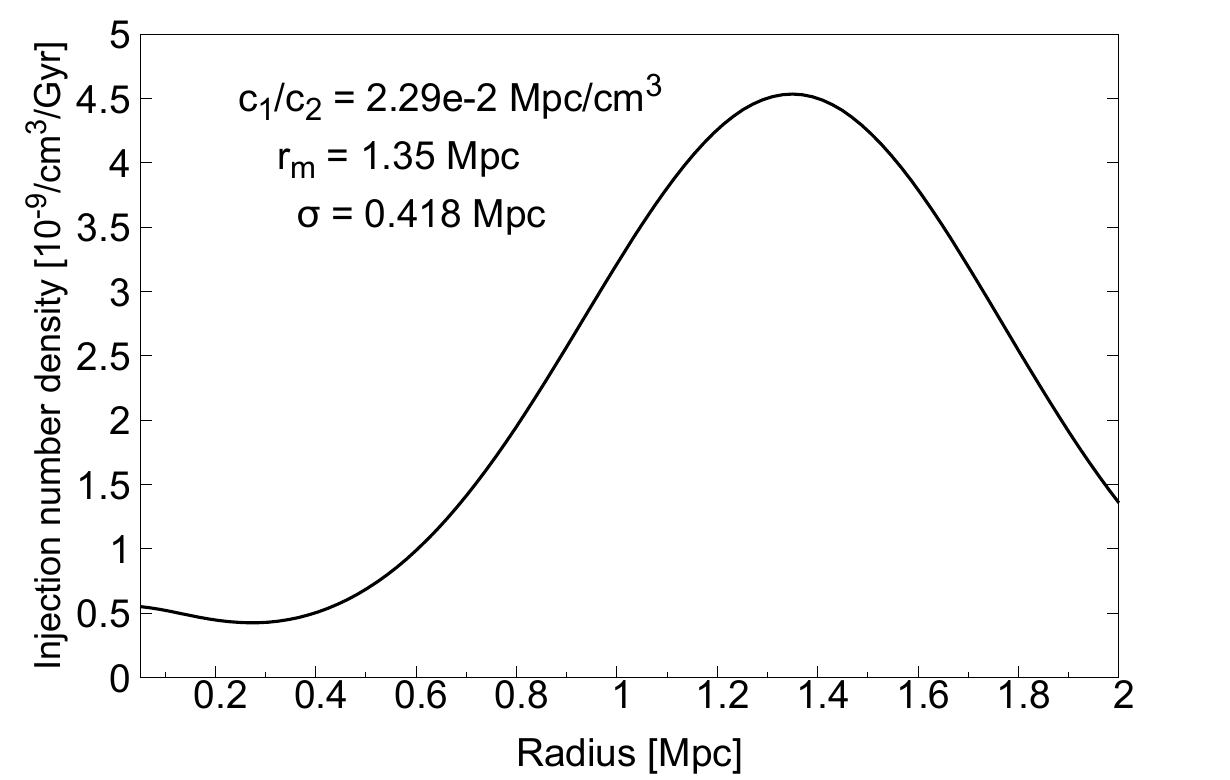}{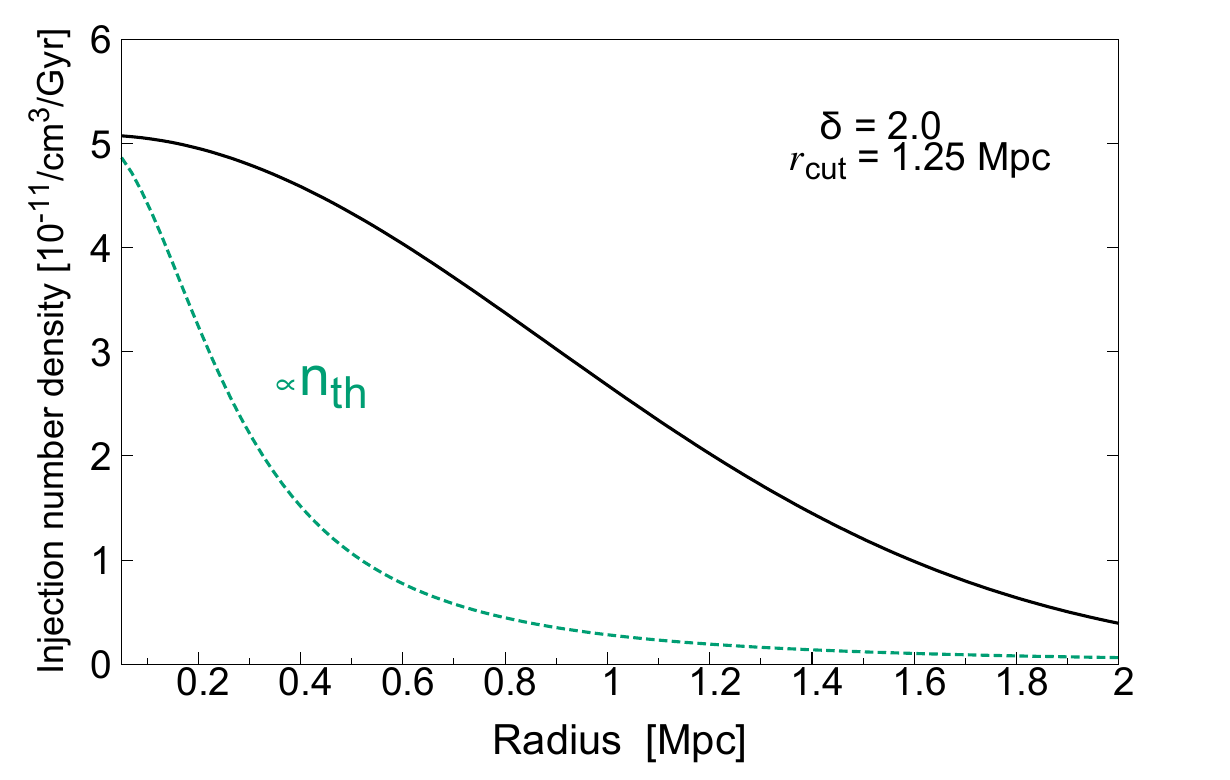}
    \caption{Primary CRP injection profiles as a function of radial distance for $f_{\mathrm{ep}}=0$ ({\it left}) and $f_{\mathrm{ep}}=0.01$ ({\it right}). The hard-sphere reacceleration ($q=2$) with $B_0=4.7$ $\mu$G and $\alpha=2.45$ are assumed here. The dashed line in the right figure shows a profile proportional to the thermal gas density. \label{fig:inj_alpha2}}
   \end{figure*}

We have tested various injection profiles and confirmed that those biased to the center, for example, $K(r)\propto \delta(r)$ or $K(r)\propto 4\pi r^2n_\mathrm{th}$, are rejected, if $D_{pp}$ is constant with radius. Such profiles do not produce extended halo emission but small core emission. Figure \ref{fig:surface} shows the surface brightness profile of the RH and the corresponding CRE distribution. The CRE distribution needs to be roughly uniform within $r\lesssim1$ Mpc. The two-peaked feature in the spatial distribution of the CREs is caused by a combination of the injection profile of primary CRPs (Figure~\ref{fig:inj_alpha2} left), and the cored profile of the ICM (Eq.~(\ref{eq:beta-model})).  \par
Since the giant RH extends up to 1 Mpc from the center, a sufficient amount of primary CRs should be supplied outside the cluster core. 
Especially in the secondary-dominant scenario, the density of the primary CRP should increases with $r$ and has a peak at $\sim 1$ Mpc to realize the CRE distribution shown in Figure \ref{fig:surface}.
Such a profile of CRPs could originate from an injection from the shock waves induced by the cluster formation process, such as mergers of clusters or mass accretion. Considering the injection from a shock front located at $r \sim r_\mathrm{m}$ and internal sources such as AGNs, we use the following expression of $K(r)$ for the secondary-dominant model ($f_{\mathrm{ep}}=0$):
\begin{eqnarray}
K_1(r)=4\pi r^{2}&&\biggl[\frac{c_1}{\sqrt{2\pi}\sigma}\exp\left[-\frac{(r-r_\mathrm{m})^2}{2\sigma^2}\right]\nonumber\\
&&+c_2n_{\mathrm{th}}(r)\biggr].
\label{eq:inj1}
\end{eqnarray}
The component proportional to $n_{\mathrm{th}}$ represents the injection from internal sources. The factor $4\pi r^2$ is introduced to convert the volume density into the linear density. We find that typical values of the parameters are $c_1/c_2\simeq$ 3.5$\times10^{-2}$ Mpc/cm$^3$, $r_\mathrm{m}\simeq1.3$ Mpc, and $\sigma\simeq 0.45$ Mpc. The values of those parameters adopted in our calculations are summarized in Table \ref{tab:sec}. \par

\begin{table*}[tbh]
\centering
\caption{Parameters for the secondary-dominant models ($f_\mathrm{ep}=0$) \label{tab:sec}}
\begin{tabular}{ccccccc}
\hline\hline
$q$&$\alpha$&$t_\mathrm{R}$&$c_1/c_2$&$r_\mathrm{m}$&$\sigma$&$L_\mathrm{p}^\mathrm{inj}(>10\; \mathrm{GeV})$\\
&&[Myr]&[Mpc/cm$^3$]&[Mpc]&[Mpc]&[erg/s]\\ \hline
2& 2.0&400&$3.65\times10^{-2}$&1.39&0.418&$1.9\times10^{45}$ \\
(hard-sphere)&2.1&400&$3.65\times10^{-2}$&1.38&0.418&$7.4\times10^{44}$ \\
&2.2&400&$3.32\times10^{-2}$&1.35&0.418&$5.4\times10^{44}$\\
&2.45&500&$2.29\times10^{-2}$&1.35&0.418&$1.7\times10^{44}$\\
\hline
5/3&2.0&180&$4.11\times10^{-2}$&1.55&0.474&$9.7\times10^{44}$\\
(Kolmogorov)&2.1&160&$4.11\times10^{-2}$&1.55&0.474&$8.4\times10^{44}$\\
&2.2&160&$4.63\times10^{-2}$&1.55&0.474&$5.9\times10^{44}$\\
&2.45&160&$2.29\times10^{-2}$&1.25&0.418&$1.7\times10^{44}$\\
\hline
\end{tabular}
\\{\bf Note}: $c_1$, $c_2$, $r_\mathrm{m}$ and $\sigma$ are the parameters for $K_1(r)$ (Eq. (\ref{eq:inj1})). 
\end{table*}

The appropriate choice of $t_\mathrm{R}$ and $K(r)$ should be changed when primary electrons are present. In the primary-dominant case, $K(r)$ is roughly proportional to the current distribution of the CREs (Figure \ref{fig:surface} (right)), since the radial diffusion of $\sim$ GeV CREs is not efficient. Hence, the injection profile needs to be nearly uniform within $\sim1$ Mpc :

\begin{eqnarray}
K_2(r)&=&r^{\delta}\exp\left[-\left(\frac{r}{r_{\mathrm{cut}}}\right)^\delta\right].
\label{eq:inj2}
\end{eqnarray}
This functional shape implies $\frac{dK_2}{dr}=0$ at $r_\mathrm{cut}$, which means that CRs are typically injected around $r\sim r_\mathrm{cut}$ even in the primary-dominant model. This may suggest that primary sources are distributed over the halo volume, or the injection radius shifts with time to achieve the above functional shape just before the onset of the reacceleration. The typical values of the parameters are $\delta\simeq2$ and $r_{\mathrm{cut}}\simeq 1.2$ Mpc (Table \ref{tab:pri}).

\begin{table}[htb]
\begin{center}
\caption{Parameters for the primary-dominant models ($f_\mathrm{ep}=0.01$) \label{tab:pri}}
\begin{tabular}{cccccc}
\hline\hline
$q$&$\alpha$&$t_\mathrm{R}$&$r_\mathrm{cut}$&$\delta$&$L_\mathrm{p}^\mathrm{inj}(>10\;\mathrm{GeV})$\\
&&[Myr]&[Mpc]&&[erg/s]\\ \hline
2& 2.0&600&1.70&2.3&$5.6\times10^{43}$ \\
(hard-sphere)&2.1&600&1.70&2.1&$2.0\times10^{43}$ \\
&2.2&600&1.60&2.1&$8.7\times10^{43}$\\
&2.45&800&1.25&2.0&$5.9\times10^{41}$\\
\hline
5/3&2.0&240&1.40&2.0&$1.0\times10^{43}$\\
(Kolmogorov)&2.1&240&1.30&1.9&$1.6\times10^{42}$\\
&2.2&240&1.25&1.9&$5.6\times10^{41}$\\
&2.45&240&1.15&1.8&$5.1\times10^{40}$\\
\hline
\end{tabular}
\end{center}
{\bf Note}: $r_\mathrm{cut}$ and $\delta$ are the parameters for $K_2(r)$ (Eq. (\ref{eq:inj2})). 
\end{table}

Figure \ref{fig:inj_alpha2} shows the radial dependence of the injection density of primary CRPs for models with hard-sphere reacceleration and $\alpha=2.45$. Those profiles are derived under the assumption that the efficiency of the acceleration, or $\tau_\mathrm{acc}$, is constant with $r$. If $\tau_\mathrm{acc}$ decreases with radius, the CR injection profile can be more concentrated on the center \citep[e.g.,][]{Pinzke2017,2021A&A...648A..60A}.  \par
The normalization of the injection is determined from the observed radio flux at 350 MHz. Once the model parameters, $\alpha$, $K(r)$ and $C^\mathrm{inj}_\mathrm{p}$, are given, the luminosity of the CR injection can be calculated by integrating Eq. (\ref{eq:pri}) over $r$ and $p$. The injection luminosity above 10 GeV is also shown in Tables \ref{tab:sec} and \ref{tab:pri}. Above that energy, CRPs are not significantly affected by the Coulomb cooling, so they lose their energies mainly through $pp$ collisions and the diffusive escape. The required luminosity ranges from $L_\mathrm{p}^\mathrm{inj}\sim10^{41}$ erg/s to $10^{45}$ erg/s.  \par

The fiducial value of the luminosity adopted in previous studies \citep[e.g.,][]{Murase2008,2009ApJ...707..370K,Kushnir_2010,Fang2016,2018NatPh..14..396F,Hussain:2021dqp} is $L_\mathrm{p}^\mathrm{inj}\sim10^{44}-10^{45}$ erg/s. The injection power of our secondary-dominant model ($f_\mathrm{ep}=0$) is comparable to those values. 
Note the additional energy injection from the turbulent reacceleration in our model. The hard-sphere reacceleration makes the energy density of CRs about ten times larger (see also Figure \ref{fig:CR}), so we need about ten times larger injection power in the pure-secondary model, where the reacceleration is absent.
\par
The injections required in primary-dominant models ($f_\mathrm{ep}=0.01$) are much smaller. The injection luminosity of $L_\mathrm{p}^\mathrm{inj}\gtrsim10^{45}$ erg/s overproduces the observed radio luminosity even without the reacceleration.

\subsection{Hard-sphere type acceleration \label{subsec:hard}}

\begin{figure*}[tbh]
 	\centering
   	 \plottwo{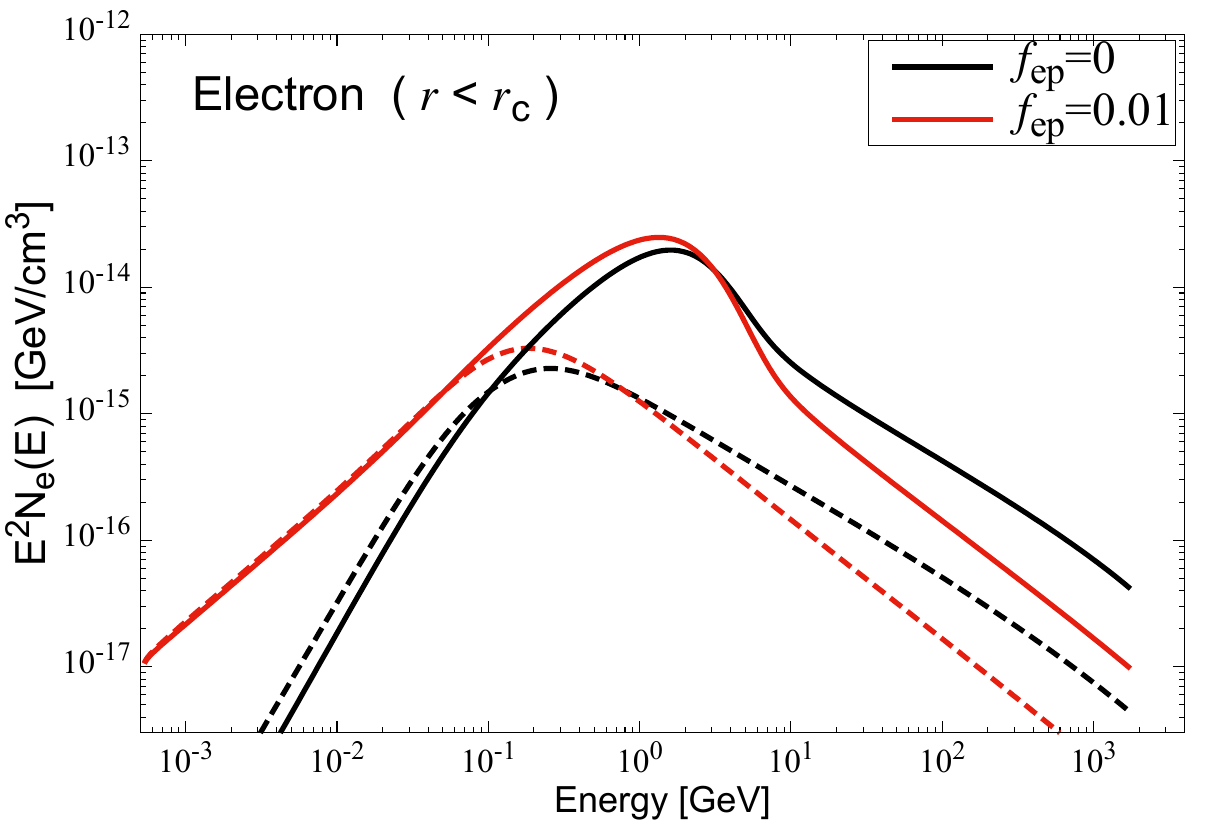}{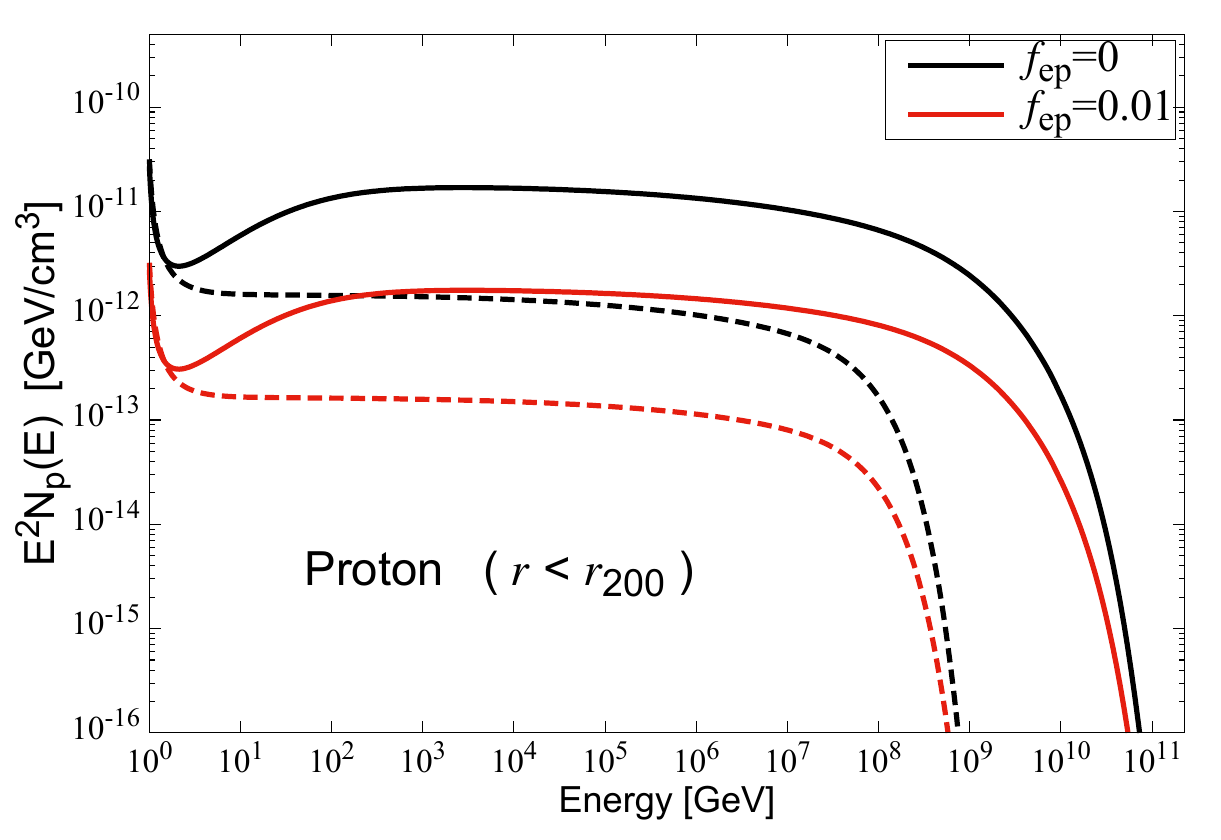}
	 \plottwo{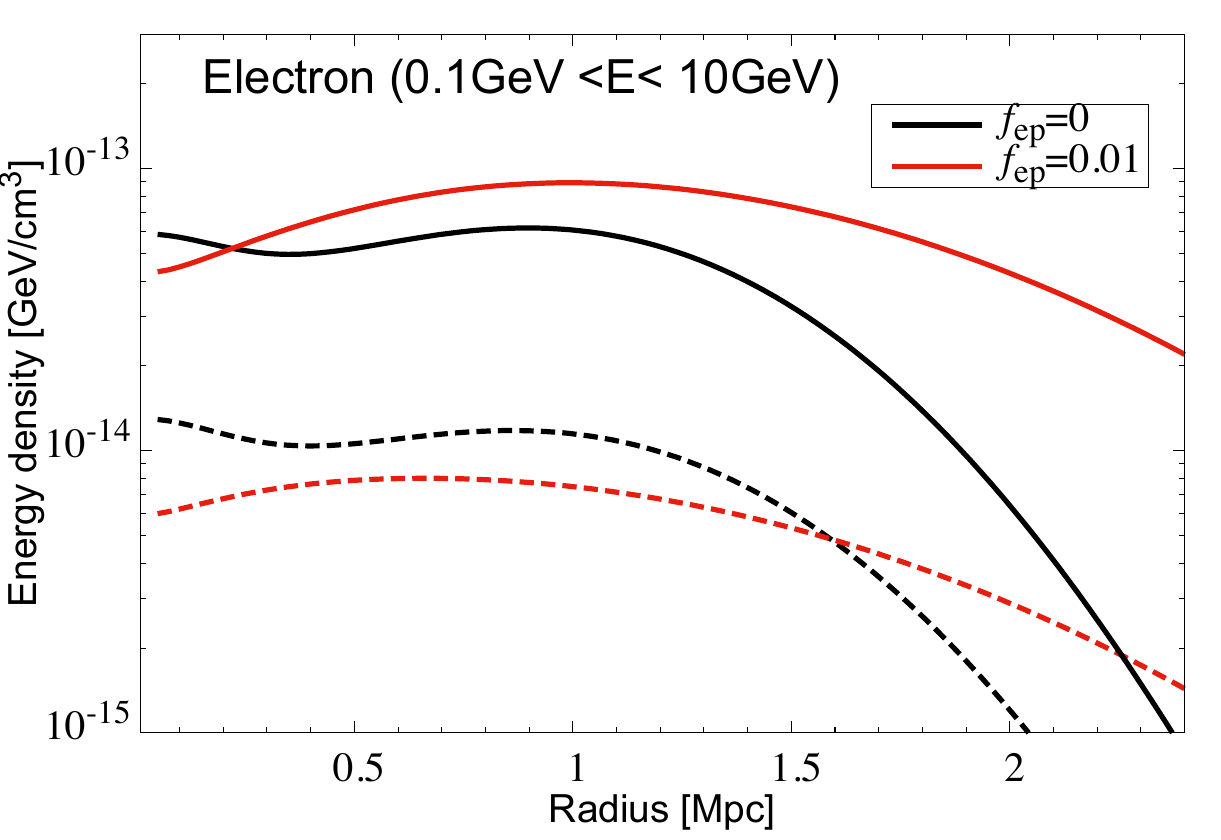}{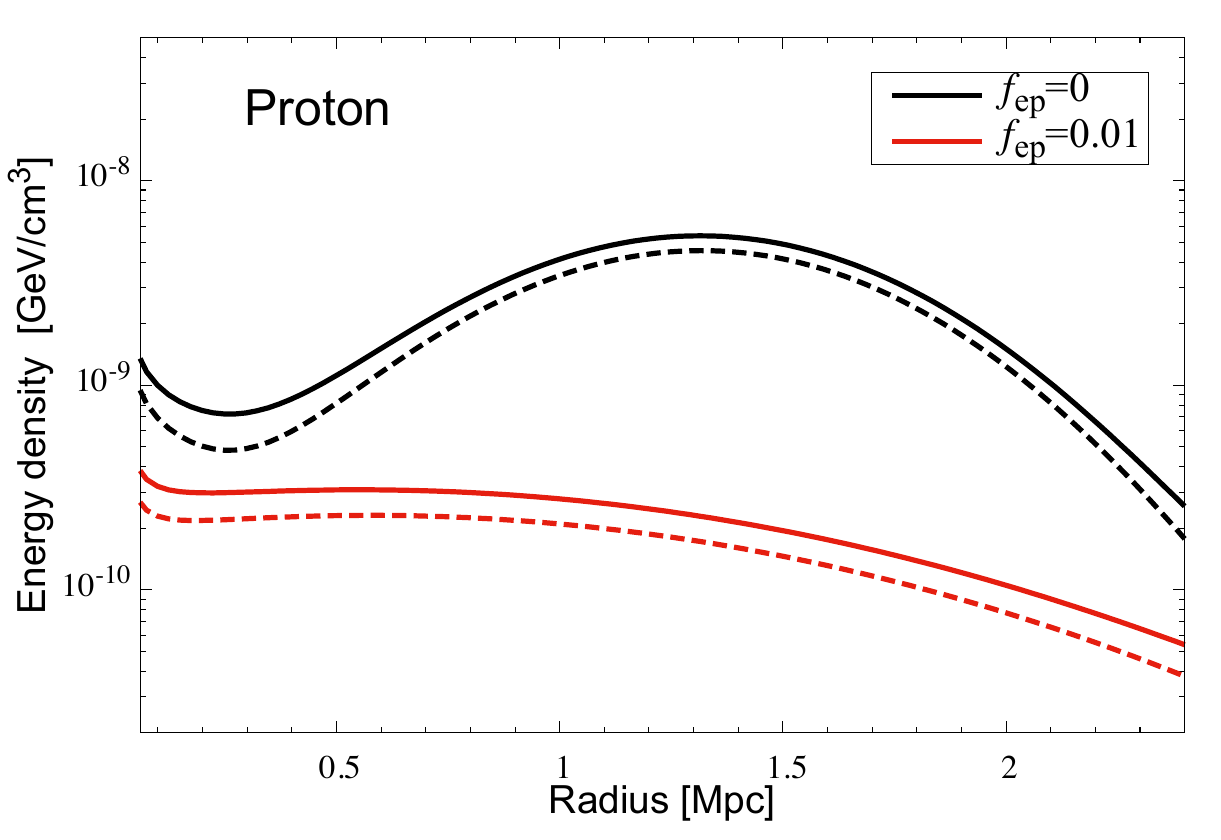}
    \caption{CR energy density distribution as a function of energy ({\it top}) and radial distance ({\it bottom}) for the hard-sphere model with $B_0=4.7$ $\mu$G and $\alpha=2.0$. The results for CREs and CRPs are shown in the left and right panels, respectively. The results for the secondary-dominant case ($f_\mathrm{ep}=0$) are shown in the black and those for the case with $f_\mathrm{ep}=0.01$ are in red. The ``loud" and ``quiet" state correspond to the solid and dashed curves, respectively. The CRE spectra ({\it top left}) are averaged within $r\le r_\mathrm{c}$, while the CRP spectra ({\it top right}) are averaged within $r\le r_{200}$. In the {\it left bottom} panel, the spectrum of CREs is integrated between $0.1 \le E \le10$ GeV to show the radial distribution of radio-emitting CREs.  \label{fig:CR}}
\end{figure*}

Hereafter in this section, we discuss multi-wavelength and neutrino emission from the Coma cluster based on the model for the RH explained above. The results for each model are summarized in the tables below (Tables \ref{tab:Hard} and \ref{tab:Kol}). First, we show the results for the hard-sphere type acceleration; $q = 2$. In this case, all CRs have the same $t_{\mathrm{acc}}$ regardless of their energies. That reacceleration produces high-energy CRPs more efficiently, and the emissivities of the hadronic emission become larger than the Kolmogorov reacceleration.\par

The resulting CR distributions are shown in Figure \ref{fig:CR}. The top panels show the energy spectra of CREs and CRPs, while the bottom panels show their spatial distributions. 

Primary CREs are distributed down to trans-relativistic energies ($E_\mathrm{e}\simeq 0.1m_\mathrm{e}c^2$), while secondary CREs with energies less than 1/10 of the $pp$ threshold energy $E_\mathrm{th}$ are hardly produced. That causes the difference in the CRE spectra (top left panel) below 100 MeV between $f_\mathrm{ep}=0$ and $f_\mathrm{ep}=0.01$.

We normalize the results using the synchrotron flux at 350 MHz, and this frequency corresponds to the electron energy of $E_\mathrm{e}\simeq 2.6$ GeV for $B=4.7$ $\mu$G. Thus, the amount of CREs at 2.6 GeV should be the same in all models at the radio-loud state. In reality, there is a small deviation at that energy, which may arise from the difference in the radial distribution of CREs (left bottom panel, see also Figure \ref{fig:surface}). The energy density of $\approx5\times10^{-14}~{\rm GeV}/{\rm cm}^3$ is in good agreement with other studies with the same assumption on the magnetic field \citep[e.g.,][]{2021A&A...648A..60A}.

\begin{figure*}[htp]
    \centering
  \includegraphics[width=\linewidth]{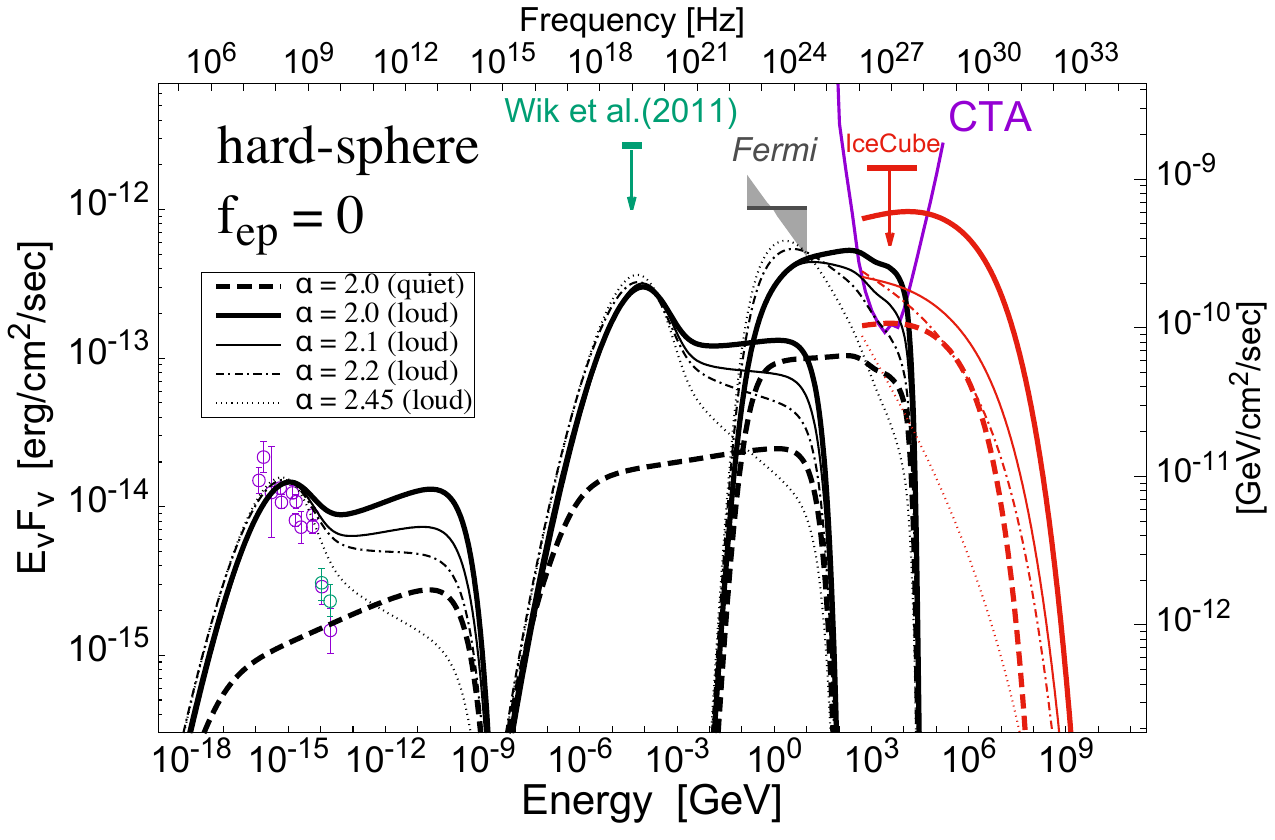}
    \caption{Non-thermal electromagnetic and all-flavor neutrino $\nu F_\nu$ spectra for the most optimistic cases; $q=2$ and $f_{\mathrm{ep}}=0$. From left to right, synchrotron, inverse-Compton and $\pi^0$ gamma emission is shown in black curves. The red curves show neutrino fluxes. The``loud" and ``quiet" state are shown in the thick solid and thick dashed curves, respectively. The loud state spectra for different $\alpha$ are also shown: $\alpha=2.1$ (thin solid), 2.2 (thin dot-dashed), and 2.45 (thin dotted). Note that each spectra is calculated with different aperture radius $r_\mathrm{ap}$ to compare with observations: $r_\mathrm{ap}= 0.525$ Mpc and $= 1.2$ Mpc for synchrotron and ICS spectra, respectively, while that for gamma-ray is 2.0 Mpc. The high-frequency cutoffs of the leptonic radiation is artificial ones. The upper limits for hard-X ray, gamma-ray and neutrinos are taken from \citet{Wik_2011}, \citet{Ackermann2015} and \citet{2020PhRvL.124e1103A}, respectively. The point source sensitivity of the Cherenkov Telescope Array (CTA) with 50 h observation is drawn with a magenta line (adopted from https://www.cta-observatory.org/science/cta-performance/).}
    \label{fig:nFn_H}
\end{figure*}

\begin{figure*}
\centering
    \includegraphics[width=\linewidth]{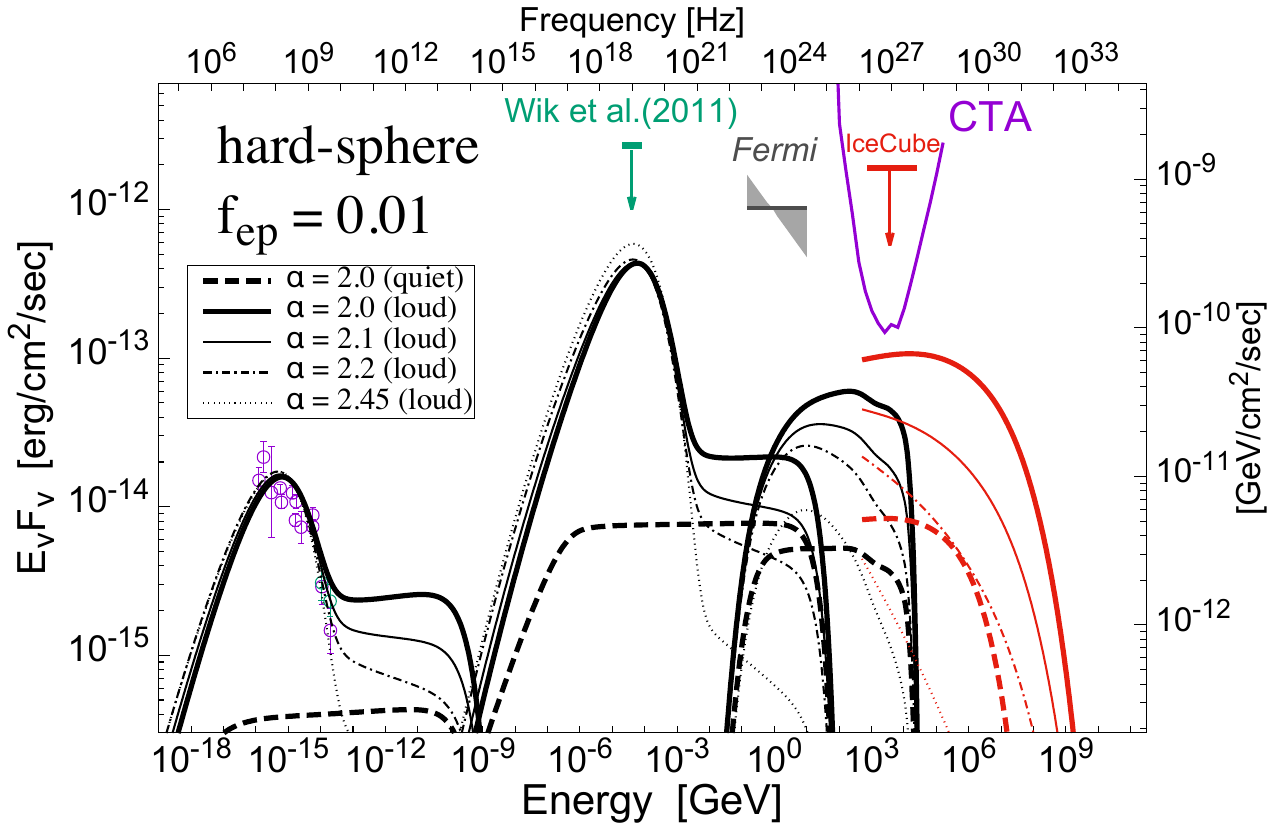}
    \caption{Multi-wavelength and neutrino $\nu F_\nu$ spectra for $f_{\mathrm{ep}}=0.01$. Other parameters are the same as Figure \ref{fig:nFn_H}. \label{fig:nFn_H_pri}}
\end{figure*}

The radial diffusion slightly flattens the distributions of CRPs (bottom right) compared to the injection profile, $K(r)$. On the other hand, CREs are more concentrated towards the cluster center for $f_\mathrm{ep}=0$, because the production of the secondary CREs is more efficient at smaller radius. This difference in radial distribution between CRPs and CREs is relatively small for $f_\mathrm{ep}=0.01$, since the distribution of primary CREs is not affected by the density profile of ambient ICM.
Figure \ref{fig:nFn_H} shows the overall spectrum of the non-thermal electromagnetic and all-flavor neutrino emission together with the observational data and upper limits. In the secondary-dominant models ($f_\mathrm{ep}=0$), we can expect larger fluxes of hadronic emission than the primary-dominant models ($f_\mathrm{ep}=0$, Figure \ref{fig:nFn_H_pri}). Gamma rays above $\sim$TeV energies are attenuated by interactions with the EBL (Sect. \ref{subsec:emiss}). The cutoff shape appearing in the neutrino fluxes simply reflects the cutoff in the CRP spectra, so the flux above 1 PeV is sensitive to $E_\mathrm{p}^\mathrm{max}$ (Eq. (\ref{eq:pri}), see also Section \ref{subsec:diffuse}). The upper limit on the neutrino flux in this figure is given by the point source search with ten years of IceCube data \citep{2020PhRvL.124e1103A}. That shows the median upper limit of the flux from the direction of the Coma cluster at a 90\% confidence level. The angular extension of the Coma cluster is not considered here, because the extension is comparable to the angular resolution of muon track events ($\sim1$~{\rm deg} at TeV energies)~\citep{Murase:2012rd}. Our result is consistent with the lack of a significant excess of neutrino events from Coma \citep[e.g.,][]{2014ApJ...796..109A}. \par

A unique feature of our 1D calculation appears in the spectra of the leptonic radiation in Figure \ref{fig:nFn_H}. In this model, the synchrotron spectrum does not fit the data well for any choice of $t_\mathrm{R}$ within $150\le t_{\rm R}\le 500$ Myr. The resulting spectra are clearly harder than the observational data, especially at higher frequencies above 1.4 GHz. In the case of $\alpha=2.0$, the radio spectral index at the quiet state (black dashed line) becomes $\alpha_{\rm syn}\approx-0.9$, while it is expected to be $\alpha_{\rm syn}=-1.0$, when the injection spectrum of secondary CREs follows $Q_{\rm e}^{\rm sec}\propto p^{-2.0}$. One of the possible causes of this spectral hardening is the energy-dependent diffusion of parental CRPs. Since most of the CRPs are injected outside the core (Figure \ref{fig:inj_alpha2}) and the diffusion is faster for higher-energy CRPs ($t_\mathrm{diff}\propto E^{-1/3}$), the CRP spectra are harder in the core region than in the injection region. Secondary CREs also show hard spectra in the core region (Figure \ref{fig:CR} top left), where the magnetic field is strong. To confirm this, we tested the case without radial diffusion ($D_{rr}=0$, not shown in the figure) and found that the spatial diffusion actually hardens the CRE spectral index by $\approx 0.05$. The weak energy dependence in the $pp$ cross section is another cause of the spectral hardening (Sect.~\ref{subsec:secondary}). That makes the spectral index of $Q_{\rm e}^{\rm sec}$ above $\gamma_{\rm e}>10^3$ harder by $\approx0.05$, compared to the case of $\sigma_{pp}={\rm Const.}$ \citep[e.g.,][]{Kelner2006} \par
Note that the brightness profile of the RH can also be explained by, e.g., the ``{\it M}-turbulence" model of \citet{Pinzke2017}, where the efficiency of the reacceleration increases with radius and the CR distribution is more concentrated towards the central region. In such models, the spectral hardening due to the radial diffusion is not effective, and the tension between observed and calculated RH spectra could be relaxed (see also Sect.~\ref{subsec:caveats}). \par

In the reacceleration phase, the flux above $\sim$ 1 GHz, where the cooling timescale becomes shorter than the reacceleration timescale, also increases with time, because the injection rate of the secondary CRE increases due to the reacceleration of CRPs.\par 

The emissivity of ICS is independent of the cored structure of the galaxy cluster, so the resulting spectrum is softer than synchrotron. ICS from CREs contributes to the gamma-ray flux at 1 GeV up to 0.3 times as much as the emission from the decay of $\pi^0$. We did not solve the evolution of high-energy CREs with Lorentz factor larger than $\gamma_\mathrm{e}=10^7$ to save the computation time, so the high-frequency cutoff shown in the leptonic radiation is an artificial one.\par

The radial and spectral distributions of CRs for the primary-dominant case ($f_\mathrm{ep}=0.01$) are shown in Figure \ref{fig:CR} with red lines. In this case, the CRE spectrum becomes softer than $f_\mathrm{ep}=0$, since the injection profile is nearly uniform within the RH and the CRE spectrum is not significantly affected by the spatial diffusion of CRPs. For all values of $\alpha=$2.0, 2.1, 2.2, and 2.45, the radio spectrum can be reproduced with optimal values of $t_\mathrm{R}$ listed in Table \ref{tab:pri}. Therefore, the primary-dominant model is more preferable to the secondary-dominant model unless steeper index $\alpha\gtrsim2.5$ or radially increasing $D_{pp}$ is adopted. \par

The $\nu F_\nu$ fluxes for $f_{\mathrm{ep}}=0.01$ are shown in Figure \ref{fig:nFn_H_pri}. The fluxes of hadronic emission is about one order of magnitude smaller than the secondary-dominant cases. They are well below the upper limits, so $t_\mathrm{R}$ is constrained solely from the shape of the radio spectrum. The ICS spectrum around 100 keV is almost the same as the secondary-dominant case, since the leptonic radiation is constrained by the radio flux (see also Section \ref{subsec:hardX}). \par

In Table \ref{tab:Hard}, we summarize PeV neutrino fluxes for each $\alpha$ together with the fluxes of $\pi^0$ gamma rays and escaping CRPs (Sect. \ref{subsec:esc}). 

The models with smaller $\alpha$, i.e., harder injection, naturally predict larger neutrino fluxes. When the amount of CREs is constrained by the luminosity of the RH, less hadronic emission is predicted with larger $f_{\mathrm{ep}}$. The neutrino flux for $f_\mathrm{ep}=0.01$ is smaller by about one order of magnitude than $f_\mathrm{ep}=0$ (Table \ref{tab:Hard}). The luminosity of escaping CRs shown here is the one from the loud state. This luminosity is powered by the reacceleration, and it strongly depends on $t_\mathrm{R}/\tau_\mathrm{acc}$ and $E_\mathrm{p}^\mathrm{max}$ (see Section \ref{subsec:esc}). 
This luminosity can be comparable to the injection luminosity $L_\mathrm{p}^\mathrm{inj}$ listed in Tables \ref{tab:pri} and \ref{tab:sec}, since the power of the reacceleration $P_\mathrm{reacc}\sim10^{42}$-$10^{47}$ erg/s, depending on the parameters, can dominate the injection power.

 \begin{table*}[tb]
    \centering
    \caption{Predicted neutrino and gamma-ray fluxes together with the luminosity of escaping CRPs, for the hard-sphere models ($q = 2$) with $\tau_\mathrm{acc}=260$ Myr and $(B_0,\eta_B)=(4.7 \mu\mathrm{G},0.5)$. The maximum energy of primary protons is assumed to be $E_\mathrm{p}^\mathrm{max}=100$~PeV. \label{tab:Hard}}
    \centering
        \begin{tabular}{cccccccccc}
    \hline\hline
      $f_\mathrm{ep}\tablenotemark{a}$&$\alpha$  &$E_\nu F_\nu$(loud)\tablenotemark{b}&$E_\nu F_\nu$(quiet)\tablenotemark{c}&$E_\gamma F_\gamma$(loud)\tablenotemark{d}&$L_\nu$\tablenotemark{e}&$L_{\rm CRP}(>10^{17}~{\rm eV})$\tablenotemark{f}&$L_\mathrm{CRP}(>10^{18.5}~{\rm eV})$\tablenotemark{g}\\ 
      &&[GeV/cm$^2$/s]&[GeV/cm$^2$/s]&[GeV/cm$^2$/s]&[erg/s]&[erg/s]&[erg/s]\\
      \hline
     0& 2.0 &$4.1\times10^{-10}$&$4.3\times10^{-11}$&$2.2\times10^{-10}$&$6.6\times10^{42}$&$6.4\times10^{44}$&$1.3\times10^{43}$  \\
&2.1&$1.3\times10^{-10}$&$7.1\times10^{-12}$&$2.0\times10^{-10}$&$2.5\times10^{42}$&$1.7\times10^{44}$&$3.1\times10^{42}$\\
      &2.2&$5.3\times10^{-11}$&$8.2\times10^{-12}$&$3.1\times10^{-10}$&$1.0\times10^{42}$&$4.7\times10^{43}$&$4.8\times10^{41}$ \\
      &2.45&$2.7\times10^{-12}$&$6.3\times10^{-14}$&$3.7\times10^{-10}$&$1.1\times10^{41}$&$1.9\times10^{42}$&$3.9\times10^{40}$ \\ 
        \hline
      	0.01&2.0&$5.7\times 10^{-11}$&$2.4\times 10^{-12}$&$2.5\times10^{-11}$&$8.4\times10^{41}$&$8.9\times10^{43}$&$7.6\times10^{42}$\\
	&2.1&$1.1\times 10^{-11}$&$3.7\times 10^{-13}$&$1.4\times10^{-11}$&$2.1\times10^{41}$&$1.6\times10^{43}$&$1.2\times10^{42}$\\
	 &2.2 &$2.3\times 10^{-12}$&$6.1\times 10^{-14}$&$1.1\times10^{-11}$&$5.7\times10^{40}$&$3.0\times10^{42}$&$1.7\times10^{41}$\\
	&2.45&$9.0\times 10^{-14}$&$2.2\times 10^{-16}$&$5.6\times10^{-12}$&$3.3\times10^{39}$&$6.1\times10^{40}$&$8.6\times10^{38}$\\  
	\hline
    \end{tabular}
   \flushleft
   \tablecomments{$^a$ The ratio of Primary CREs to primary CRPs defined in Eq. (\ref{eq:Qepri}); $^b$ All flavor neutrino flux at 1 PeV for the radio-loud state; $^c$ The neutrino flux at 1 PeV for the radio-quiet state; $^d$ Gamma-ray flux at 1 GeV for the radio-loud state; $^e$ All flavor neutrino luminosity integrated above 10 PeV; 
   $^{f,g}$ Luminosity of CRPs escaping from the virial radius of the cluster $r=r_{200}\approx2.3$ Mpc of energy $E>100$ PeV and $E>10^{18.5}$ eV, respectively, at the radio-loud state.
}
   \end{table*}

\begin{figure}[tb]
\centering
    \includegraphics[width=8.6cm]{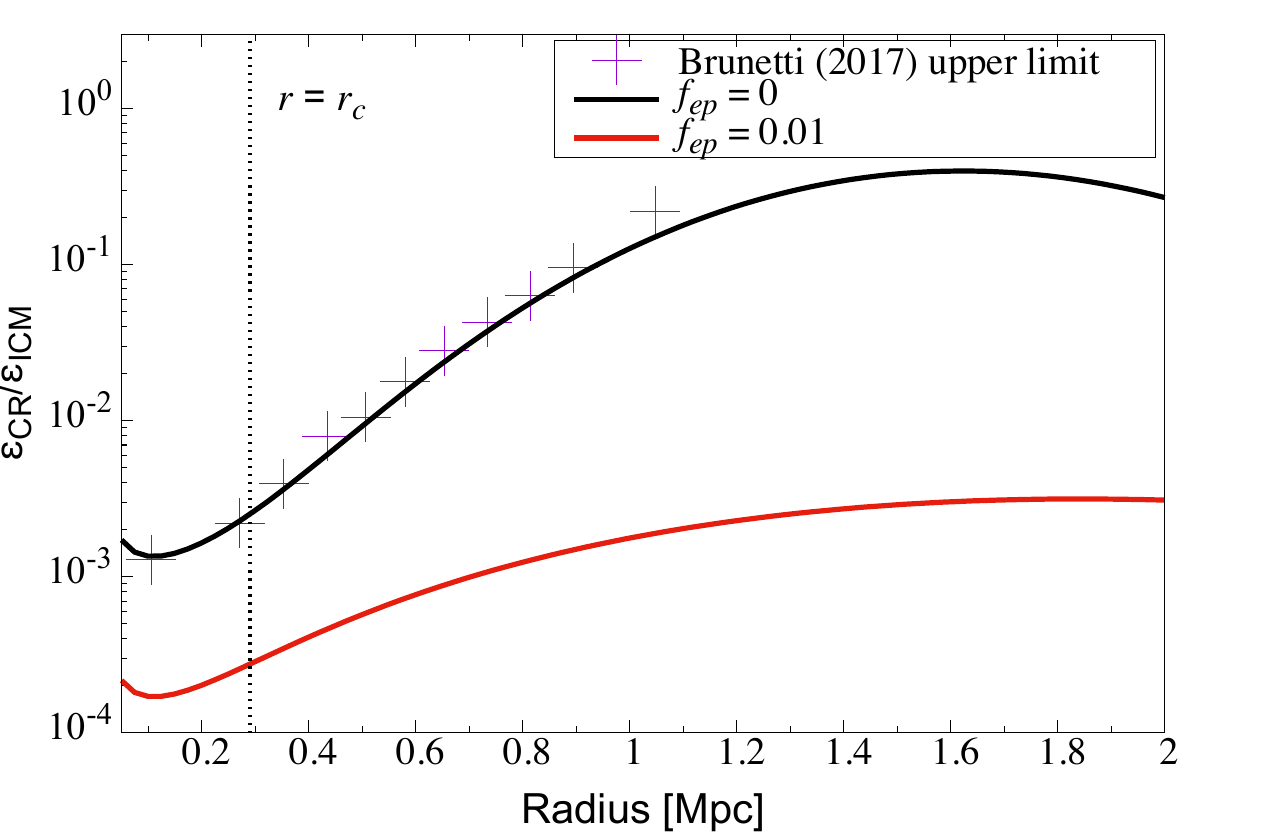}
    \caption{Radial dependence of the energy density ratio of CRPs $\epsilon_\mathrm{CR}$ to the thermal ICM $\epsilon_{\mathrm{ICM}}$. The results for the secondary-dominant model ($f_\mathrm{ep}=0$) and the primary-dominant model ($f_\mathrm{ep}=0.01$) are shown with black and red solid lines, respectively. We assumed $q=2$ and $\alpha=2.45$ in both models. The dotted line denotes the core radius $r_c=290$ kpc. For comparison, we plot the upper limit for the CRP energy density given by \citet{Brunetti2017}, which is derived from the gamma-ray upper limit for the model with $\tau_\mathrm{acc}=260$ Myr, $t_\mathrm{R}=720$ Myr, $B_0$=4.7 $\mu$G, $\eta_B=0.5$ and $\alpha=2.45$.  \label{fig:epeicm}}
   \end{figure}

Figure \ref{fig:epeicm} shows the radial dependence of the energy density ratio of CRPs to the thermal ICM, $\epsilon_{\mathrm{CR}}/\epsilon_{\mathrm{ICM}}$, for models with two different $f_{\mathrm{ep}}$. In both cases, the ratio increases with radius. Many previous studies have pointed out the similar trend, using a semi-analytical argument \citep{Keshet:2010aq,2013ApJ...767L...4F} or by post-processing cosmological simulations \citep{Pfrommer2008}. The purple points in Figure \ref{fig:epeicm} show the upper limits given by \citet{Brunetti2017} using the {\it Fermi} upper limit \citep{Ackermann2016}, and our results are consistent with these limits. The model of \citet{Brunetti2017} is basically similar to our secondary-dominant hard-sphere model (their parameters are summarised in the caption of Figure~\ref{fig:epeicm}.), but the spatial diffusion of CRs was not included there. This figure suggests that the evolution of the CRP density should be very different from that of the thermal components and disfavors the so-called isobaric model for the CRP distribution. To study that point in more detail, we need a more detailed calculation that can simulate both the cosmological evolution of the cluster and the injection of primary CRs during the evolution. The radial profiles of CR injection obtained in this study should provide some hints for such studies. As we have mentioned in Section \ref{subsec:halo},
if the ratio between the turbulent energy and thermal energy increases with radius, the CRP distribution can be more concentrated towards the center \citep[e.g.,][]{Pinzke2017}.\par
\citet{2021A&A...648A..60A} claimed the detection of diffuse gamma-ray emission from the Coma cluster and constrained the CRP energy density. They defined $X_{\rm CRp}=U_{\rm CRp}/U_{\rm th}$, where $U_{\rm CRp}$ and $U_{\rm th}$ are the energy densities enclosed within $r_{500}$ for CRPs and the thermal gas, respectively. Their best fit value from {\it Fermi} data is $X_{\rm CRp}\approx1~\%$, while our secondary-dominant model shown in Figure~\ref{fig:epeicm} predicts $X_{\rm CRp}=6~\%$. On the other hand, the expected gamma-ray flux below 10 GeV is comparable to the data of the possible detection \citep{2021A&A...648A..60A} (see also Sect.~\ref{subsec:gamma}). The smaller $X_{\rm CRp}$ in their analysis would be due to the steeper spectral indices of $\alpha=2.6$ - $2.8$. On the other hand, in our primary-dominant model, $X_{\rm CRp}=0.2~\%$ and the gamma-ray flux is one order of magnitude smaller than the possible detection.

\subsection{Kolmogorov type acceleration \label{subsec:Kol}}
\begin{figure*}[tb!]
    \centering
          \plottwo{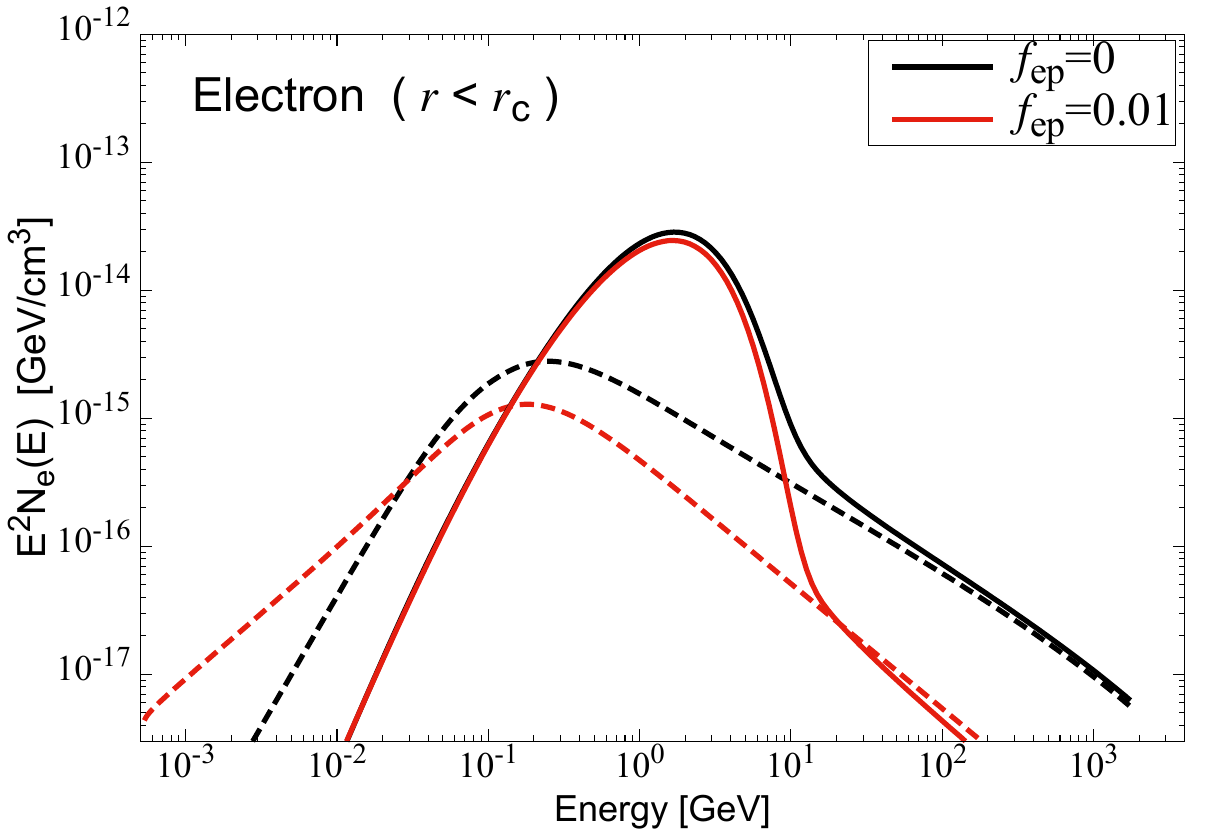}{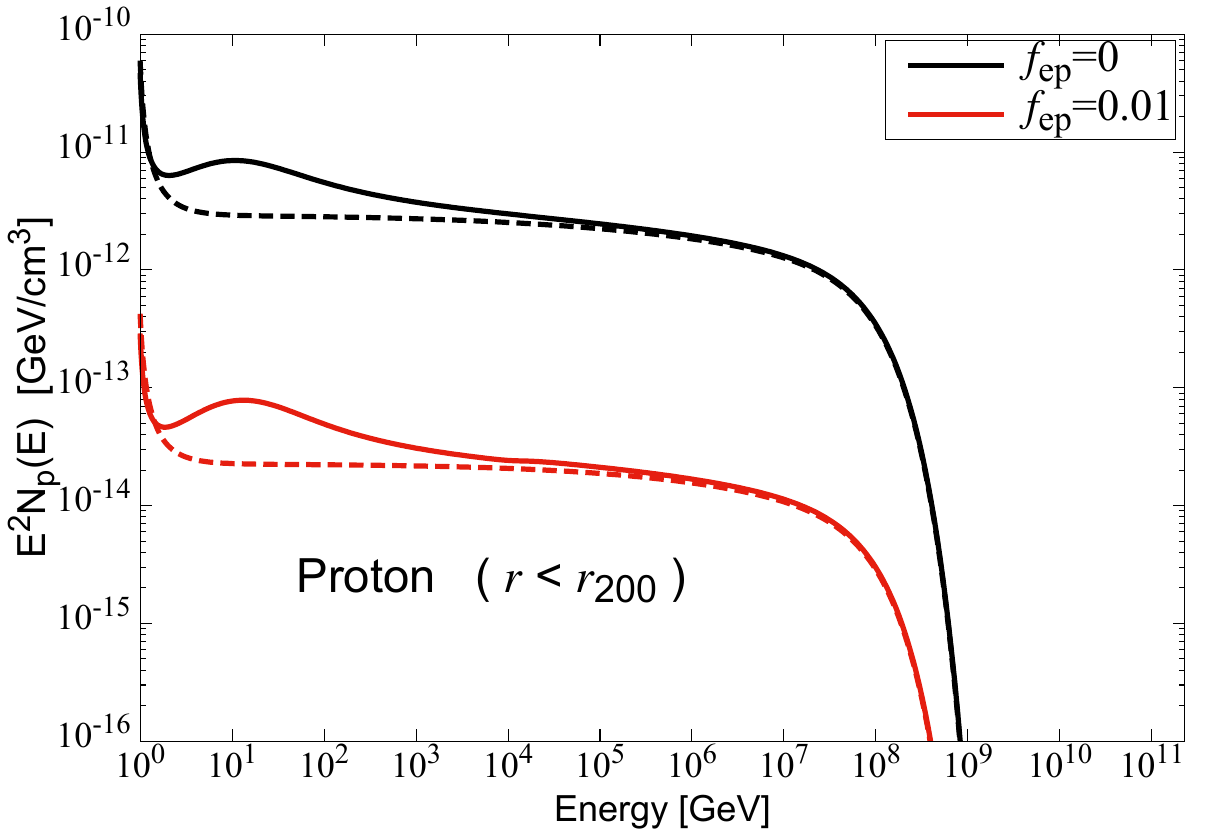}
	 \plottwo{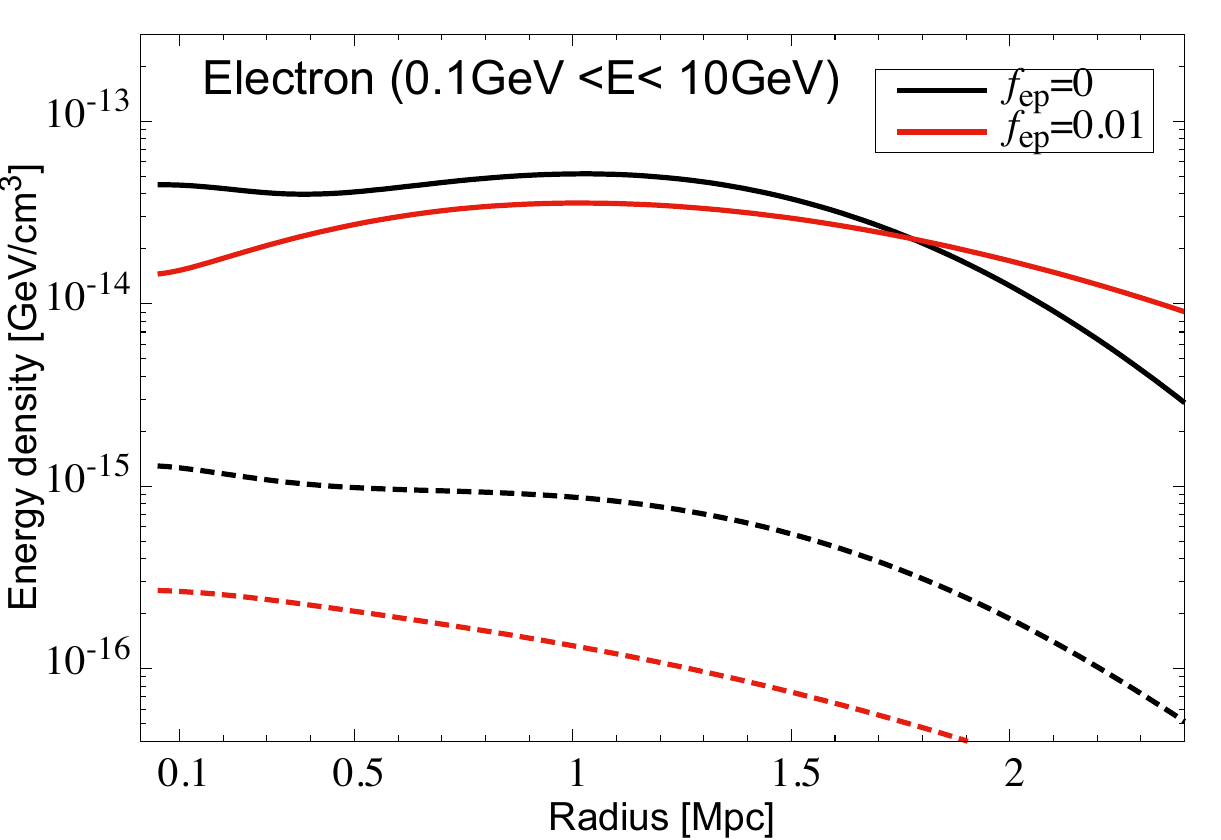}{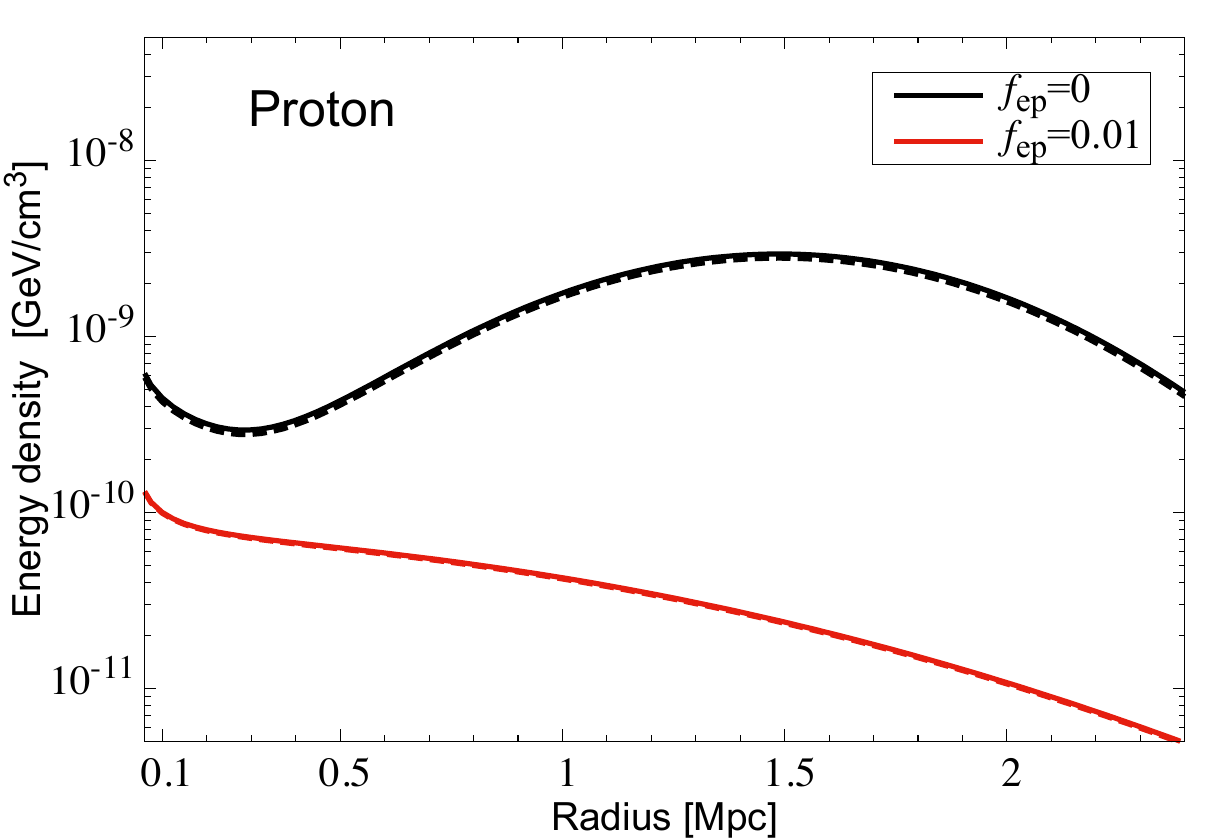}
    \caption{Same as Figure \ref{fig:CR}, but for the Kolmogorov models \label{fig:CR_K} with $\tau_\mathrm{acc}=100$ Myr. $t_\mathrm{R}=180$ Myr and $t_\mathrm{R}=240$ Myr are chosen for $f_\mathrm{ep}=0$ and $f_\mathrm{ep}=0.01$, respectively (see Table \ref{tab:Kol}). }
   \end{figure*}

In this section, we show the results for the Kolmogorov type reacceleration, $q = 5/3$. In this case, the acceleration time becomes shorter for lower energy CRs: $t_{\mathrm{acc}}\propto E^{-1/3}$.  As shown in Figure \ref{fig:CR_K}, the energy distributions of CRs are fairly different from the hard-sphere case. The maximum energy of CRPs does not increase by the reacceleration since $t_\mathrm{acc}$ above $E_\mathrm{p}^{\mathrm{max}}$ is longer than the calculation time ($\sim$ Gyr). All of the low-energy CREs with $E\le$ GeV have $t_\mathrm{acc}$ shorter than the cooling timescale, so they are efficiently reaccelerated (Figure \ref{fig:time}). This makes a bump around GeV in the CRE spectrum sharper than the hard-sphere case (Figure \ref{fig:CR}). The bump may be tested by dedicated higher frequency observations. 
\par
\begin{figure*}[tb!]
\centering
    \plottwo{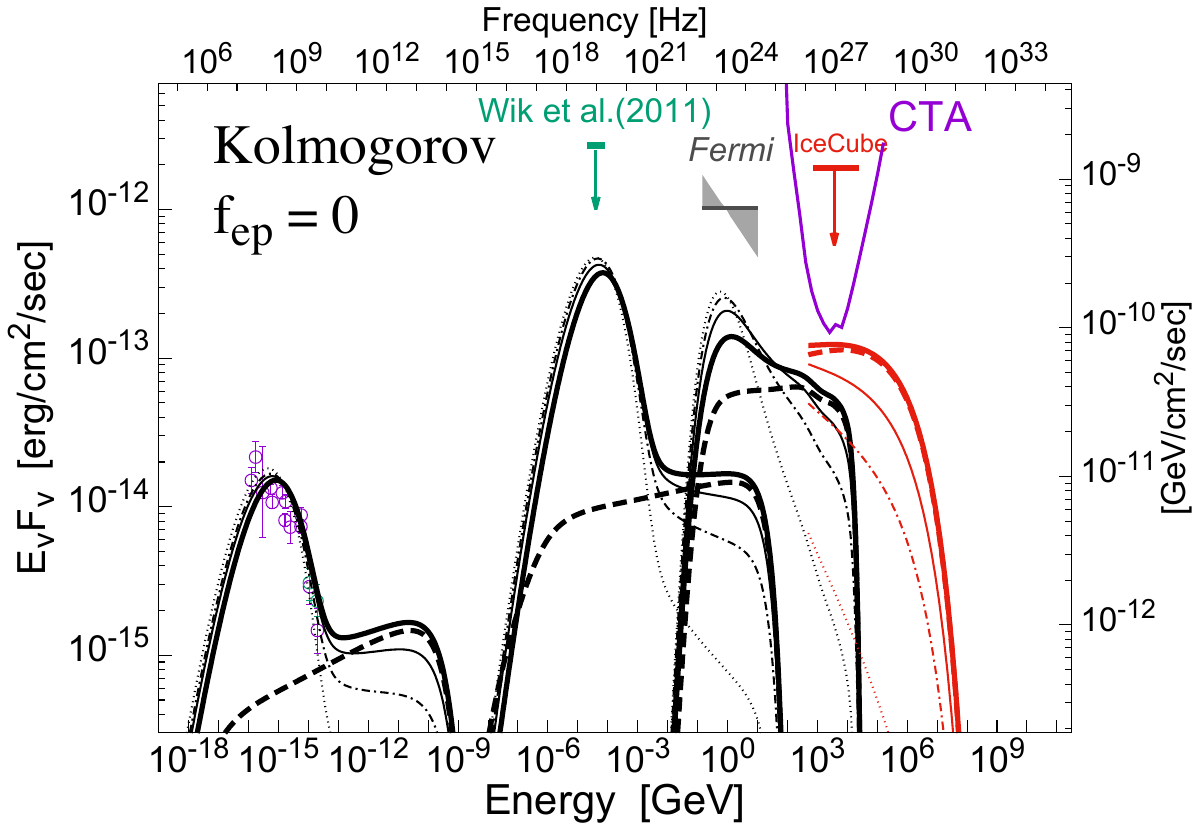}{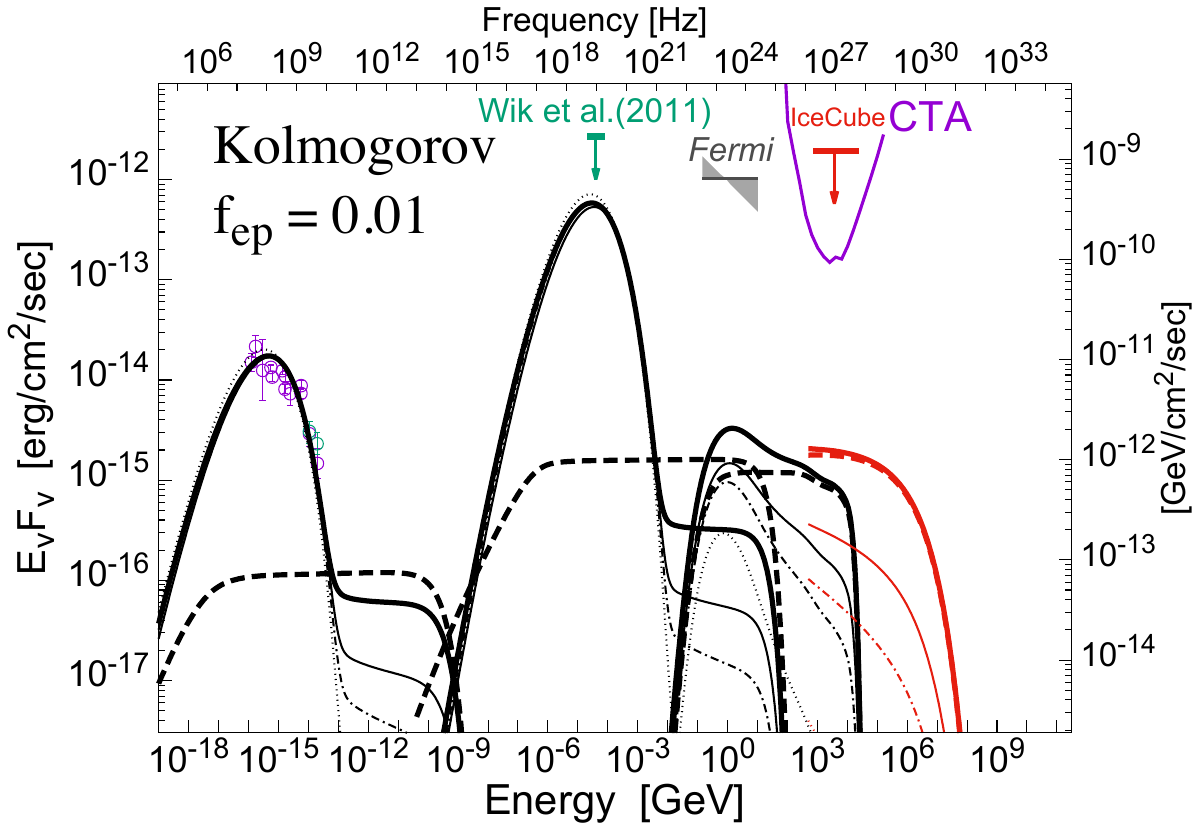}
    \caption{Same as Figure \ref{fig:nFn_H}, but for the Kolmogorov reacceleration models ($q=5/3$). Thick lines are the results for $\alpha=2.0$. The results are reported for $f_\mathrm{ep}=0$ (left) and $f_\mathrm{ep}=0.01$ (right). \label{fig:nFn_K}}
\end{figure*}

Notably, the radio spectrum can be reproduced with reasonable values of $(\tau_{\mathrm{acc}},t_\mathrm{R})$ even in the secondary-dominant models ($f_\mathrm{ep}=0$). As with the hard-sphere case, the radio spectra before the reacceleration is too hard when $f_\mathrm{ep}=0$ (Figure \ref{fig:nFn_K} (left)). However, the Kolmogorov reacceleration efficiently accelerate low-energy CREs and the shape of the observed spectrum can be reproduced well. Unlike the hard-sphere case, we use $\tau_{\mathrm{acc}}=100$ Myr, because the reacceleration with $\tau_\mathrm{acc}\ge200$ Myr causes too sharp bumps in CRE spectra and the resulting synchrotron spectrum does not match the data well.  
Since $t_{\mathrm{acc}}$ for radio-emitting CREs is shorter than that of gamma-emitting CRPs in this case, the increase of $L_{\mathrm{radio}}/L_\gamma$ due to the reacceleration is larger than the hard-sphere case. As a result, the {\it Fermi}-LAT upper limit is not yet sufficient to give a meaningful constraint on $t_\mathrm{R}$, except for a soft injection with $\alpha=2.45$. \par

We find that the duration of the reacceleration need to be $t_\mathrm{R}\approx180$~Myr to reproduce the convexity in the radio spectrum. The acceleration timescale $t_\mathrm{acc}$ for CRPs above 10 PeV is longer than 1 Gyr, so there is not so much difference between the ``loud" and ``quiet" states regarding the PeV neutrino flux (Figure \ref{fig:nFn_K}). 
The CR injection power is close to the hard-sphere case (Table \ref{tab:sec}), so the quiet-state fluxes are similar for both reacceleration models. \par
The predicted neutrino fluxes are mostly determined by the injection index $\alpha$ (Table \ref{tab:Kol}) and the maximum energy, $E_\mathrm{p}^\mathrm{max}$. Note that the cutoff shape appears in the neutrino spectrum above 100 TeV directly reflects the cutoff in the primary injection (see Eq. (\ref{eq:pri})). When $E_{p}^{\mathrm{max}}=10$ PeV, the PeV neutrino flux becomes smaller by a factor of 2.\par 

Our most pessimistic result is obtained when $f_\mathrm{ep}=0.01$ (Figure \ref{fig:nFn_K} (right)). That model predicts the fluxes of hadronic radiations more than two orders of magnitude below the IceCube limit. Currently, we cannot exclude those pessimistic scenarios from the radio observations. 

\begin{table*}[tb!]
    \centering
    \caption{Same as Table \ref{tab:Hard}, but for $q = 5/3$ (Kolmogorov), $\tau_\mathrm{acc}=100$ Myr  \label{tab:Kol}}
    \begin{tabular}{cccccccc}
    \hline\hline
      	$f_\mathrm{ep}$&$\alpha$  &$E_\nu F_\nu$(loud)&$E_\gamma F_\gamma$(loud)&$L_\nu$&$L_\mathrm{CRP}(>10^{17}~{\rm eV})$&$L_\mathrm{CRP}(>10^{18.5}~{\rm eV})$\\ 
	&&[GeV/cm$^2$/s]&[GeV/cm$^2$/s]&[erg/s]&[erg/s]&[erg/s]\\
	\hline
	0&2.0 &$3.5\times 10^{-11}$&$8.4\times10^{-11}$&$3.1\times10^{41}$&$1.1\times10^{43}$&$1.7\times10^{28}$\\
	&2.1&$1.4\times 10^{-11}$&$1.3\times10^{-10}$&$1.0\times10^{41}$&$3.7\times10^{42}$&$3.9\times10^{27}$ \\
	&2.2 &$3.5\times 10^{-12}$&$1.6\times10^{-10}$&$3.3\times10^{40}$&$8.3\times10^{41}$&$6.4\times10^{26}$\\
	&2.45&$4.0\times 10^{-14}$&$1.6\times10^{-10}$&$1.1\times10^{39}$&$6.1\times10^{39}$&$3.0\times10^{24}$\\ 
	\hline
	0.01&2.0&$4.8\times 10^{-13}$&$2.0\times10^{-12}$&$2.7\times10^{40}$&$1.2\times10^{41}$&$2.0\times10^{26}$\\
       &2.1&$3.6\times 10^{-14}$&$9.2\times10^{-13}$&$4.8\times10^{39}$&$7.7\times10^{39}$&$9.7\times10^{24}$\\
       &2.2&$4.6\times 10^{-15}$&$6.0\times10^{-13}$&$9.6\times10^{38}$&$8.3\times10^{38}$&$7.8\times10^{23}$\\
       &2.45 &$2.7\times 10^{-17}$&$1.8\times10^{-13}$&$1.6\times10^{37}$&$1.6\times10^{36}$&$1.2\times10^{21}$\\ 	
   \hline
    \end{tabular}
    \flushleft
    \tablecomments{See Table \ref{tab:Hard} for the detail of each quantity. The neutrino fluxes for the radio-quiet states are the same in two significant digits as radio-loud states.}
\end{table*}

\subsection{Escaping cosmic rays \label{subsec:esc}}
\begin{figure}[tb!]
\centering
    \includegraphics[width=8.6cm]{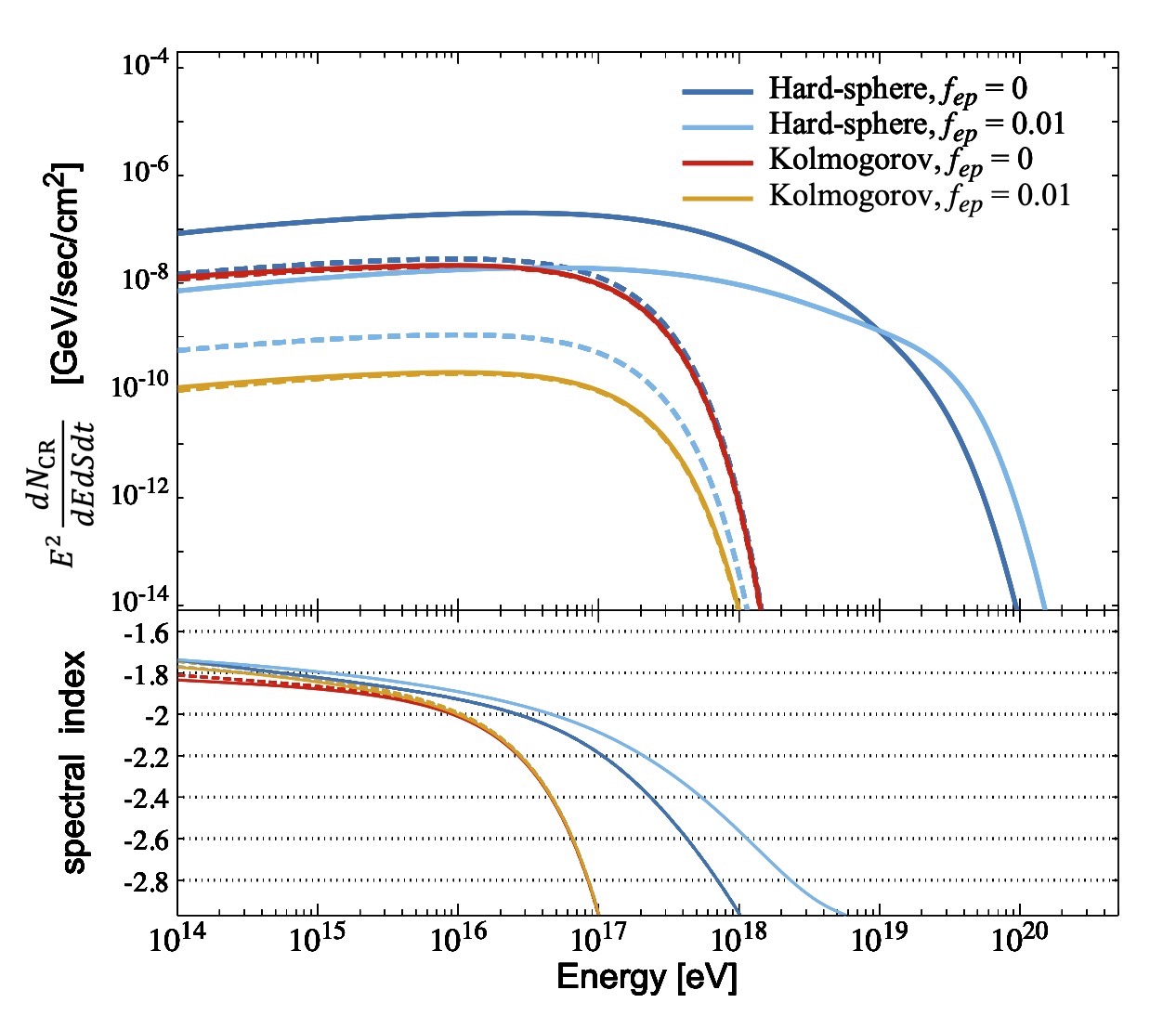}
    \caption{Spectra of CRPs escaping from the virial radius $r_{200}$ of the Coma cluster at the radio-loud state. We show with different colors the results for the hard-sphere acceleration ($q=2$) with $f_\mathrm{ep}=0$ (blue) and $f_\mathrm{ep}=0.01$ (light blue), and the Kolmogorov acceleration ($q=5/3$) with $f_\mathrm{ep}=0$ (red) and $f_\mathrm{ep}=0.01$ (orange). In each model, the injection spectral index is $\alpha=2.0$. Bottom panel shows the spectral indexes of the number fluxes in the unit of [/GeV/cm$^{2}$/sec] (Eq. (\ref{eq:CRp})). Spectra for the ``loud" and ``quiet'' states are shown with solid and dashed lines, respectively. The maximum energy of the primary CRPs is assumed to be $E_\mathrm{p}^\mathrm{max}=100$ PeV.\label{fig:CRp}}
\end{figure}

We can calculate the number flux of CRs that diffuse out from the cluster with the diffusion coefficient $D_{rr}$ and the gradient of particle number density at the boundary of the cluster, 
\begin{eqnarray}
\frac{dN_\mathrm{CR}}{dEdSdt}=-\left.D_{rr}\frac{dn_\mathrm{p}}{dr}\right|_{r=r_{200}}.\label{eq:CRp}
\end{eqnarray}
In Figure \ref{fig:CRp}, we show the energy fluxes of CRPs escaped from the Coma cluster for different reacceleration models with $\alpha=2.0$. We can see a clear difference between two types of the reacceleration. The hard-sphere type can accelerate CRPs up to ultrahigh energies, while the Kolmogorov type never produces CRPs of energies higher than $E_\mathrm{p}^\mathrm{max}$.\par

Note that the fluxes before reacceleration, i.e., at $t=t_0$, differ in each model (dashed lines in the figure). That is simply because we take different injection rates of primary CRs for each model to reproduce the RH at the radio-loud state. In other words, we take different $t_\mathrm{R}/\tau_\mathrm{acc}$ for each model, which regulates the energy injection from the turbulent reacceleration. The normalization of the CR injection depends on the choice of $t_\mathrm{R}$, as smaller $t_\mathrm{R}/\tau_{\mathrm{acc}}$ results in a larger injection power of primary CRs, i.e., larger flux of escaping CRs.  \par

The maximum energy of CRPs increases with $t_\mathrm{R}/\tau_\mathrm{acc}$. In the hard-sphere case, that can be expressed as 
\begin{eqnarray}
E_\mathrm{p}^\mathrm{max}(t_\mathrm{R})=E_\mathrm{p}^\mathrm{max}(0)\exp\left(\frac{t_\mathrm{R}}{\tau_\mathrm{acc}}\right),
\label{eq:Epmaxhard}
\end{eqnarray} 
where $E_\mathrm{p}^\mathrm{max}(0)$ is the maximum energy before the reacceleration. When $E_\mathrm{p}^\mathrm{max}(0)= 100$ PeV, the maximum energy could reach ultrahigh energies ($>10^{18.5}$ eV) in $t_\mathrm{R}/\tau_\mathrm{acc}=3.44$ or  $t_\mathrm{R}= 895$ Myr for $\tau_\mathrm{acc}=260$ Myr. All of $t_\mathrm{R}$ in our calculation are smaller than that value (Table. \ref{tab:Hard}). Thus, our model is not sensitive to the parameter $E_{\rm c}^{\rm max}(r)$ in Eq. (\ref{eq:Dpp}), unless $l_{\rm c}^{\rm F}\ll 0.1$ Mpc. Assuming $n_\mathrm{GC}\sim10^{-6}$ ${\rm Mpc}^3$ for the local number density of galaxy clusters, our model would not overproduce the observed UHECR intensity significantly.
\par
The spectral index of the escaping CRPs is also shown in Figure \ref{fig:CRp} below. It becomes harder than the injection index $\alpha$ for $E_\mathrm{p}\lesssim10^{16}-10^{17}$ eV, which means CRPs below that energy are well confined within the cluster. \par

\section{Discussion \label{sec:discussion}}

\subsection{{\rm Caveats} \label{subsec:caveats}}
In this section, we discuss various limitations in our assumptions and their potential impacts on our conclusions. We have neglected the uncertainties in strength and profile of the magnetic field. The constraints on $(B_0, \eta_B)$ from the RM measurement \citep{Bonafede2010} are not much stringent, ranging from $(B_0,\eta_B)=(3.0~\mu {\rm G},0.2)$ to  $(7.0~\mu{\rm G}, 1.0)$ within 3$\sigma$. In addition, \citet{2020ApJ...888..101J} pointed out that the magnetic field estimated with RM can include an irreducible uncertain factor of $\approx$3. On the other hand, the non-detection of IC radiation provides the lower limit ($>0.25~\mu{\rm G}$) of the magnetic field \citep{Wik_2011}. \par

As reported in \citet{Brunetti2017}, the ratio of the radio flux to gamma-ray and neutrino fluxes becomes larger for flatter profiles of the magnetic field, i.e. $\eta_B<0.5$. For example, when $\eta_B=0.2$ and other parameters including $B_0$ are unchanged, the predicted fluxes of gamma-ray and neutrinos decrease by a factor of $\approx2$. This means that the constraint on $t_{\rm R}/\tau_{\rm acc}$ from the gamma-ray upper limit becomes less stringent; $t_{\rm R}>200$ Myr for the secondary-dominant hard-sphere model with $\tau_{\rm acc} =260$ Myr (see also Figure~\ref{fig:halo}). The injection profile of primary CRs $K(r)$ also depends on the magnetic field profile, as smaller $\eta_B$ requires more centrally concentrated $K(r)$. However, in the secondary-dominant scenario, we have confirmed that $D_{rr}$ of Eq.~(\ref{eq:Drr2}) requires an injection profile with a peak at $r\ge0.8$ Mpc even in the extreme case of uniform magnetic field ($\eta_B=0$). \par

Although we have fixed $D_{rr}$ as Eq.~(\ref{eq:Drr2}), smaller values of $D_{rr}$ could be possible, depending on the turbulent nature of the ICM. The radial diffusion of parent CRPs is one of the causes of the hard synchrotron spectrum in the secondary dominant model (Figure~\ref{fig:nFn_H}). For $\alpha=2.45$, we confirmed that the RH spectrum actually fits well when $D_{rr}$ is 1/10 times smaller than Eq.~(\ref{eq:Drr2}). The spectral hardening due to the energy dependence in $pp$ cross section is unavoidable even for $D_{rr}\equiv0$, which makes it difficult to fit the RH spectrum with harder injections ($\alpha\le2.2$) in the hard-sphere model. A smaller $\tau_{\rm acc}$ would result in better match at higher frequencies around $\sim1~{\rm GHz}$, but worsen the fit at lower frequencies around $\sim20~{\rm MHz}$. Note that the Kolmogorov model does not have to suffer from the difficulty due to hard indices (Figure~\ref{fig:nFn_K}).
\par

We have also assumed that $f_\mathrm{ep}$ and $D_{pp}$ are constant with the radius. Under this condition, we showed that the profile of the RH can be reproduced by the stable injection profile shown in Figure~\ref{fig:inj_alpha2} combined with the spatial diffusion of CRPs. However, the secondary-dominant model results in the hardening in radio spectrum (Figure~\ref{fig:nFn_H}). On the other hand, some numerical simulations suggest that the ratio of turbulent pressure to the thermal one increases with the distance from the cluster center \citep[e.g.,][]{Nelson2014,2018MNRAS.481L.120V}. \citet{Pinzke2017} showed that the profile of the RH can also be reproduced when the efficiency of the reacceleration with radius and the CR distribution is more concentrated towards the central region. In such models, the tension in the RH spectrum would be partially relaxed.\par
In our calculation, the time dependence of some quantities, such as $n_{\rm th}(r)$, $B(r)$, $f_{\rm ep}$, and $Q_{\rm p}$, is not taken into account. The statistical properties of RHs would give constraints on the time evolution of those quantities, which should be studied in future studies. \par

\subsection{Future prospects for detecting high-energy emission\label{subsec:obs}}
\subsubsection{Gamma rays \label{subsec:gamma}}
Observations of gamma-ray photons from the ICM provide important constraints on the amount of CRPs. In this paper, we have adopted the upper limit given by \citet{Ackermann2016}, taking a conservative approach. However, there are three recent studies that claimed possible detection in the direction of the Coma cluster using the {\it Fermi} data \citep{2017ifs..confE.151K,2018PhRvD..98f3006X,2020ApJS..247...33A,2020arXiv200511208B,2021A&A...648A..60A}. 
Although a point source (4FGL~J1256.9+2736) may account for most of the signal, \citet{2021A&A...648A..60A} showed that models including extended components match the data better.\par
\begin{figure}[htb]
 \centering
 \includegraphics[width=8.5cm]{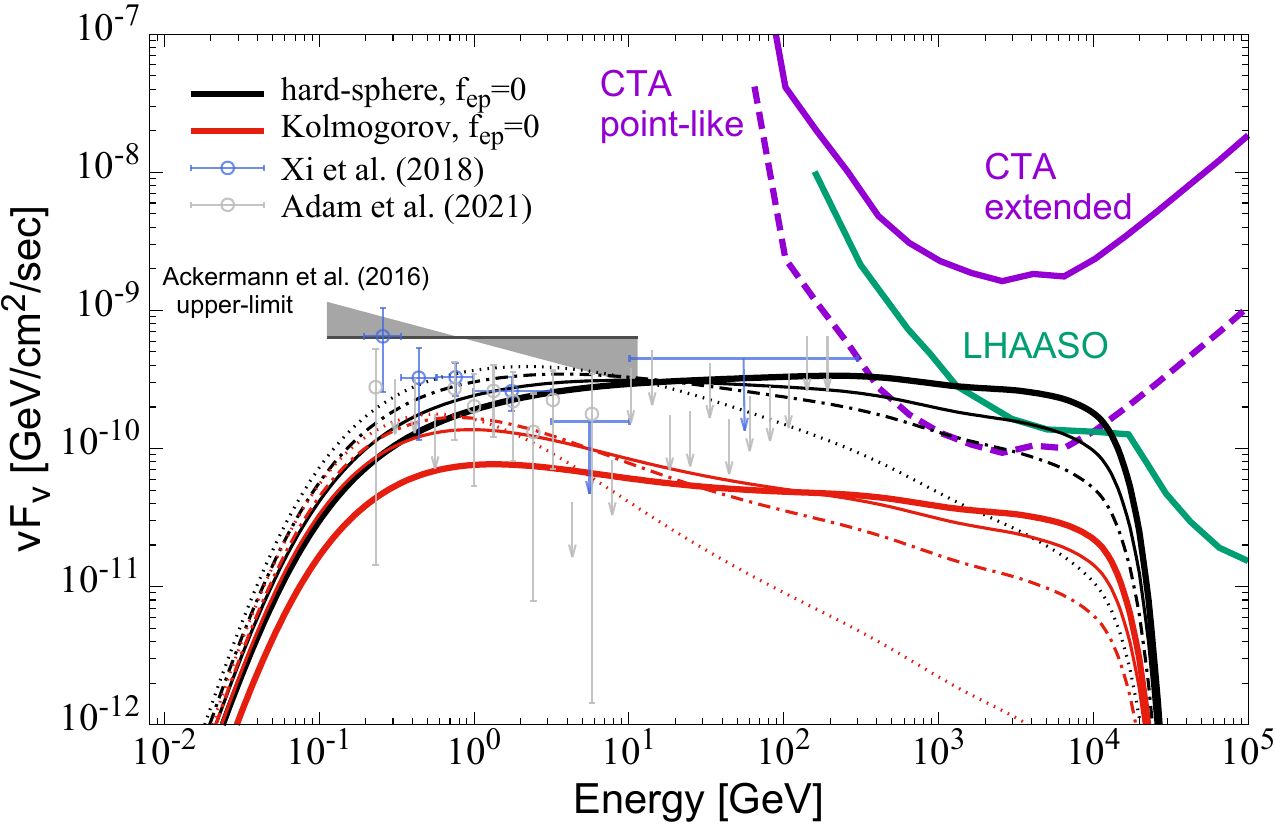}
 \caption{Expected gamma-ray fluxes. As a reference, we show the spectra of the possible extended component of the gamma-ray source obtained by \citet{2018PhRvD..98f3006X} (radio+p1 model, blue points) and \citet{2021A&A...648A..60A} (scenario 3 extended model, gray points). The arrows show the upper limits. The magenta and green lines show the point source sensitivity of CTA North (50 hours) and LHAASO (five years), respectively. The dashed magenta line shows the CTA sensitivity for extended sources with a subtended angle of $1\rlap{.}^\circ2$ (see text). The black and red lines are the fluxes for the radio-loud state. In each model, we assume the secondary-dominant model ($f_\mathrm{ep}=0$). The results for four different injection indexes $\alpha=2.0,2.1,2.2$, and $2.45$ (thick solid, thin solid, thin dot-dashed, thin dotted) are shown.\label{fig:CTA}}
 \end{figure}

In Figure~\ref{fig:CTA}, we plot the gamma-ray spectrum given by \citet{2018PhRvD..98f3006X} (blue points) and \citet{2021A&A...648A..60A} (gray points). Here, we have assumed $r_{\rm ap}=2.0$~Mpc to calculate the expected fluxes, same as Figure~\ref{fig:nFn_H}. Our secondary-dominant models are in good agreement with the data around $\sim$GeV, except for the Kolmogorov model with $\alpha=2.0$, where the relatively large value of $t_{\rm R}$ results in small values of $F_\gamma/F_{\rm radio}$ (Sect.~\ref{subsec:halo}). It is worth noting that the upper limit given by \citet{2018PhRvD..98f3006X} in $[3,10]~{\rm GeV}$ is incompatible with our secondary-dominant hard-sphere models (black lines) with $\alpha\leq2.45$. However, \citet{2021A&A...648A..60A} obtained different constraints in the same energy range, and the deep limit obtained by \citet{2018PhRvD..98f3006X} is still controversial. Future observations or analysis around this energy range are necessary to give robust constraints on the reacceleration, primary CREs, and the injection indexes. \par
We also show the sensitivities of the future TeV gamma-ray telescopes: Cherenkov Telescope Array (CTA) and Large High Altitude Air Shower Observatory (LHAASO). The dashed magenta line shows the point-source sensitivity of the CTA North site with 50~hr observation. With this sensitivity, TeV gamma rays from the Coma cluster can be accessible only for the optimistic hard-sphere model with $f_\mathrm{ep}=0$. 
The flux becomes $\sim1/10$ in the primary-dominant scenario ($f_\mathrm{ep}=0.01$, see Figure~\ref{fig:nFn_H_pri}).\par
Since the RH of Coma is an extended source for CTA, its sensitivity should be modified for its extension. The angular resolution of the instrument at 1 TeV is $\theta_\mathrm{inst}\approx4\arcmin$, while the subtended angle corresponding to the radius of $r=2.0$ Mpc is $\theta_\mathrm{source}\approx1\rlap{.}^\circ2$. 
We estimate the flux sensitivity for an extended source by multiplying the point source sensitivity by the factor $\mathrm{max}[1,\theta_\mathrm{inst}/\theta_\mathrm{source}]$ (solid magenta), assuming that the sensitivity is limited by the background. We here implicitly assumed ON/OFF analysis and constant intensity over the observed extension, although these assumptions may not be optimal for instruments with better sensitivities. We caution that the diffuse sensitivity estimated here should be regarded as an upper limit. On the other hand, the angular resolution of LHAASO is $\sim1^\circ$ \citep{2019arXiv190502773B} so that the Coma cluster can be approximated as a point source (solid green). It seems challenging to detect extended gamma-ray emission from the Coma cluster with CTA, but the point-like signal can be accessible with LHAASO. 

\subsubsection{Hard X-ray\label{subsec:hardX}}
\begin{figure}[htb]
 \centering
 \includegraphics[width=8.5cm]{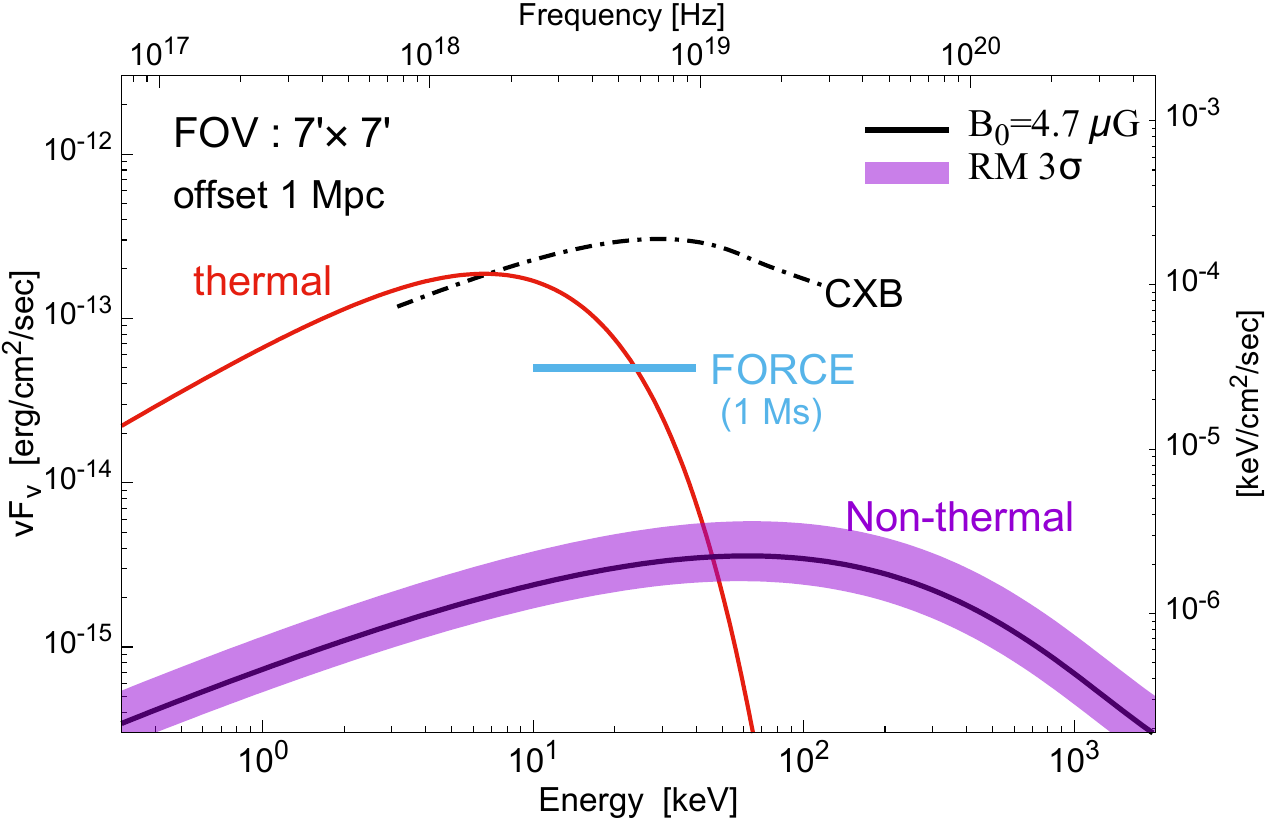}
 \caption{Expected hard X-ray fluxes. Free-free emission from the thermal ICM electron is shown in red, and ICS radiation from CREs is shown with the solid black line. The shaded region corresponds to the $3\sigma$ uncertainty in the magnetic field $B_0$ measured by \citet{Bonafede2010}. Intensities of those emission are integrated within the $7\arcmin\times7'$ region, which corresponds to the FOV of FORCE. Note that the center of the FOV is shifted by 1 Mpc ($\approx$ 36') from the cluster center.  
 The black dot-dashed line shows the cosmic X-ray background (CXB), which is evaluated from the CXB intensity measured with {\it INTEGRAL} \citep{2007A&A...467..529C}.
 As a reference, the target sensitivity of FORCE for point-like sources is shown with the thick cyan line. 
 \label{fig:Brems}}
 \end{figure}

In our calculation, the distribution of CREs at the radio-loud state is constrained by the RH observation, so the ICS flux is almost uniquely determined for a given magnetic field.
Figure~\ref{fig:Brems} shows spectra of X-ray emission from both thermal and non-thermal components in the ICM, which are integrated within the field of view (FOV) of the future X-ray mission, Focusing On Relativistic universe and Cosmic Evolution (FORCE). The free-free emission is calculated with the profile of temperature and density shown in Eqs. (\ref{eq:beta-model}) and (\ref{eq:temp}). 
The center of the FOV is shifted by 1~Mpc from the cluster center, where the thermal X-ray flux is too bright. Since the flat distribution of CREs up to $r\approx1$ Mpc is more extended than the ICM or the magnetic field, the relative strength of the non-thermal flux to the thermal X-ray increases up to this radius. In our calculation, where Eq. (\ref{eq:B}) with $B_0=4.7$ $\mu$G is adopted for the magnetic field, the predicted ICS flux is $F_{X}\approx2.0\times10^{-15}$~erg cm$^{-2}$ s$^{-1}$ at 30~keV, which is smaller than the thermal flux by almost two orders of magnitude below 10 keV. 
The non-thermal flux becomes comparable to the thermal one only above $\sim$ 50 keV, though it is still significantly smaller than the cosmic X-ray background (CXB, black dot-dashed).\par

The hard X-ray satellite FORCE is characterized by its high sensitivity and high angular resolution in a broad band of 10 - 80 keV \citep{2016SPIE.9905E..1OM}. Thanks to its high angular resolution of $15\arcsec$ and the target sensitivity within 1~Ms of $S_X = 3\times 10^{-15}$~erg cm$^{-2}$ s$^{-1}$ keV$^{-1}$ for point-like sources, FORCE is expected to resolve $\sim$80\% of the CXB emission into point sources \citep{2018SPIE10699E..2DN}, which can reduce the background flux by a factor of $\sim$3. We expect the very first detection of the ICS emission from high-redshift RHs or radio relics is expected with this instrument. Several MeV gamma-ray missions, such as the Compton Spectrometer and Imager ({\it COSI}), Gamma-Ray and AntiMatter Survey (GRAMS) and All-sky Medium Energy Gamma-ray Observatory (AMEGO), are being planned \citep{2019BAAS...51g..98T,2020APh...114..107A,2019BAAS...51g.245M}. Those instruments will constrain the ICS components from clusters above $\sim$100 keV energies.\par

\subsection{Contributions to Cumulative Neutrino and Gamma-Ray Backgrounds\label{subsec:background}}
Once the luminosity of high energy emission and the number density of the sources are specified, we can evaluate the intensity of the background emission. The cumulative background intensity is estimated by~\citep[e.g.,][]{Waxman1999},
\begin{eqnarray}
E^2\Phi\approx\frac{c}{4\pi}E\frac{dQ}{dE}\xi_zt_\mathrm{H},
\label{eq:Waxman}
\end{eqnarray}
where the Hubble time is $t_\mathrm{H}\approx13$ Gyr, and $\xi_z\sim\mathcal{O}(1)$ is a parameter that depends on the redshift evolution of the luminosity density $Q$. 
For example, $\xi_z\sim3.0$ for the evolution of the star-formation rate, i.e., $n_\mathrm{s}\propto(1+z)^m$ with $m=3$, while $\xi_z\approx0.6$ for non-evolving sources ($m=0$). Here we take $\xi_z=1$ as a reference value. Note that $\xi_z$ is reduced only by a factor of $2-3$ even for negatively evolving ($m<0$) sources. 
The generation rate density of gamma-ray photons or high-energy neutrinos per unit comoving volume $dQ/d\ln E=E(dQ/dE)$ is evaluated from 
\begin{eqnarray}
E\frac{dQ}{dE}\approx n_\mathrm{GC}^\mathrm{eff}\left.E^2\frac{d\dot{N}}{dE}\right|_\mathrm{Coma},
\end{eqnarray}
where $\left.\frac{d\dot{N}}{dE}\right|_\mathrm{Coma}$ is the luminosity of hadronic emission from the Coma cluster at the radio-loud state, and the effective number density of the radio-loud clusters $n_\mathrm{GC}^\mathrm{eff}$ is the product of the observed radio-loud fraction $f_\mathrm{loud}\approx0.4$ \citep{2013A&A...557A..99K,2016A&A...593A..81C} and the number density of Coma-like clusters with the virial mass of $M_\mathrm{500}\approx6\times10^{14}$ $h^{-1}M_\sun$ \citep{Reiprich_2002}, $n_\mathrm{GC}\sim10^{-6}$ Mpc$^{-3}$ \citep[e.g.,][]{2001MNRAS.321..372J}. 
The virial mass $M_{\Delta}$ is defined as $M_{\Delta}=4\pi  r_{\Delta}^3 \rho_\mathrm{m}/3$ and $\rho_\mathrm{m}$ is the mean matter density of the universe.

\begin{figure*}[htb]
 \centering
  \plottwo{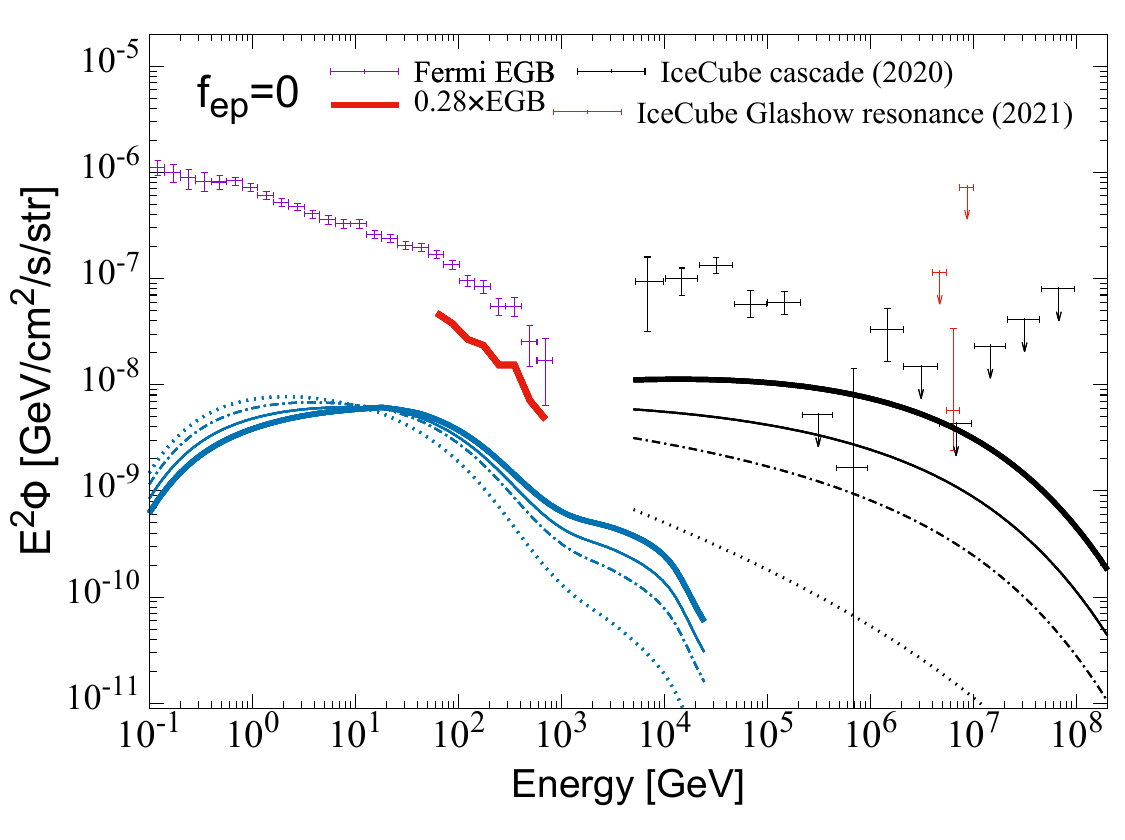}{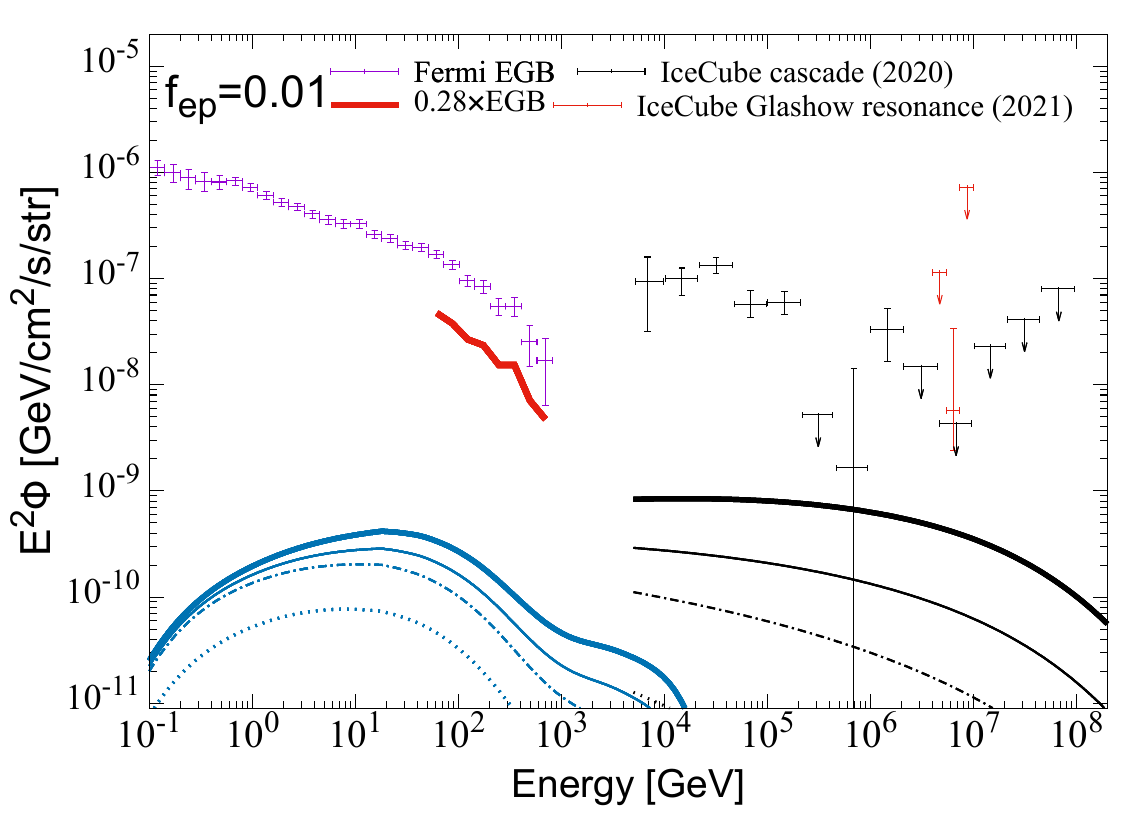}
 \caption{Background intensities of gamma-ray (blue) and neutrino (black) for hard-sphere models ($q=2$). The magenta data points are the total EGB intensity observed with {\it Fermi} \citep{Ackermann2015}. As a reference, we show the EGB intensity multiplied by 0.28 with thick red line above 50~GeV, which corresponds to the approximate upper limit for the contribution from non-blazar components. 
 The black crosses show the all-flavor neutrino intensity of the IceCube cascade events \citep{2020PhRvL.125l1104A}, while the red ones show the data at the energy of the Glashow resonance \citep{2021Natur.591..220I}. We adopt the effective optical depth for the EBL attenuation consistent with $\xi_z=1$. {\it Left}: Results for the secondary-dominant models ($f_\mathrm{ep}=0$). {\it Right}: Results for the primary-dominant models ($f_\mathrm{ep}=0.01$). The results for four different injection indexes are shown: $\alpha=2.0, 2.1, 2.2$, and $2.45$ (thick solid, thin solid, thin dot-dashed, thin dotted).\label{fig:IGRB}}
\end{figure*}
In Figure \ref{fig:IGRB}, we show high-energy neutrino and gamma-ray background intensities from Coma-like clusters estimated from Eq. (\ref{eq:Waxman}). Following \citet{Murase_2012}, we estimate the effective optical depth for the EBL attenuation consistent with $\xi_z=1$. 
Note that the mass function of the dark matter halo is quite sensitive to the mass around $M\sim10^{15}$ $h^{-1}M_\odot$. Our treatment of $n_\mathrm{GC}$ is so rough that the estimate of the background intensities includes the ambiguity of a factor of $2-5$.\par

According to \citet{PhysRevLett.116.151105}, about 70\% of the extragalactic gamma-ray background (EGB) above 50 GeV is likely to originate from resolved and unresolved point sources like blazars, so the gamma-ray background from clusters would be smaller than $\sim$30\% of the observed EGB intensity~\citep[see also][]{2016ApJ...832..117L}. 
The red line in Figure \ref{fig:IGRB} shows the 1$\sigma$ upper limit for the non-blazar component \citep{PhysRevLett.116.151105}. The gamma-ray backgrounds of both secondary-dominant and primary-dominant models are well below this limit even when the ambiguity in our estimate is taken into account.\par

We see that the neutrino intensity for the secondary-dominant model with $\alpha=2.0$ (left panel, thin solid) becomes $\lesssim3-30$\% to the observed one, which is consistent with previous results on the typical accretion shock scenario~\citep{Murase2008,Fang2016,Hussain:2021dqp}. Note that the neutrino luminosity function has not been considered in our estimate. By including the cluster mass function and its redshift evolution as well as the CR distribution, the intensities can be further enhanced by a factor of $\sim\mathcal{O}(1)-\mathcal{O}(10)$, in which $\sim100$\% of the IceCube intensity can be explained. 
For example, if CRPs are mainly injected from internal sources like AGNs, the redshift evolution of the source density can be as large as $\xi_z\sim3$, instead of $\xi_z\sim1$ \citep[e.g.,][]{2018NatPh..14..396F,Hussain:2021dqp}. The neutrino luminosity scales
with $L_\nu\propto M_{500}$ (that is different from the accretion shock scenario),
the background intensity mainly originates from more clusters lighter than Coma \citep{Murase2013}. Such a contribution is not constrained from current radio observations \citep{Zandanel2015}, and larger effective number densities are consistent with the present IceCube limit from multiplet searches~\citep{Murase2016}. On the other hand, the contribution to the IceCube neutrino intensity can be as much as $\sim3$\% in the primary-dominant model, and we caution many uncertainties in the estimate.\par

Here we remark an important constraint that is applied when the injected CR luminosity density is normalized by the IceCube data. In this case, softer spectral indexes with $\alpha\gtrsim2.2$ are excluded, because such models inevitably overproduce the gamma-ray background around $10-100$ GeV \citep{Murase2013}.
However, in our calculation, the normalization is given by the radio luminosity, and their contribution to the EGB is minor. This result (see Figure 15) is also consistent with the previous work~\citep{2009IJMPD..18.1609M}. 
The intensity of such cosmogenic gamma-rays, which is not shown in Figure \ref{fig:IGRB} is compatible with the non-blazar EGB. 
The contribution of gamma rays from clusters can be only $\sim1-10$\% of the IGRB, and smaller especially in the central source scenario if CRs are confined inside radio lobes.


\subsection{Comparison with previous studies \label{subsec:diffuse}}
There are several studies that calculated the cumulative neutrino intensity. \citet{Murase2008} calculated the neutrino background by convolving the neutrino luminosity with the mass function of dark matter halos assuming that galaxy clusters are the sources of CRs above the second knee, and predicted that the all-sky neutrino intensity is $E_\nu^2\Phi_\nu\sim(0.3-3)\times{10}^{-8}~{\rm GeV}~{\rm cm}^{-2}~{\rm s}^{-1}~{\rm sr}^{-1}$ for $\alpha=2.0$, considering both accretion shock and AGN scenarios~\citep{Murase:2015ndr}. The similar neutrino intensity was found by \citet{2009ApJ...707..370K} that assumes an AGN as a central source. \cite{Murase2013} showed that these models are viable for the IceCube data if the CR spectrum is hard and low-mass clusters are dominant, and steep CR spectra lead to negligible contributions~\citep{2020ApJ...892...86H}.
\citet{Fang2016} estimated the contribution to the IceCube intensity from galaxy clusters, taking into account 1D spatial diffusion of CRPs. They adopted the CRP injection luminosity similar to our secondary-dominant models ($L_\mathrm{p}^\mathrm{inj}\sim 10^{45} \mathrm{erg/s}$),
and concluded that the accretion shock scenario could explain only $\lesssim20$\%, while the central source scenario could explain both the flux and spectrum of the IceCube data above $\sim100$~TeV. 

Including the radio constraints, \citet{Zandanel2015} evaluated the gamma-ray and neutrino background with both phenomenological and semi-analytical approaches. They obtained the maximum neutrino fluxes for nearby clusters at 250 TeV assuming a simple relation between the gamma-ray luminosity and cluster mass; 
\begin{eqnarray}
\log_{10}\left[\frac{L_\gamma(100\; \mathrm{MeV})}{\mathrm{s}^{-1} \mathrm{GeV}^{-1}}\right]=P_1+P_2\log_{10}\left(\frac{M_{500}}{M_\sun}\right),
\end{eqnarray}
where $P_1\approx 20$ is determined so that the cumulative number of radio-loud clusters counts does not overshoot the observed counts from National Radio Astronomy Observatory Very Large Array sky survey (NVSS), which is found in, e.g., \citet{Cassano2010}. They fixed $P_2=5/3$ assuming that the luminosity of hadronic emission scales as the cluster thermal energy, i.e., $L_\gamma\propto M_{500}^{5/3}$ according to the accretion shock scenario. They also assumed that the radio luminosity $L_\mathrm{radio}$ linearly scales with $L_\gamma$, so they implicitly assumed $L_\mathrm{radio}\propto M_{500}^{5/3}$. Their models with the magnetic field of $B=1$ $\mu$G typically predicts the gamma-ray flux from Coma-like clusters to be $F_\gamma\sim10^{-10}$ GeV cm$^{-2}$s$^{-1}$ at 100 MeV, which is similar to our results in secondary-dominant models. They concluded that the contribution to the IceCube flux from all clusters is at most $10\%$ in their phenomenological modeling, which is inline with our results in Section \ref{subsec:background}.
On the other hand, these radio constraints are much weaker for the central source scenario, where lower-mass and higher-redshift sources are important as discussed above. Along this line, \cite{2018NatPh..14..396F} investigated the AGN scenario, in which the all-sky UHECR, neutrino, and non-blazar EGB fluxes are explained simultaneously, and the similar flux level is obtained by \citet{Hussain:2021dqp}. 
\par

\section{Conclusions\label{sec:conclusion}}
In this paper, we have studied the CR distribution in the giant RH of the Coma cluster. Our model includes most of the physical processes concerning CRs in galaxy clusters: turbulent reacceleration, injection of both primary and secondary CREs, and diffusion of parent CRPs. We have followed the turbulent reacceleration scenario \citep[e.g.,][]{1987AA...182...21S} and modeled the multi-wavelength and neutrino emission from the RH by solving the one-dimensional FP equations (Eqs. (\ref{eq:Np}) and (\ref{eq:Ne})) numerically. \par

We have modeled the spatial evolution of the CRs with the diffusion approximation (Eq. (\ref{eq:Drr2})) and non-uniform injections (Eqs. (\ref{eq:inj1}) and (\ref{eq:inj2})). Secondary CREs are injected through inelastic $pp$ collisions (Eq. (\ref{eq:Qe})). 
A merging activity of the cluster suddenly turns on the reacceleration (Eq. (\ref{eq:Dpp})), and CREs are reaccelerated up to $\sim$1 GeV to form the RH. CRPs are also reaccelerated and power the associated emission of gamma rays and neutrinos (Eqs. (\ref{eq:egamma}), (\ref{eq:enumu}), and (\ref{eq:enue})). We have assumed the radial dependence of the magnetic field (Eq. (\ref{eq:B})) and ICM density (Eq. (\ref{eq:beta-model})), and adopted the best fit parameters from a RM measurement \citep{Bonafede2010}. \par

The detailed nature of turbulence is still unknown. 
We have examined two types of reacceleration: the hard-sphere type ($q=2$) and the Kolmogorov type ($q=5/3$). We adopted $\tau_\mathrm{acc}=260$ Myr and $\tau_\mathrm{acc}=100$ Myr for $q=2$ and $q=5/3$, respectively. 
We have tested two extreme cases for the amount of primary CREs: the secondary dominant model ($f_\mathrm{ep}=0$) and the primary dominant model ($f_\mathrm{ep}=0.01$). The observed radio spectrum and the gamma-ray upper limit give constraints on the duration of the reacceleration $t_R$. The radial dependence of the injection $K(r)$ is constrained by the surface brightness profile of the RH. Note that those quantities are constrained under the assumption that $\tau_{\rm acc}$ and $f_{\rm ep}$ are constant with radius. \par

The main results of this work are summarized below:
\begin{itemize}
\item The secondary dominant models ($f_\mathrm{ep}=0$) with the hard-sphere reacceleration ($q=2$) produce hard synchrotron spectra compared to the observation even for $\alpha=2.45$ (Figure \ref{fig:nFn_H}). That hardness is caused by the energy-dependent diffusion of parent CRPs together with the weak energy dependence in the $pp$ cross section.  \par
\item The CRE distribution is required to be nearly uniform within the RH under the assumption that the reacceleration timescale does not depend on the radius. That requirement disfavors centrally concentrated injections, such as the delta-functional injection from the center. \par
\item The required injection profiles of primary CRs significantly differ between the secondary dominant and primary dominant models. The injection should occur at the edge of the RH in the former case, while the injection itself needs to be uniform in the latter case.\par
\item  The radio spectrum can be reproduced in both hard-sphere and Kolmogorov models by adjusting the value of $t_\mathrm{R}/\tau_\mathrm{acc}$. The Kolmogorov models are more compatible with the secondary dominant scenario. \par
\item Regarding hadronic emission, the most optimistic results are obtained in the case of the hard-sphere reacceleration ($q=2$). Neutrino and gamma-ray fluxes can optimistically be as large as $\sim10^{-10}$ GeV/cm$^2$/s, and the next generation of TeV gamma-ray telescope like LHAASO may detect gamma-rays from the Coma RH.
\item The models with pessimistic assumptions on hadronic emission, such as the primary dominant scenario ($f_\mathrm{ep}=0.01$) or the Kolmogorov reacceleration ($q=5/3$) from the radio constraints can also reproduce the observed radio properties.  
\end{itemize}
 
We discussed the detectability of gamma-ray and hard X-ray with future experiments in Section \ref{subsec:obs}. 
As seen in Figure \ref{fig:Brems}, hard X-ray emission is dominated by free-free emission from thermal electrons below $\sim$ 50 keV. The deviation from the thermal spectrum is pronounced only above $\sim$50 keV, so new instruments with better sensitivities in the 10 -100 keV band are necessary.\par
We have estimated high-energy neutrino and gamma-ray backgrounds from Coma-like clusters. Notably, we have taken into account effects of CRP and CRE reacceleration such that the models are consistent with the radio observations of Coma. 
Our estimate suggests that the contribution from the radio-loud massive galaxy clusters in the local universe is $\lesssim3-30$\% of the observed neutrino intensity, which is consistent with previous results~\citep[e.g.,][]{Fang2016,Murase2016,Hussain:2021dqp}, although a larger contribution may come from lower-mass/higher-redshift clusters.
\par 

\acknowledgments
K.N. acknowledges the support by the Forefront Physics and Mathematics Program to Drive Transformation (FoPM). Numerical computations were in part carried out on Cray XC50 at Center for Computational Astrophysics, National Astronomical Observatory of Japan. This work is supported by the joint research program of the Institute for Cosmic Ray Research (ICRR), the University of Tokyo.
The work of K.M. is supported by NSF Grant No.~AST-1908689, and KAKENHI No.~20H01901 and No.~20H05852. 

\appendix

\section{Inverse-Compton radiation}
In this paper, we adopt the formula for the inverse-Compton radiation given in \citet{1996ApJ...463..555I}, which is accurate enough in both the Thomson and Klein-Nishina regimes. The energies of soft (i.e., CMB) photons, scattered photons and CREs are denoted as $\epsilon_0m_\mathrm{e}c^2$, $\epsilon m_\mathrm{e}c^2$ and $\gamma m_\mathrm{e}c^2$, respectively. The photon production rate can be expressed as
\begin{eqnarray}
q(r,\epsilon)=\int d\epsilon_0n(\epsilon_0)\int d\gamma n_\mathrm{e}(r,\gamma)C(\epsilon,\gamma,\epsilon_0),
\label{eq:qIC}
\end{eqnarray}
where $n_0(\epsilon_0)$ is the number density of CMB photons per $d\epsilon_0$, which is equivalent to that of the black body radiation with the temperature $T=T_0(1+z)$;
\begin{eqnarray}
n_0(\epsilon_0)=\frac{8\pi}{c^3}\left(\frac{m_\mathrm{e}c^2}{h}\right)^3\frac{\epsilon^2_0}{\exp(\frac{\epsilon_0 m_\mathrm{e}c^2}{k_BT})-1},
\end{eqnarray}
where $T_0=2.757$ K, and $z=0.0232$ is the redshift of Coma cluster.
The function $C$ in Eq.(\ref{eq:qIC}) is called the Compton kernel, which is \citep[see][]{PhysRev.167.1159}
\begin{eqnarray}
C(\epsilon,\gamma,\epsilon_0)=\frac{2\pi r_\mathrm{e}^2c}{\gamma^2\epsilon_0}&&\left[2\kappa\ln\kappa+(1+2\kappa)(1-\kappa)+\frac{(4\epsilon_0\gamma\kappa)^2}{2(1+4\epsilon_0\gamma\kappa)}(1-\kappa)\right],
\end{eqnarray}
where $\kappa=\frac{\epsilon}{4\epsilon_0\gamma(\gamma-\epsilon)}$. The emission coefficient can be written as
\begin{eqnarray}
\varepsilon_\nu^{\mathrm{IC}}(\nu,r)=\frac{h}{4\pi}\epsilon q(r,\epsilon),\; \nu=\frac{m_\mathrm{e}c^2}{h}\epsilon.
\label{eq:eIC}
\end{eqnarray}
Also, the momentum loss rate used in Eq. (\ref{eq:Ne}) is written as
\begin{eqnarray}
b_{\mathrm{IC}}(\gamma)=\int d\epsilon_0 \int d\epsilon (\epsilon m_\mathrm{e}c^2) n_0(\epsilon_0)C(\epsilon,\gamma,\epsilon_0).
\label{eq:bIC}
\end{eqnarray}

\section{Pion and secondary neutrino spectra}
In this paper, we adopt the approximate expression for the spectra of pions and neutrinos given in \citet{Kelner2006}. For pion production (Eq. (\ref{eq:Qpi})) from the inelastic $pp$ collision, we apply their QGSJET model, regardless of the energy of pions. The spectrum of pions produced by the $pp$ collision of a CRP of the energy $E_\mathrm{p}$ is approximated as
\begin{eqnarray}
F_\pi(E_\pi,E_\mathrm{p})=4\phi B_\pi x^{\phi-1}\left(\frac{1-x^\phi}{(1+ux^\phi)^3}\right)^4\left(\frac{1}{1-x^\phi}+\frac{3u}{1+ux^\phi}\right)\left(1-\frac{m_\pi}{xE_\mathrm{p}}\right)^{1/2},
\label{eq:QGSJET}
\end{eqnarray}
with the best fit parameters
\begin{eqnarray}
B_\pi=5.58+0.78L+0.10L^2,\\
u=\frac{3.1}{B_\pi^{3/2}},\; \;\; \phi=\frac{0.89}{B_\pi^{1/2}(1-e^{-0.33B_\pi})},
\end{eqnarray}
where $x=E_\pi/E_\mathrm{p}$ and $L=\ln(E_\mathrm{p}/10^3 \mathrm{GeV})$. We use this expression for the pion production by all CRPs above the threshold energy $E_\mathrm{th}\approx1.32$ GeV.\par
The neutrino spectra from the decay of charged ultra-relativistic pions are also given in \citet{Kelner2006}. The spectra of muonic neutrinos from the decay of secondary muon used in Eq. (\ref{eq:enumu}) is written as 
\begin{eqnarray}
f_{\nu_\mu^{(2)}}=g_{\nu_\mu}(x)\Theta(x-\eta)+(h^{(1)}_{\nu_\mu}+h_{\nu_\mu}^{(2)})\Theta(\eta-x),
\end{eqnarray}
where $\eta=1-\zeta=(m_\mu/m_\pi)^2=0.573$, and
\begin{eqnarray}
g_{\nu_\mu}(x)&=&\frac{3-2\eta}{9(1-\eta)^2}(9x^2-6\ln x-4x^3-5),\\
h_{\nu_\mu}^{(1)}(x)&=&\frac{3-2\eta}{9(1-\eta)^2}(9\eta^2-6\ln \eta-4\eta^3-5),\\
h_{\nu_\mu}^{(2)}(x)&=&\frac{(1+2\eta)(\eta-x)}{9\eta^2}\left[9(\eta+x)-4(\eta^2+\eta x+x^2)\right].
\end{eqnarray}
Similarly, for electron neutrinos
\begin{eqnarray}
f_{\nu_\mathrm{e}}=g_{\nu_\mu}(x)\Theta(x-\eta)+(h^{(1)}_{\nu_\mathrm{e}}+h_{\nu_\mathrm{e}}^{(2)})\Theta(\eta-x),
\end{eqnarray}
where
\begin{eqnarray}
g_{\nu_\mathrm{e}}(x)&=&\frac{2(1-x)}{3(1-\eta)^2}\left[6(1-x)^2+\eta(5+5x-4x^2)+\frac{6\eta\ln x}{1-x}\right],\label{eq:gnue}\\
h_{\nu_\mathrm{e}}^{(1)}(x)&=&\frac{2}{3(1-\eta)^2}\left[(1-\eta)(6-7\eta+11\eta^2-4\eta^3)+6\eta\ln \eta\right],\\
h_{\nu_\mathrm{e}}^{(2)}(x)&=&\frac{2(\eta-x)}{3\eta^2}(7\eta^2-4\eta^3+7x\eta-4x\eta^2-2x^2-4x^2\eta).
\end{eqnarray}
Note that a minor typo in Eq. (\ref{eq:gnue}) in \citet{Kelner2006} is fixed here.

\bibliography{Coma}{}

\bibliographystyle{aasjournal}

\end{document}